\documentclass{jfm}
\usepackage{graphicx}
\usepackage{epstopdf, epsfig}

\usepackage{setspace} 
\usepackage{graphicx}
\usepackage{amssymb}
\usepackage{epstopdf}
\usepackage{color,graphicx}
\usepackage{color, xcolor}
\usepackage{tikz}
\usetikzlibrary{shapes}
\usepackage{psfrag}
\usepackage{subfigure}
\usepackage{amsmath}
\usepackage{adjustbox}
\usepackage{fancyhdr}
\usepackage{fancybox}
\usepackage{wrapfig}
\usepackage[version=4]{mhchem}
\usepackage{tabularx}
\usepackage{array} 
\usepackage{multirow}

\pdfpageattr {/Group << /S /Transparency /I true /CS /DeviceRGB>>} 

\def\be{\begin{equation}}
\def\ee{\end{equation}}
\def\beno{\begin{eqnarray}}
\def\eeno{\end{eqnarray}}
\def\beqno{\begin{eqnarray*}}
\def\eeqno{\end{eqnarray*}}

\def\del{\partial}

\newcommand{\blkcircle}{\raisebox{0.5pt}{\protect\tikz{\protect\node[draw,scale=0.7,circle,fill=none](){};}}}
\newcommand{\darkgraycircle}{\raisebox{0.5pt}{\protect\tikz{\protect\node[draw=black!60,scale=0.7,circle,fill=none](){};}}}
\newcommand{\graysquare}{\raisebox{0.5pt}{\protect\tikz{\protect\node[draw=gray,scale=0.7,regular polygon, regular polygon sides=4,fill=none,rotate=0](){};}}}
\newcommand{\redsquare}{\raisebox{0.5pt}{\protect\tikz{\protect\node[draw=red,scale=0.7,regular polygon, regular polygon sides=4,fill=none,rotate=0](){};}}}
\newcommand{\redtriangle}{\raisebox{0.5pt}{\protect\tikz{\protect\node[draw=red,scale=0.6,regular polygon, regular polygon sides=3,fill=none,rotate=0](){};}}}
\newcommand{\blktriangle}{\raisebox{0.5pt}{\protect\tikz{\protect\node[draw,scale=0.6,regular polygon, regular polygon sides=3,fill=none,rotate=0](){};}}}
\newcommand{\blkdiamond}{\raisebox{0pt}{\protect\tikz{\protect\node[draw,scale=0.6,diamond,fill=none](){};}}}

\newcommand{\blkline}{\raisebox{2pt}{\protect\tikz{\protect\draw[-,black!40!black,solid,line width = 1.0pt](0,0) -- (3.0mm,0);}}}
\newcommand{\greenline}{\raisebox{2pt}{\protect\tikz{\protect\draw[-,green!40!green,solid,line width = 1.0pt](0,0) -- (3.0mm,0);}}}
\newcommand{\redline}{\raisebox{2pt}{\protect\tikz{\protect\draw[-,red!40!red,solid,line width = 1.0pt](0,0) -- (3.0mm,0);}}}
\newcommand{\cyanline}{\raisebox{2pt}{\protect\tikz{\protect\draw[-,blue!0.5!cyan,fill=blue!0.5!cyan,solid,line width = 1.0pt](0,0) -- (3.0mm,0);}}}

\newcommand{\blackdottedline}{\raisebox{2pt}{\protect\tikz{\protect\draw[-,black!40!black,dotted,line width = 1.2pt](0,0) -- (3.0mm,0);}}}
\newcommand{\reddottedline}{\raisebox{2pt}{\protect\tikz{\protect\draw[-,red!40!red,dotted,line width = 1.2pt](0,0) -- (3.0mm,0);}}}
\newcommand{\blkdashdotline}{\raisebox{2pt}{\protect\tikz{\protect\draw[-,black!40!black, dashdotted,line width = 1.1pt](0,0) -- (4.0mm,0);}}}
\newcommand{\reddashdotline}{\raisebox{2pt}{\protect\tikz{\protect\draw[-,red!40!red, dashdotted,line width = 1.1pt](0,0) -- (4.0mm,0);}}}
\newcommand{\blkdashline}{\raisebox{2pt}{\protect\tikz{\protect\draw[-,black!40!black, dashed,line width = 1.1pt](0,0) -- (3.0mm,0);}}}
\newcommand{\bluedashline}{\raisebox{2pt}{\protect\tikz{\protect\draw[-,blue!40!blue, dashed,line width = 1.1pt](0,0) -- (3.0mm,0);}}}
\newcommand{\grayline}{\raisebox{2pt}{\protect\tikz{\protect\draw[-,gray!40!gray,solid,line width = 1.0pt](0,0) -- (3.0mm,0);}}}
\newcommand{\graydottedline}{\raisebox{2pt}{\protect\tikz{\protect\draw[-,gray!40!gray,dotted,line width = 1.2pt](0,0) -- (3.0mm,0);}}}
\newcommand{\graydashdotline}{\raisebox{2pt}{\protect\tikz{\protect\draw[-,gray!40!gray, dashdotted,line width = 1.1pt](0,0) -- (4.0mm,0);}}}
\newcommand{\graydashline}{\raisebox{2pt}{\protect\tikz{\protect\draw[-,gray!40!gray, dashed,line width = 1.1pt](0,0) -- (3.0mm,0);}}}

\shorttitle{Turbulent flow over permeable and impermeable sediment beds}
\shortauthor{S. K. Karra, S. V. Apte, X. He and T. D. Scheibe}

\title{Pore-resolved simulations of turbulent boundary layer flow over permeable and impermeable sediment beds
}
\author{Shashank K. Karra\aff{1},
 Sourabh V. Apte\aff{1} \corresp{\email{sourabh.apte@oregonstate.edu}}, \\ 
 Xiaoliang He\aff{2}, \and Timothy D. Scheibe\aff{2}\\}

\affiliation{\aff{1}School of Mechanical, Industrial and Manufacturing Engineering, Oregon State University, Corvallis, OR, 97331
\aff{2}Pacific Northwest National Laboratory, Richland, WA, 99354}

\begin{document}

\maketitle

\begin{abstract}
Pore-resolved direct numerical simulations of turbulent open channel flow are performed comparing the structure and dynamics of turbulence over impermeable rough and smooth walls to a porous sediment bed at permeability Reynolds number ($Re_K$) of 2.6, representative of aquatic beds.
Four configurations are investigated; namely, (i) permeable bed with randomly packed sediment grains, (ii) an impermeable-wall with full layer of roughness elements matching the top layer of the sediment bed, (iii) an impermeable-wall with half layer of roughness elements , and (iv) a smooth wall. A double-averaging methodology is used to compute the mean velocity, Reynolds stresses, form-induced stresses, and turbulent kinetic energy budget. It is observed that the mean velocity, Reynolds stresses, and form-induced pressure-velocity correlations representing upwelling and downwelling fluxes are similar in magnitude for the permeable-bed and impermeable-full layer wall cases.  However, for the impermeable-half layer case, the wall blocking effect results in 
higher streamwise and lower wall-normal stresses 
compared to the permeable-bed case. Bed roughness increases Reynolds shear stress whereas permeability has minimal influence. Bed permeability, in contrast to Reynolds stresses, significantly influences form-induced shear stress. 
Bed shear stress statistics show that probability of extreme events increases in permeable bed and rough wall cases as compared to the smooth wall. Findings suggest that bed permeability can have significant impact on modeling of hyporheic exchange and its effect on bed shear and pressure fluctuations are better captured by considering at least the top layer of sediment.

\end{abstract}

\section{Introduction}\label{sec:intro}
The interchange of mass and momentum between surface water and ground water occurs in the porous bed beneath the streams, termed as the hyporheic zone. Hyporheic transient storage or retention and transport of solutes such as chemicals and pollutants, dissolved oxygen, nutrients, and heat across the sediment-water interface (SWI) is one of the most important concepts for stream ecology, and has enormous societal value in predicting source of fresh drinking water, transport and biogeochemical processing of
nutrients, and sustaining diverse aquatic ecosystems~\citep{bencala1983rhodamine,d1993transient,valett1996parent,harvey1996evaluating,anderson2008groundwater,briggs2009method,grant2018hypor}.

A broad range of spatio-temporal scales corresponding to disparate physical and chemical processes contribute to mixing within the hyporheic zone~\citep{hester2017importance}. Turbulent transport across the sediment-water interface, coherent flow structures, and non-Darcy flow within the sediment bed have been hypothesized as critical mechanisms impacting transient storage. However, fundamental understanding of turbulence over a {\it stationary permeable bed} is still in its infancy. Turbulence characteristics over a permeable bed are different compared to a rough, impermeable wall~\citep{jimenez2004turbulent}; impacting long time-scales of retention within the bed. The importance of penetration of turbulence within the bed and near bed pressure fluctuations are crucial and their impact on the hyporheic transient storage is poorly understood~\citep{hester2017importance}. 

\begin{figure}
   \centering
     \subfigure[]{
   \includegraphics[width=8cm,height=5cm,keepaspectratio]{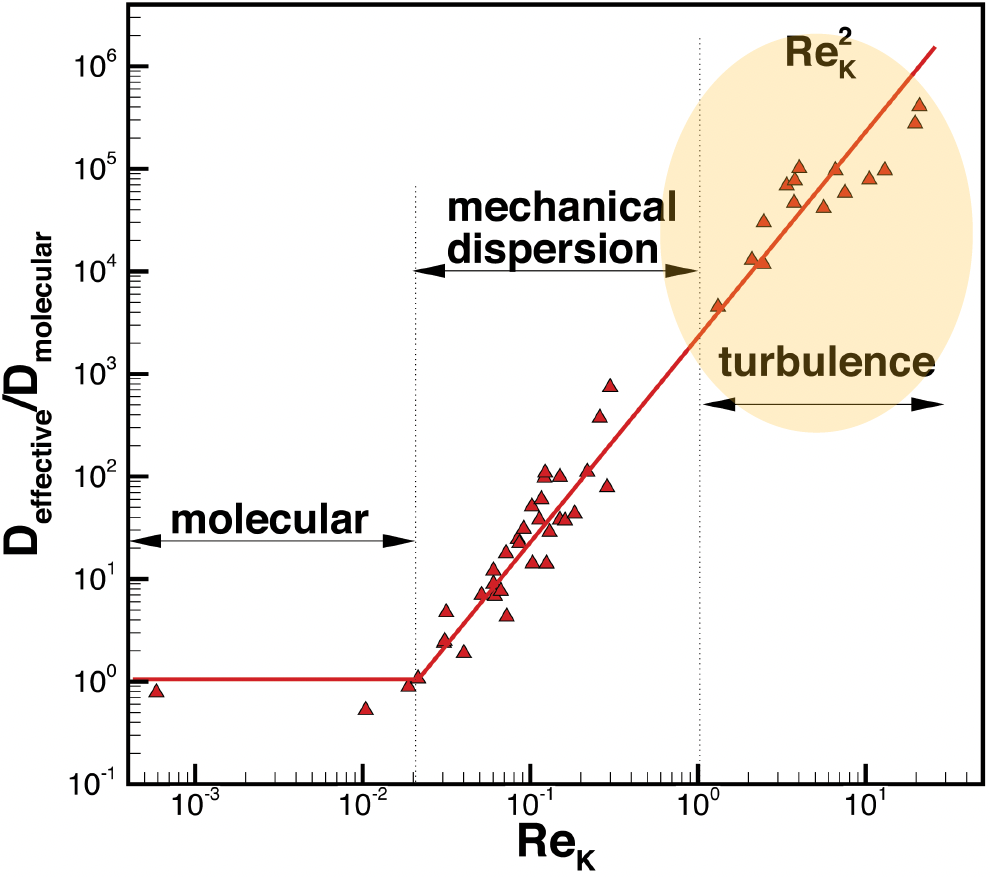}}
      \subfigure[]{
   \includegraphics[width=7cm,height=5cm,keepaspectratio]{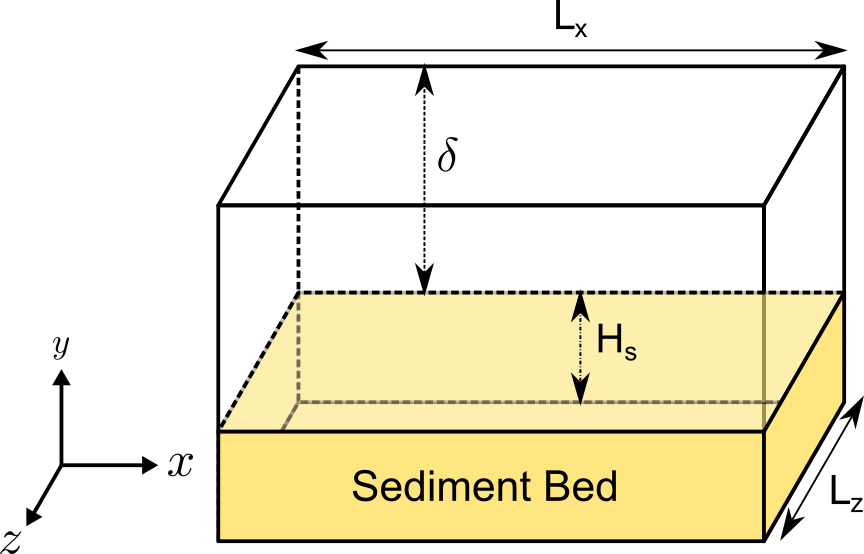}}
\caption{\small (a) Effective dispersion coefficient versus $Re_{k}$~(based on~\citet{voermans2017variation,grant2018hypor}), (b) schematic of permeable bed/impermeable rough wall schematic showing height of roughness element layers, $H_s$, and boundary layer thickness, $\delta$.}
\label{fig:regime_bedschem}
\end{figure}

Mass and momentum transport in turbulent boundary layer over a naturally occurring permeable sediment bed (figure~\ref{fig:regime_bedschem}b) are characterized by bed permeability, $K$ (which depends upon its porosity and average grain size according to the Carman-Kozeny relation), sediment bed arrangement (flat versus complex bedforms), bulk velocity, $U_b$, friction velocity, $u_{\tau}$, boundary layer thickness, $\delta$, sediment depth, $H_s$, size of the roughness element, $k_s$ ($k_s/\sqrt{K}\approx 9$ for monodisperse, spherical particles~\citep{wilson2008grain,voermans2017variation}. Free surface flow and waves typically do not affect the hyporheic exchange in natural stream and river flows under subcritical conditions with small Froude numbers. Different Reynolds numbers can be used to characterize the flow: $Re=U_b \delta/\nu$, $Re_{\tau}=u_{\tau}\delta/\nu$, and permeability Reynolds number, $Re_K = u_{\tau} \sqrt{K}/\nu$~\citep{voermans2017variation}. The permeability Reynolds number, representing the ratio between the permeability scale to the viscous scale ${\sqrt{K}}/({\nu/u_{\tau}})$, is typically used to identify different flow regimes based on the dominant transport mechanisms across the SWI. The three flow regimes characterized by~\citet{voermans2017variation,voermans2018model,grant2018hypor} can be summarized as (figure~\ref{fig:regime_bedschem}a):
(i) the molecular regime $Re_K<0.01$, where the bed is nearly impermeable and the transport is governed by molecular diffusion; (ii) the dispersive regime $0.01<Re_K<1$, where dispersive transport associated with laminarization of stream turbulence is important; and (iii) the turbulent regime $Re_K>1$, where turbulence is dominant near the highly permeable interface.
Bedforms can induce static and dynamic pressure variations with vertical dispersive mixing due to flow separations; however, in the present work, only flat beds with random arrangement of sediment grains is investigated.
Aquatic beds typically yield $1\leq Re_K < 10$ with turbulence penetrating the interfacial layer even for flat beds. Based on data collected from several local streams and rivers near Oregon State University~\citep{jackson2013fluid,jackson2015flow}, it is found that the gravel grain sizes varied over the range 5--70mm, with friction velocity, $u_{\tau}=0.004$--$0.088~\rm m/s$, and $Re_K=2$--70~(figure~\ref{fig:regime_bedschem}a). Those observations provide the basis for the parameters used in the present work.

A few experimental and numerical studies in the literature have evaluated turbulence characteristics over a permeable bed. \citet{zagni1976channel} conducted experiments with open channel flows over permeable beds composed of spheres. They varied bed permeability by varying the size and grading of the bed material. They found that friction factor was higher for permeable beds compared to impermeable ones with the same roughness characteristics. Similar observations were made by \citet{zippe1983turbulent}, who conducted wind tunnel experiments on boundary layer flow over smooth walls, permeable, and non-permeable rough beds with same roughness characteristics in order to isolate the effect of bed permeability on turbulence characteristics. Rise in friction factor was observed for permeable beds which was attributed to additional energy dissipation caused by exchange of momentum across the SWI. \citet{manes2009turbulence} studied the interactions between surface and subsurface flows in an open channel for bottom sediment layers arranged in a uniform cubic pattern with (i) a single layer of spheres above an impermeable wall, and (ii) five layers of spheres, considered to be a permeable bed. The experiments were carried out at $Re_K$ of 31.2 and 44.6. Apart from increase in flow resistance they also found an intense transport of turbulent kinetic energy (TKE) occurring from the surface to subsurface flow. This transport of TKE was hypothesized to be mainly driven by pressure fluctuations; however, a direct correlation was not obtained as it is difficult to measure pressure field in experimental setups.

\citet{kim2020experimental} also investigated, through experimental observations, the dynamic interplay between surface and subsurface flow in the presence of a smooth permeable wall and rough permeable wall, composed of a uniform cubic arrangement of packed spheres. The experiments where carried out at $Re_K$ of 50, corresponding to highly turbulent transport regime (figure~\ref{fig:regime_bedschem}a). They confirmed the existence of amplitude modulation in a turbulent boundary layer over a permeable wall, a phenomenon typically identified in impermeable boundaries, whereby the outer large scales modulate the intensity of the near-wall small scale turbulence. They also found that bed roughness and permeability intensify the strength and penetration of flow into the permeable bed, modulated by large-scale structures in the surface flow, and linked to possible roughness-formed channelling effects and shedding of smaller-scale flow structures from the roughness elements. Based on the observation that the form induced Reynolds stress, $\langle\widetilde{u}\widetilde{v}\rangle^{+}$, decays quickly within the permeable beds, they postulated that amplitude modulation of subsurface flow may be driven by large-scale pressure fluctuations at the interface that are generated by the passage of large-scale motions
in the log region of the surface flow. 

\citet{voermans2017variation} studied the influence of different $Re_K$ on the interaction between surface and subsurface flows at the SWI of a synthetic sediment composed by randomly-arranged mono-dispersed spheres. Their experiments covered a wide range of $Re_K=0.36$--$6.3$. The results demonstrated a strong relationship between the structure of the mean and turbulent flow at the SWI and $Re_K$. Hydrodynamic characteristics, such as the interfacial turbulent shear stress, velocity, turbulence intensities and turbulence anisotropy tend towards those observed in flows over impermeable boundaries as  $Re_K \longrightarrow 0$ and towards those seen in flows over highly permeable boundaries as $Re_K \longrightarrow \infty$. $Re_K \sim 1-2$ was found to be the threshold beyond which the turbulent shear stress starts to dominate the total shear stress at the interface and the turbulent penetration into the sediment becomes comparable to that of the mean flow. Their results suggest that both the mean flow penetration also known as the Brinkman layer thickness and turbulent shear penetration are a function of $Re_K$. 

A few direct numerical studies (DNS) of turbulent boundary layers over permeable flat beds or porous walls have also been performed.
\citet{breugem2006influence} performed DNS of turbulent flow in channel bounded by a smooth wall at the top and a permeable wall, defined by a packed bed, at the bottom to study the influence of wall permeability on the structure and dynamics of turbulence for $Re_K = 0$--$9$. The flow inside their permeable bed was described by means of volume-averaged Navier-Stokes equations. For a highly permeable wall longitudinal low and high speed streaks, which are common near a smooth wall, were absent. This resulted in a decrease in peak of streamwise intensity and increase in spanwise and wall-normal intensities. They also observed that an exchange of momentum between flow in the channel and the top layer of the permeable wall induces a strong increase in Reynolds-shear stress relative to the smooth wall resulting in an increase in skin friction coefficient for a highly permeable wall compared to an impermeable wall. \citet{kuwata2016lattice} studied the effects of wall permeability and roughness in porous rough walls formed by interconnected staggered cube arrays with a single and multiple layers of cube arrays. Presence of pressure perturbation with alternating patterns of crests and troughs induced by the Kelvin–Helmholtz instability over the porous and rough walls was observed. This intensified the wall-normal and spanwise velocity fluctuations leading to a strong turbulent shear. \citet{singh2007numerical} conducted DNS simulations of turbulent flow over a rough impermeable bed composed of hexagonal arrangement of spheres. Data from their DNS simulation confirmed the wall similarity hypothesis of \citet{raupach1991rough}, which states that above the roughness layer, rough bed flow is structurally similar to the smooth wall. They were able to observe vigorous turbulent mixing between the fluid layer trapped below the midplane of the roughness elements and fast moving fluid in the channel.

In addition, \citet{shen2020direct} conducted pore-resolved DNS studies for turbulent flow over a (i) uniform or regular grain packing at the sediment-water interface, and (ii) random grain packing at the sediment-water interface to isolate the impact of bed roughness, from that of permeability, on mean flow and turbulence structure. In both cases the layers underneath the top layer were arranged in a random packing. Comparing the results for the two packings, at $Re_K$ = 2.56, they found that the random interface produces intense form-induced or dispersive pressure fluctuation, $\widetilde{p}^{+}$, resulting in higher vertical form-induced velocity fluctuation, $\widetilde{v}^{+}$, contributing to strong vertical flux of momentum across the SWI. More intense mixing was observed near the random interface due to increased Reynolds and form-induced stresses which resulted in a deeper penetration of turbulence (44\% higher) than the uniform interface case. 


Although there have been a few experimental and computational studies analyzing turbulent boundary layer flows over randomly distributed sediment beds, there is lack of detailed studies and data directly comparing turbulence structure over randomly arranged permeable beds to an impermeable, rough surface with similar roughness characteristics as the top layer of the permeable bed.
To authors best knowledge such realistic sediment particle arrangements have not been used in prior studies to compare turbulent flow statistics between permeable beds and impermeable rough walls, which is one of the main goals of the present study. 

In reach-scale modeling of hyporheic exchange in rivers and streams, the turbulent surface flow and underlying ground-water flow in porous media are typically modeled separately in a sequential manner~\citet{chen2018hyporheic}, or through a coupled approach~\citet{li2020flexible}. For such very large scale studies, the surface flow is obtained from Reynolds-averaged Navier-Stokes (RANS) model with the bottom bed topography treated as a {\it smooth}, impermeable wall. The mean pressure and turbulent kinetic energy distributions at the SWI obtained from the surface flow simulation are then used as boundary conditions at the top surface of the ground-water flow model domain in a decoupled, sequential approach. However, such modeling approach neglects any effects on the pressure and TKE from the rough permeable bed. Even for nearly flat beds, it is hypothesized that the hydrodynamics at the SWI will be substantially different for a smooth impermeable wall compared to that at the rough, permeable bed. The residence time calculations based on the above sequential approach can thus be inaccurate. 
On the other hand, in a fully-coupled modeling approach, the surface flow computations and the ground-water flow models are strongly coupled with iterative updates of pressure and mass flux across the interface to ensure flux consistency. However, such coupled approach can be computationally intensive. \citet{chen2022modeling} in their large scale computational modeling study highlighted the importance of calibrating sediment bed roughness in accurately estimating water surface elevation, which is used to estimate river discharge over a period of months and years. Pore scale experimental measurements~\citep{manes2009turbulence, voermans2017variation, kim2020experimental} have shown that both bed roughness and permeability enhance flow fluctuations (and hence pressure fluctuations) at the SWI. 

Therefore, it can be stated that prediction of pressure distribution and flux exchange of a reach-scale, sequential computational model can be significantly improved by considering the sediment-grain scale surface roughness and bed permeability. Characterizing the differences and similarities of turbulent pressure fluctuations and bed shear stress over a rough, impermeable and permeable beds is thus a critical first step in improved modeling of reach scale transport. In addition, the challenge remains on the thickness of the sediment layers that should be considered, whereby the trade-off between accuracy and computational expenses can be satisfied. From prior experimental and numerical studies~\citep{goharzadeh2005transition, shen2020direct,fang2018influence} it can be hypothesized that both mean flow and shear stress penetration depth is {$\mathcal O$}(1-2) times the diameter of the sediment roughness elements, especially for $Re_K=1$--10. Hence, the second goal of this study is to quantify the influence of the thickness of sediment layers on turbulence flow statistics and characterize the differences in turbulence structure and pressure fields at the SWI between rough impermeable and permeable flat beds.

The rest of the paper is organized as follows. Section $\S$\ref{sec:meth} summarizes the mathematical formulation and simulation setup for the present work. Detailed description of the results and data analysis including validation against experimental data is provided in $\S$\ref{sec:res} followed by major findings of the present work in $\S$\ref{sec:conclusions}.
 
\section{Simulation details}\label{sec:meth}
\begin{figure}
   \centering
   \subfigure[PB]{
   \includegraphics[width=10cm,height=10cm,keepaspectratio]{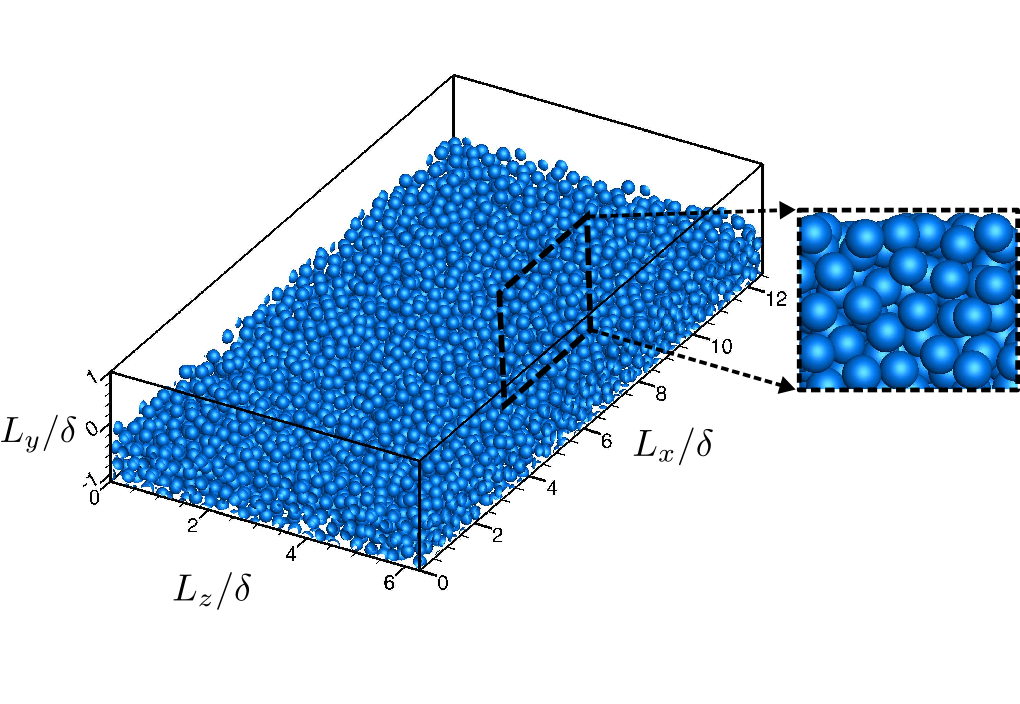}}
      \subfigure[IWF]{
    \includegraphics[width=6.5cm,height=6.5cm,keepaspectratio]{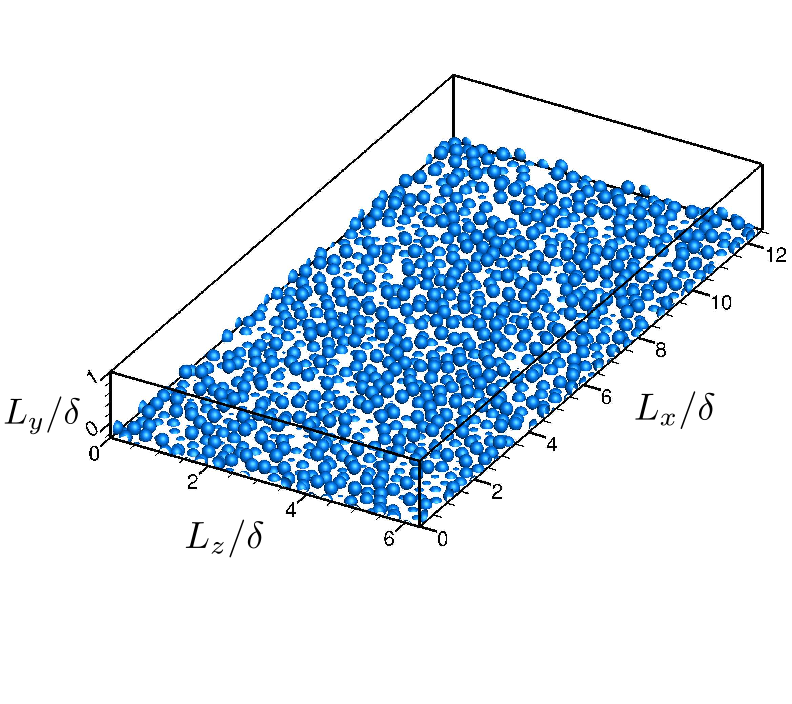}}    
      \subfigure[IWH]{
    \includegraphics[width=6.5cm,height=6.5cm,keepaspectratio]{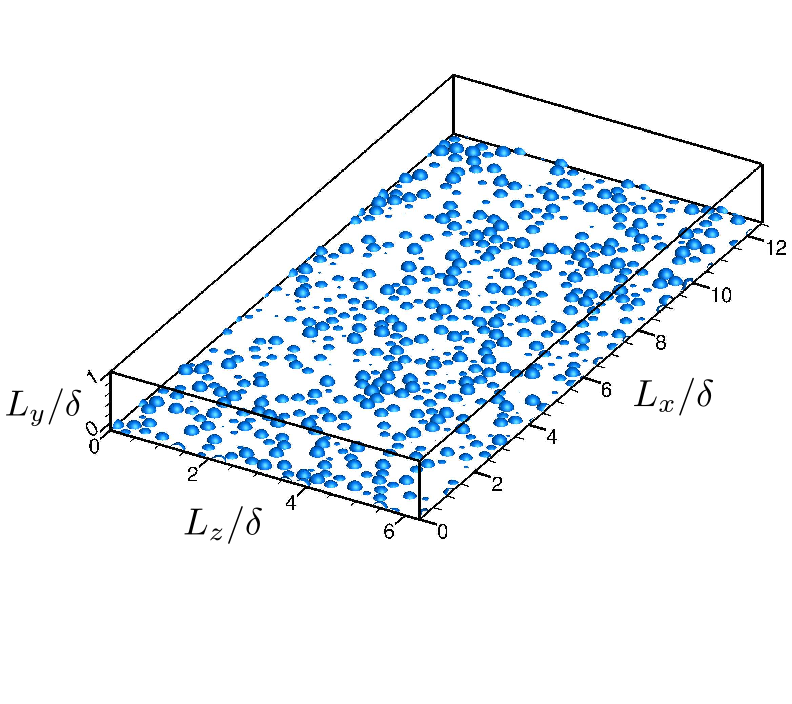}}
\caption{\small Computational domains used in present study: (a) permeable bed with four layers of sediment particles (inset shows close-up view in xy-plane),  (b) impermeable-full layer defined by matching the top layer of the permeable bed, and (c) impermeable-half-layer defined by matching the top half layer of the permeable bed.} 
\label{fig:dom}
\end{figure}

In the present work, emphasis is placed on characterizing momentum transport across the sediment-water interface over a flat, randomly packed porous bed by contrasting against the turbulent flow over rough, impermeable walls with roughness characteristics derived from the top layer of the porous sediment bed.
The computational domain for the permeable sediment bed (PB) consists of a doubly periodic domain with four layers of randomly packed, mono-dispersed sediment grains at the bottom to capture the turbulence penetration and unsteady, inertial flow (figure \ref{fig:dom}a). The random packing of mono-dispersed, spherical particles within a doubly periodic box is generated using the code developed by~\citet{dye2013description}.


To characterize the impact of bed permeability on near-bed turbulence and momentum transport, two rough, impermeable wall configurations are investigated: (i) the roughness obtained by matching the entire top layer of the sediment bed with full spherical particles (IWF: impermeable full-layer - figure \ref{fig:dom}b) and (ii) roughness obtained by matching only half of the top layer of the sediment bed (IWH: impermeable half-layer - figure \ref{fig:dom}c). In other words the IWF case has one full layer of spherical sediment particles, while the IWH case has only half a layer of hemi-spherical sediment particles, with impermeable solid wall underneath (detailed synthesis of the simulation domains is given in section~\ref{sec:simdom_synth}). Details of the flow configurations, flow parameters, computational approach, and grid resolutions used in the present work are described below.


\subsection{Simulation domain and parameters}\label{sec:sim_param}
Turbulent flow over a permeable bed can be characterized by the permeability Reynolds number ($Re_K$), the turbulent Reynolds number ($Re_{\tau}$), the ratio of sediment depth to the free-surface height ($H_s/\delta$), the ratio of the sediment grain diameter to the free-surface height ($D_p/\delta$), bed porosity ($\theta$), and the domain lengths in the axial and spanwise directions normalized by the free-surface height ($L_x/\delta$, $L_z/\delta$). Table~\ref{tab:cases1} shows detailed simulation parameters for the cases used to investigate the structure and dynamics of turbulence over porous sediment bed and impermeable rough walls. Two permeable bed cases are simulated; case VV and case PB. Case VV is used to verify and validate the DNS simulations of turbulent boundary layer flow over a sediment bed with experimental data from~\citet{voermans2017variation}. Permeable bed case with parameters porosity ($\theta) = 0.41$, $Re_K = 2.56$ and $Re_{\tau} \sim 180$ match those of Case L12 in~\citet{voermans2017variation}, and is within range for realistic aquatic sediments beds. A baseline smooth wall domain (SW) is also simulated in addition to the permeable bed and rough wall cases. The free surface height, $\delta$, for the PB, IWF and IWH cases is set be 3.5$D_{p}$ and is similar to to the experimental domains of~\citet{voermans2017variation, manes2009turbulence} and numerical simulation domains of~\citet{fang2018influence, bomminayuni2011turbulence}. 

The grid resolutions required for these configurations are based on two main considerations: (i) minimum bed-normal grid resolution near the bed, and (ii) minimum resolution required to capture flow over spherical particles. For DNS of boundary layers, the bed-normal grid resolution in wall units should be $\Delta y^{+}<1$, in order to accurately capture the bed shear stress in the turbulent flow. The grid resolutions in the axial and spanwise directions are typically 3--4 times coarser, following the smooth channel flow simulations by~\citep{moser1999direct}. Note that, the roughness features and permeability are known to break the elongated flow structures along the axial direction in smooth walls, reducing the inhomogeneity in the near-bed region~\citep{ghodke2016dns}. This flow feature is thus anticipated to alleviate the bed-normal grid resolution requirement near the bed. To capture the inertial flow features within the pore and around spherical particles, grid refinement studies conducted on flow over single sphere (not shown) are used. Accordingly, roughly 180 grid points are used in the bed-normal direction within each sediment particle in the top two layers. In the x and z directions, uniform grid with 26 grid points are used to resolve the grain geometry  for the permeable bed ($Re_K$ = 2.56) and impermeable rough wall cases. Effect of uniform, but non-cubic grids within the sediment bed was thoroughly evaluated by comparing drag coefficients to those obtained from cubic grids over a single and a layer of particles to show no discernible differences (See appendix A).


\begin{table}
\begin{center}
\caption{Parameters used in present pore-resolved direct numerical simulations where, $D_p$ is the sphere diameter, $\delta$ is the free surface height, $H_s$ is the sediment depth, and $\theta$ is the porosity.}
\def~{\hphantom{0}}
\begin{tabular}{@{}lc c c c c c c c c c}
Case & Domain &$Re_K$ & $Re_{\tau}$ & $Re_{b}$ & $\theta$ & $H_s/{\delta}$ & $D_p/{\delta}$ & ${(L_x, L_z)/ \delta}$
              & ($\Delta x^{+}, \Delta y^{+}, \Delta z^{+}) $\\ 
VV & permeable &2.56 &180  &1,886 &0.41 &1.71 &0.43 &($4\pi$,$2\pi$) &(3.01, 0.95, 3.01)  \\
PB & permeable &2.56 &270  &2,826 &0.41 &1.14 &0.29 &($4\pi$,$2\pi$) &(3.01, 0.95, 3.01)  \\
IWF & impermeable  &- &270  &2,818 &- &0.29 &0.29 &($4\pi$,$2\pi$) &(3.01, 0.95, 3.01)  \\
IWH & impermeable  &- &270  &3,338 &- &0.14 &0.29 &($4\pi$,$2\pi$) &(3.01, 0.95, 3.01)  \\
SW & impermeable  &- &270 &4,461 &- &- &- &($4\pi$,$2\pi$) &(3.01, 0.95, 3.01)   \\
\end{tabular}
\label{tab:cases1}
\end{center}
\end{table}

Deep inside the bed nearly uniform grid ($\Delta x$ = $\Delta y$ = $\Delta z$) is used as the particle Reynolds number in the bottom layers of sediment decreases significantly. From the crest of the top sediment layer, the grid is stretched, coarsening it gradually towards the free surface using a standard hyperbolic tangent function~\citep{moser1999direct}. 

For the impermeable, rough boundary cases, the number of grid points decrease because only the top (and half of the top layer) of the sediment bed is used above an impermeable wall. Based on these grid resolutions, the total grid count for the permeable bed case is $\sim$ 232 million cells, for the impermeable full-layer case is $\sim$ 158 million cells, for the impermeable half-layer case is $\sim$ 130 million cells, and for the smooth wall case is $\sim$ 107 millions cells.

The flow in the permeable bed and impermeable rough wall simulations is driven by a constant mass flow rate. A target mass flow rate is adjusted until the friction velocity, $u_{\tau}$, which results in the required $Re_K$ is obtained. $Re_{\tau}$ is then calculated based on the free surface height $\delta$. It should be noted that the friction velocity, $u_{\tau}$, is calculated from the maximum value of the double averaged total stress equation~\ref{eq:utau}, which is defined as sum of the viscous shear stress, turbulent shear stress and the form-induced shear stress~\citep{voermans2018model},
\begin{align}
     \tau(y) &= \rho\nu\del(\theta\langle\overline{u}\rangle)/\del{y} - \rho\theta\langle\overline{u^{\prime}v^{\prime}}\rangle - \rho\theta\langle\widetilde{u}\widetilde{v}\rangle.
    \label{eq:utau}
\end{align}

Following smooth wall DNS studies by~\citet{moser1999direct}, about 20 flow-through times (computed as the length over average bulk velocity $L_x/U_{avg}$) is needed for the turbulent flow to reach stationary state~\citep{moser1999direct}. Once a stationary flow field is obtained, computations are performed for another 25 flow-through times to collect single-point and two-points statistics, giving total of 45 flow-through times for the permeable and impermeable cases. 

\subsection{Double averaging (DA) procedure}\label{sec:DA_proc}
Since the flow properties are highly spatially heterogeneous near rough boundaries, double averaging procedure~\citep{raupach1981turbulence} is used, wherein spatial averaging is performed along with time averaging, 
\begin{align}
     \phi(x,t) &= \langle{\overline{\phi}}\rangle(y) + \widetilde{\phi}(x) + {\phi^{\prime}}(x,t),
    \label{eq:r_m}
\end{align}
where $\phi$ is an instantaneous flow variable, $\langle{\phi}\rangle$ is the intrinsic spatial average in the $(x,z)$ plane, $\langle{\phi}\rangle = 1/A_f \int_{A_f} \phi dA$ (where $A_f$ is the area occupied by the fluid), $\overline{\phi}$ is the temporal average, $\phi^{\prime} = \phi - \overline{\phi}$  is the instantaneous turbulent fluctuation and $\widetilde{\phi} =  \overline{\phi} - \langle\overline{\phi}\rangle $ is the form-induced or dispersive fluctuation. Accordingly, Reynolds stresses, turbulent kinetic energy budgets, pressure fluctuations and other flow statistics are computed using the double averaging procedure.

\subsection{Numerical method}
The numerical approach is based on a fictitious  domain  method  to  handle  arbitrary  shaped  immersed  objects without  requiring  the  need  for  body-fitted  grids~\citep{apte2009frs}. Cartesian grids are used in the entire simulation domain, including both fluid and solid phases. An additional body force is imposed on the solid part to enforce the rigidity constraint and satisfy the no-slip boundary condition. The absence of highly skewed unstructured mesh at the bead surface has been shown to accelerate the convergence and lower the uncertainty~\citep{finn2013relative}. 
The following governing equations are solved over the entire domain, including the region within the solid bed, and a rigidity constraint force, $\bf f$, is applied that is non-zero only in the solid region. The governing equations are given as:
	\begin{align}
		\nabla\cdot{\bf u} &= 0, \label{eq:NSa} \\
		\rho_f \bigg[\frac{\partial {\bf u}}{\partial t} + \left({\bf u}\cdot \nabla\right) {\bf u}	\bigg] &= 
		-\nabla p + \mu_f \nabla^2{\bf u} + {\mathbf f} \:, 
	\end{align}
		\label{eq:NSb}
	where $\bf u$ is the velocity vector (with components given by ${\bf u}=(u_x,u_y,u_z)$, $\rho_f$ the fluid density, $\mu_f$ the fluid dynamic viscosity, and $p$ the pressure. A fully parallel, structured, collocated grid solver has been developed and thoroughly verified and validated for a range of test cases including flow over a cylinder and sphere for different Reynolds  numbers, flow over touching spheres at different orientations, flow developed by an oscillating cylinder, among others.
The details of the algorithm as well as very detailed verification and validation studies have been published elsewhere~\citep{apte2009frs}. The solver was used to perform direct one-to-one comparison with a body-fitted solver with known second-order accuracy for steady inertial, unsteady inertial, and turbulent flow through porous media~\citep{finn2013relative} to show very good predictive capability. It has also been recently used for direct simulations of oscillatory, turbulent boundary layer over a sediment layer~\citep{ghodke2016dns,ghodke2018roughness}, and pore-resolved simulations of turbulent flow within a porous unit cell with face-centered cubic packing~\citep{he2018angular,he2019characteristics}.

\subsection{Simulation domain synthesis}\label{sec:simdom_synth}
The double averaged bed porosity profile is shown in figure~\ref{fig:bed_pdf}. 
Following the methodology by~\citet{shen2020direct} the surface roughness characteristics of the top layer of the permeable sediment bed, the impermeable full-layer domain and the impermeable half-layer domain are defined by the local roughness element height $h(x,z)$. This roughness element height is defined as the maximum $y$ location (highest point) on the surface defining the roughness element. Based on this definition of roughness height the total count of roughness elements present in the impermeable full-layer domain (as well the top layer of the permeable bed) is 1,154. Whereas the total count of roughness elements in the impermeable half-layer domain is 585. 

The roughness height fluctuations, $h^{\prime}(x,z)$, defined as the local roughness element height $h(x,z)$ minus its plane averaged value normalized by $\delta$ is plotted in figures \ref{fig:bed_pdf}a and \ref{fig:bed_pdf}b as a scatter plot for both IWF and IWH cases. Since the roughness elements in IWF case are defined by matching the top layer of the permeable bed, its surface roughness characteristics match the PB case (not shown). Each dot in scatter plots represents a roughness element (dots are not to actual size). The probability density functions $pdf$ of roughness height fluctuations are shown in figures \ref{fig:bed_pdf}c and \ref{fig:bed_pdf}d. The skewness and kurtosis of the $pdf$ of roughness height fluctuations for the IWF case is 0.008 and 1.74, respectively, whereas the IWH case shows higher skewness and kurtosis values of -0.0766 and 1.83, respectively. 

\begin{figure}
   \centering
   \includegraphics[width=4cm,height=6cm,keepaspectratio]{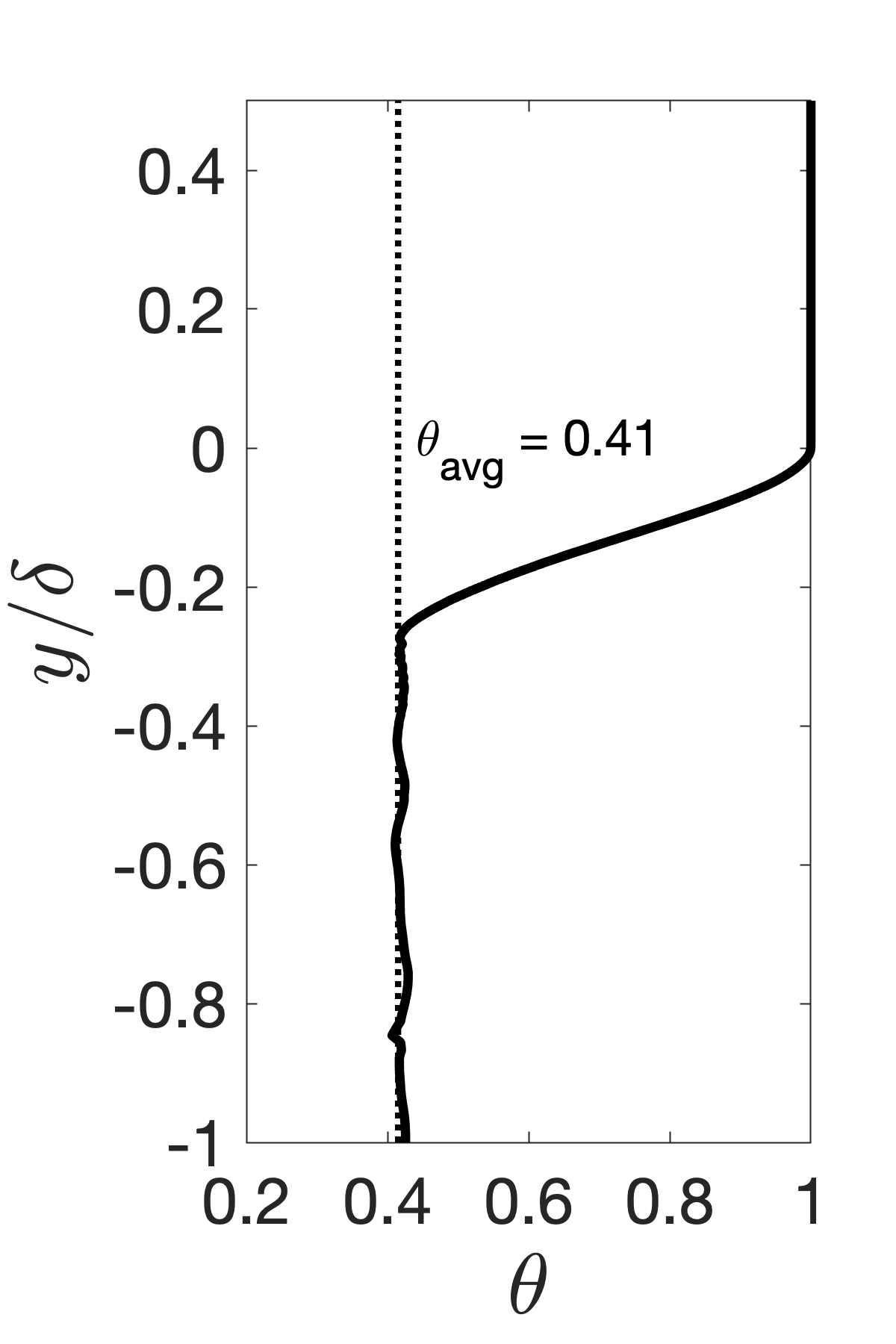}
\caption{\small Double averaged porosity profile for the permeable bed.}
\label{fig:bed_pdf}
\end{figure}

\begin{figure}
   \centering
   \subfigure[]{
   \includegraphics[width=6cm,height=4cm,keepaspectratio]{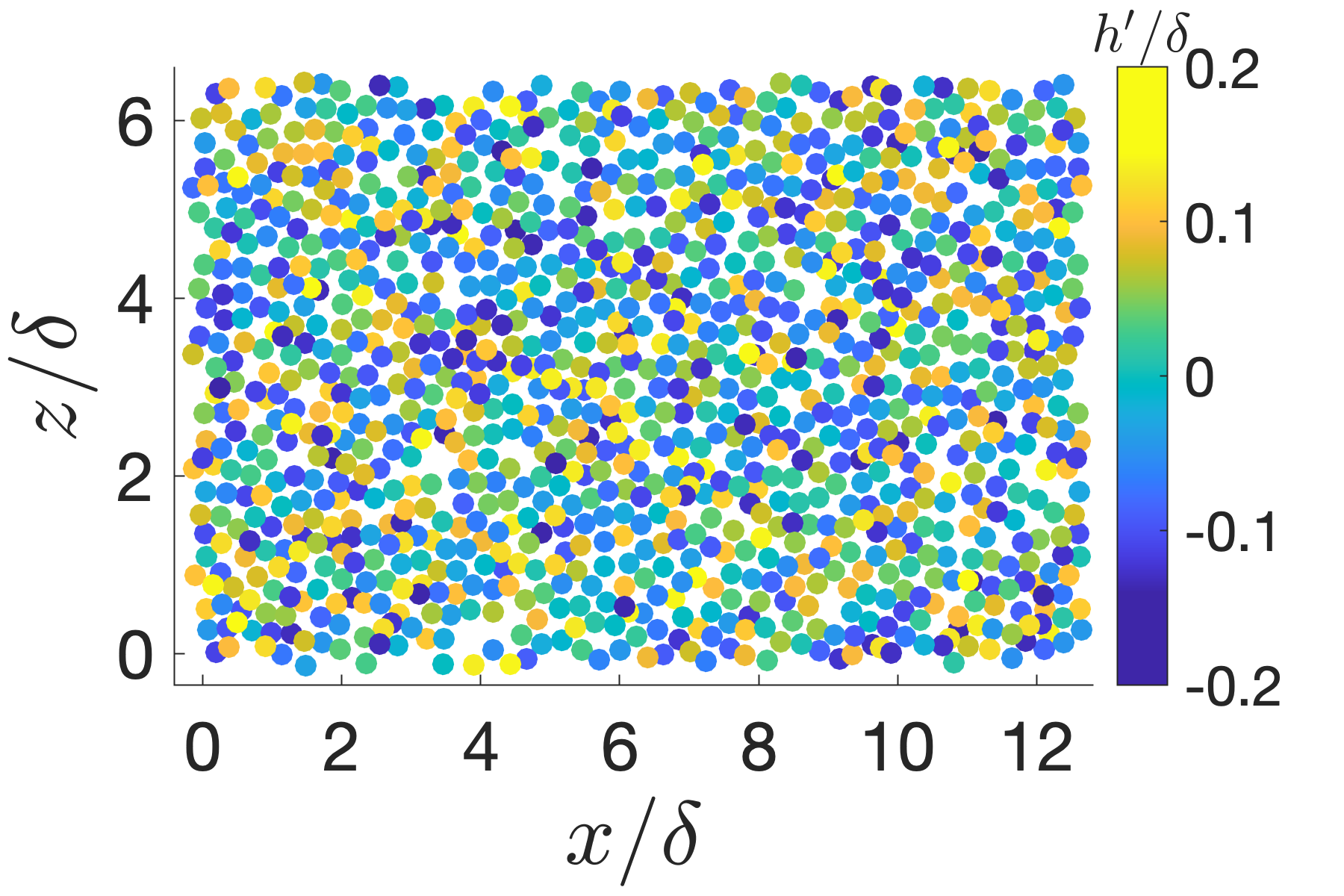}}
      \subfigure[]{
   \includegraphics[width=6cm,height=4cm,keepaspectratio]{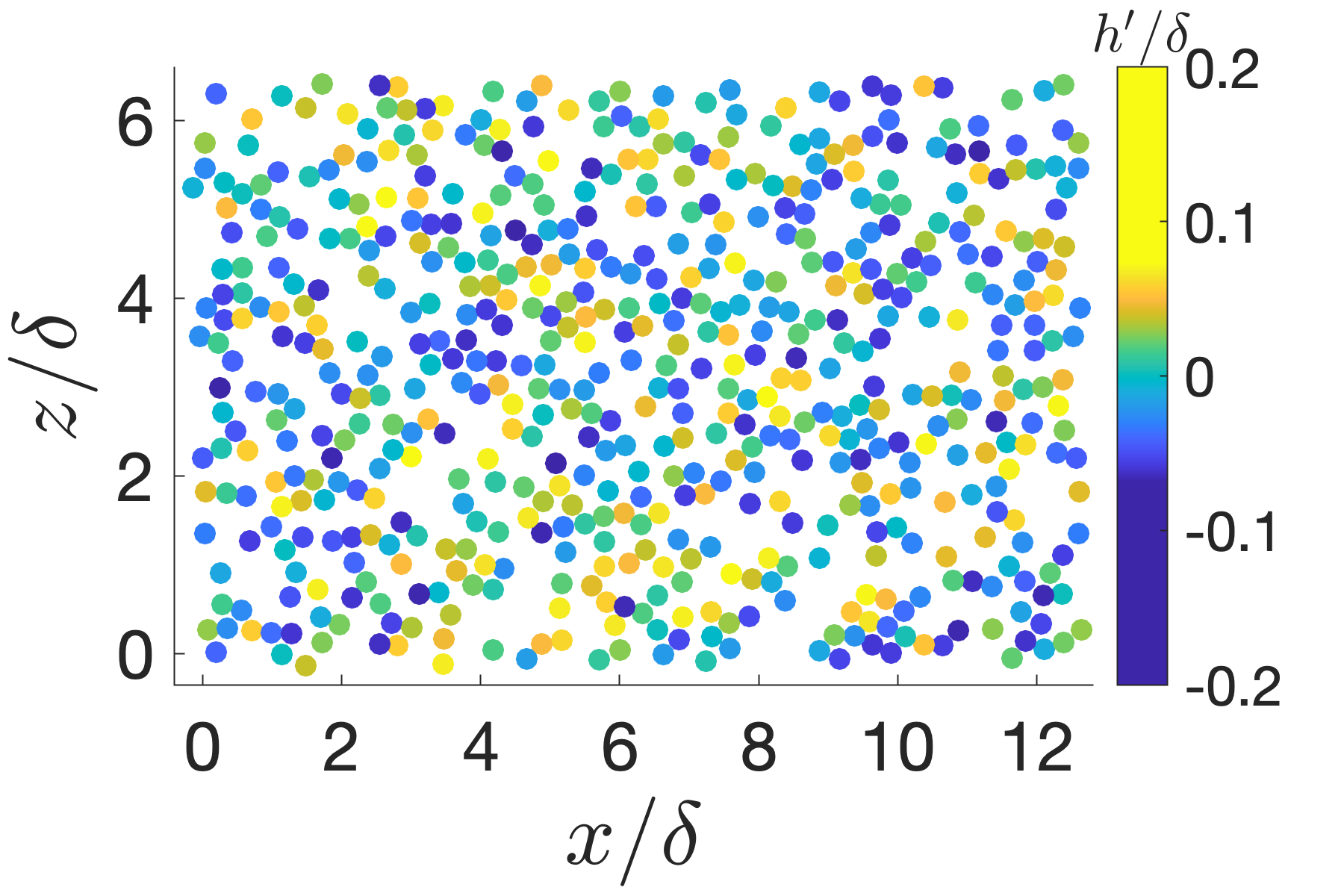}}
      \subfigure[]{
   \includegraphics[width=4cm,height=6cm,keepaspectratio]{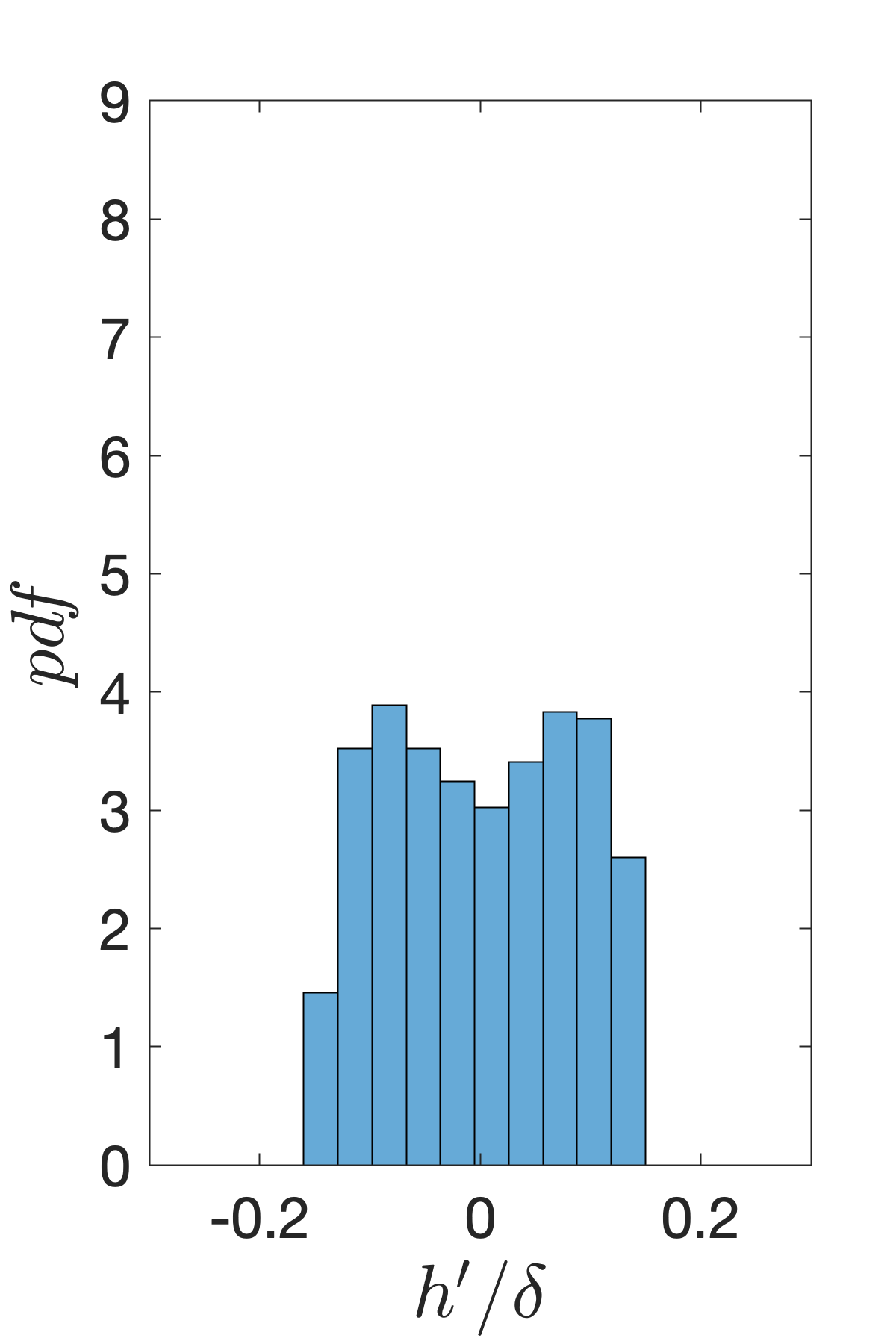}}
      \subfigure[]{
   \includegraphics[width=4cm,height=6cm,keepaspectratio]{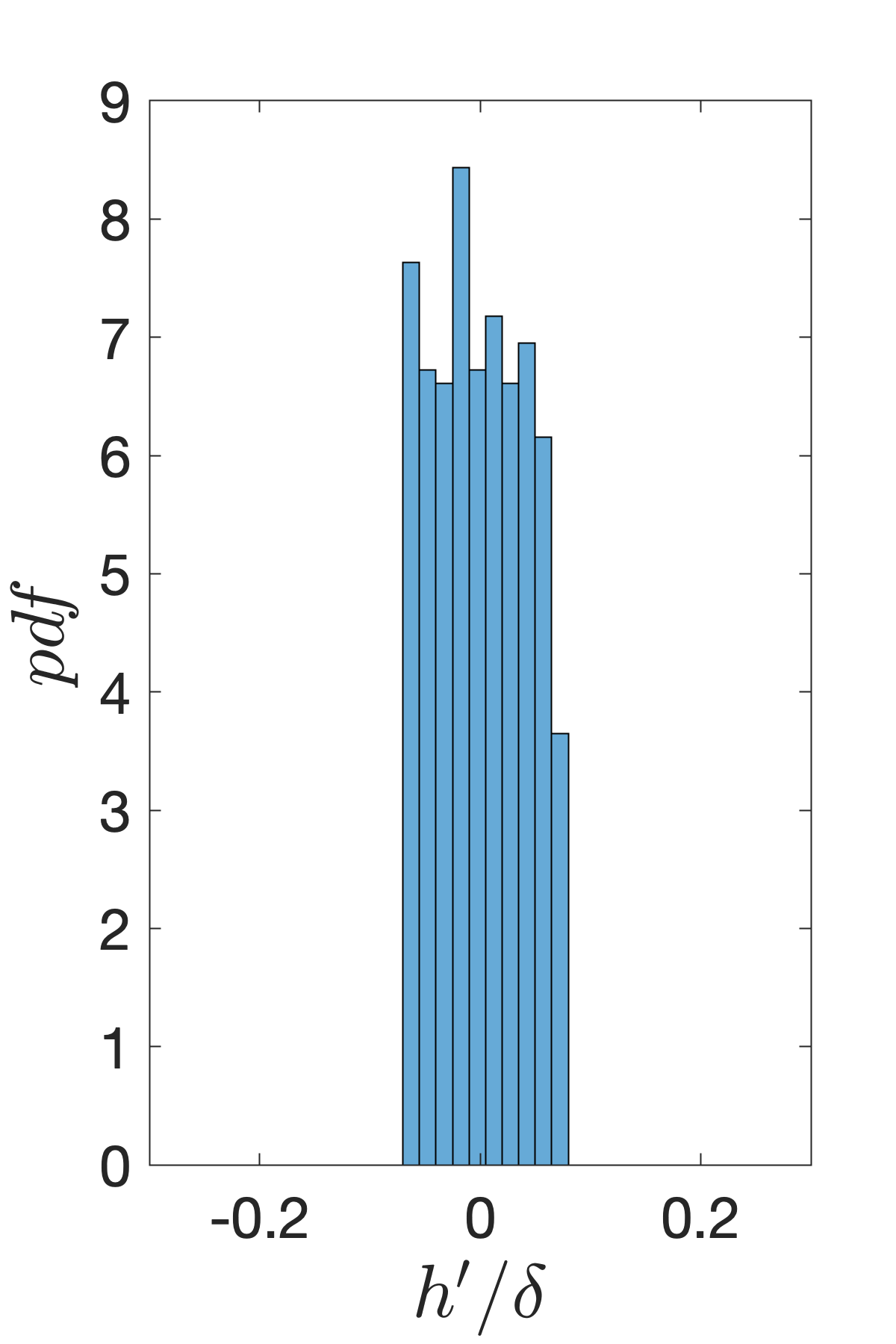}}
\caption{\small Scatter plots (dots are not to actual size) and pdfs of roughness height fluctuations: (a,c) impermeable full layer (IWF), (b,d) impermeable half layer (IWH). Plots for top layer of the permeable bed are same as that of the IWF case.}
\label{fig:bed_pdf}
\end{figure}


\section{Results}\label{sec:res}
\subsection{Comparison with experimental data} \label{sec:comp_exp_res}
Pore-resolved direct numerical simulations of turbulent boundary layer flow over a sediment bed were first validated with experimental data from~\citet{voermans2017variation}. Permeable bed case with porosity of $0.41$, $Re_K = 2.56$ and $Re_{\tau} \sim 180$ matches with Case L12 in~\citet{voermans2017variation},which are within range for realistic aquatic sediments beds. For this validation study, the simulation domain used is same as in the DNS study of~\citet{shen2020direct}. Although the averaged bed porosity is same in both studies, the approach applied to generate the randomly packed sediment particle bed is different. The code developed by~\citet{dye2013description} is used to generate a random distribution of uniform-sized spheres at a given porosity. This code applies the collective rearrangement algorithm introduced by~\citet{williams2003random}, coupled with a mechanism for controlling the system overall porosity.

\begin{figure}
   \centering
   \subfigure[]{
   \includegraphics[width=3.1cm,height=7.6cm,keepaspectratio]{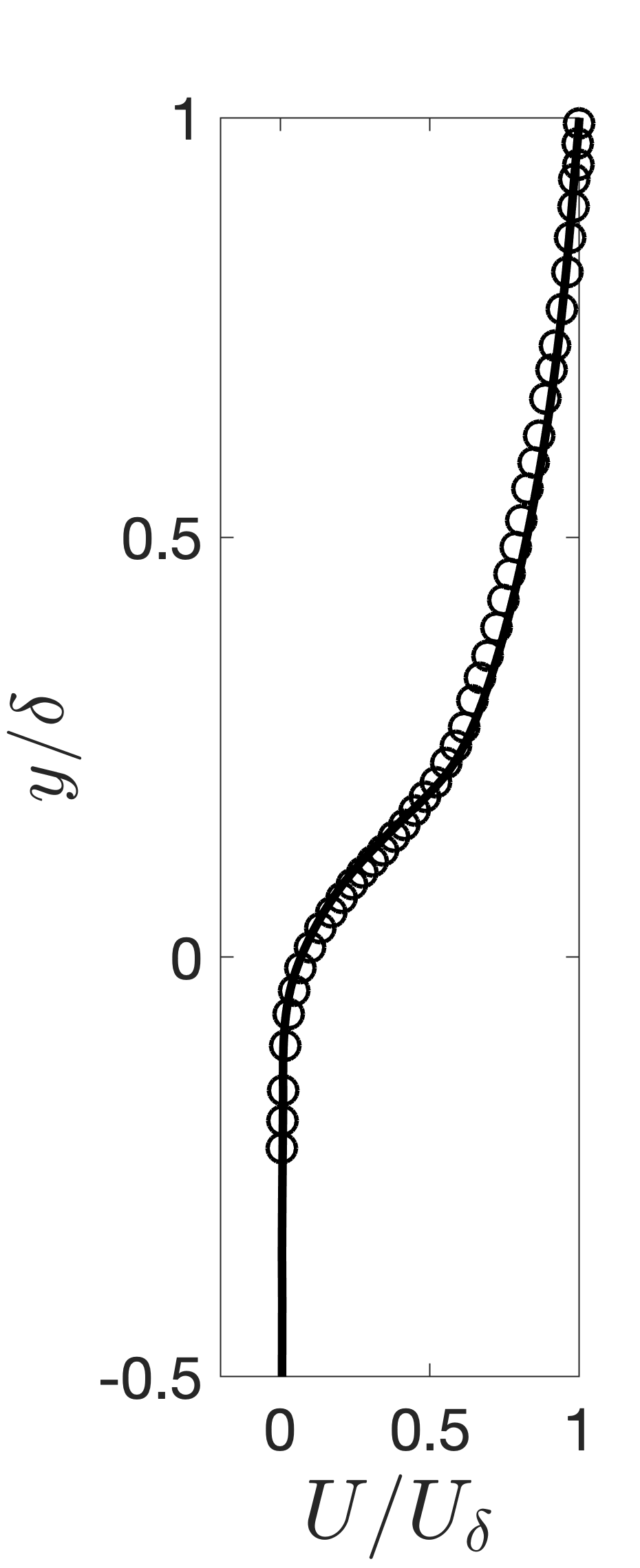}}
    \subfigure[]{
   \includegraphics[width=3.1cm,height=7.6cm,keepaspectratio]{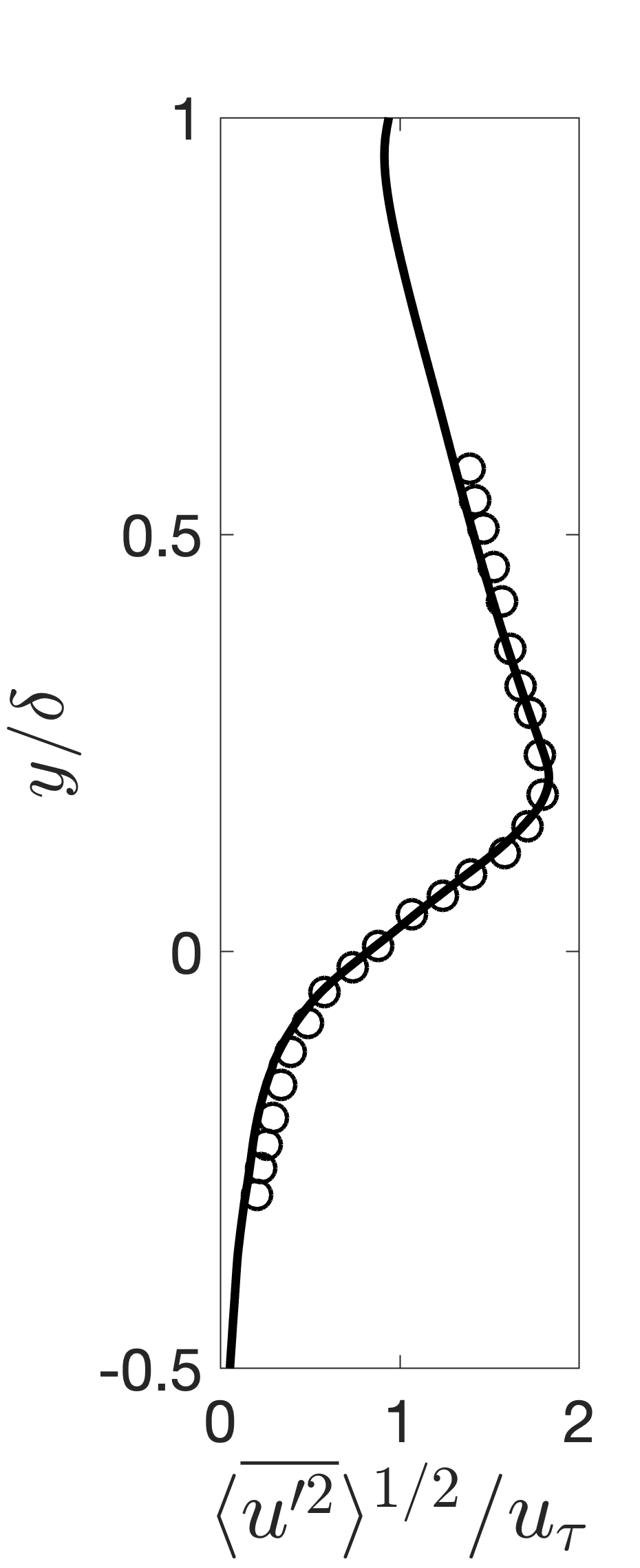}}
   \subfigure[]{
  \includegraphics[width=3.1cm,height=7.6cm,keepaspectratio]{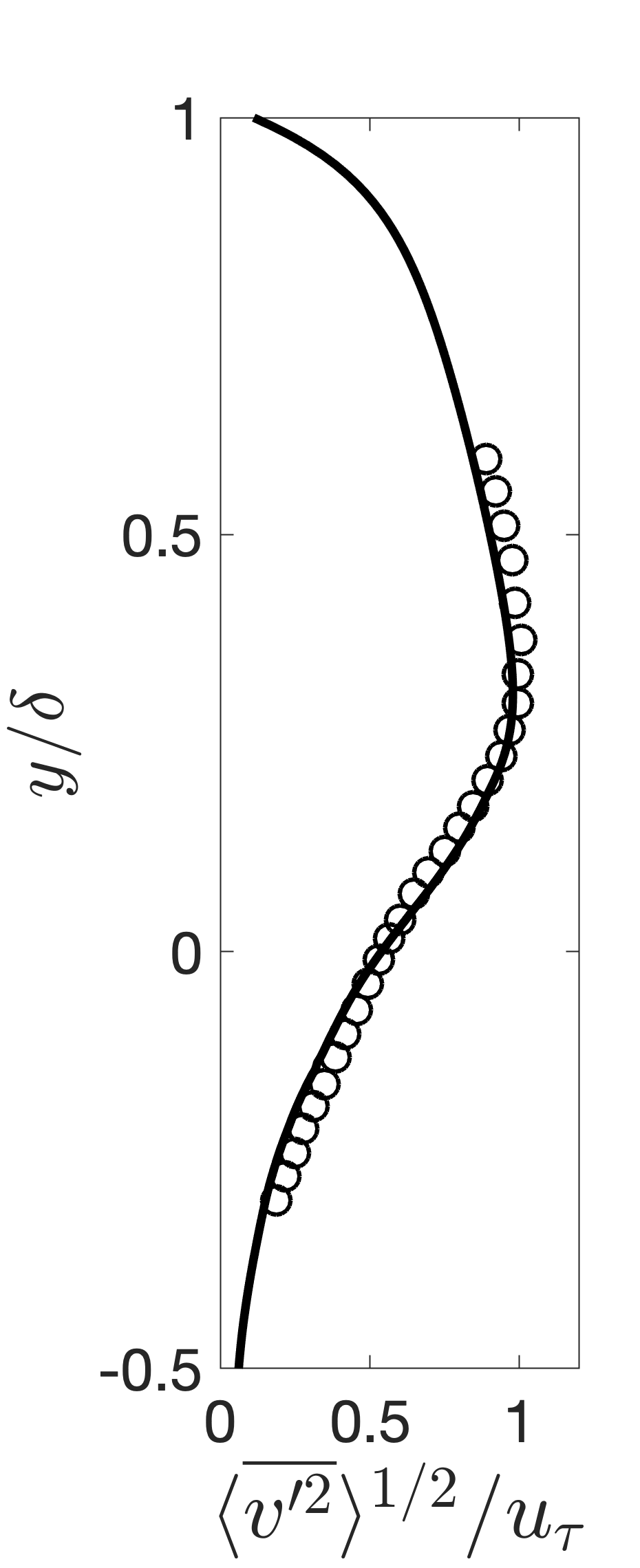}}
  \subfigure[]{
  \includegraphics[width=3.1cm,height=7.6cm,keepaspectratio]{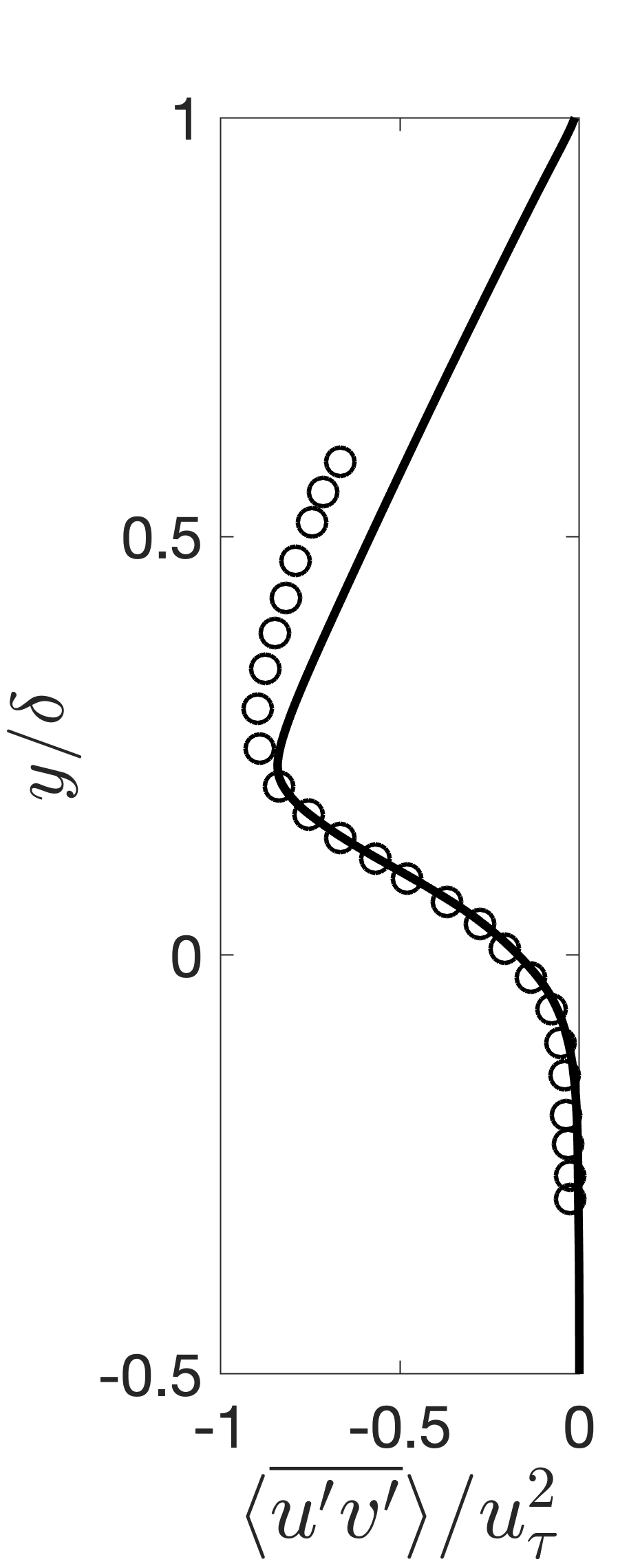}}
\caption{\small Comparison of (a) mean velocity and (b) streamwise, (c) wall-normal, and (d) shear components of Reynolds stress tensor. Experimental data by~\citet{voermans2017variation} (\blkcircle), DNS (\blkline). }
\label{fig:vald1}
\end{figure}

\begin{figure}
   \centering
   \subfigure[]{
   \includegraphics[width=3.1cm,height=7.6cm,keepaspectratio]{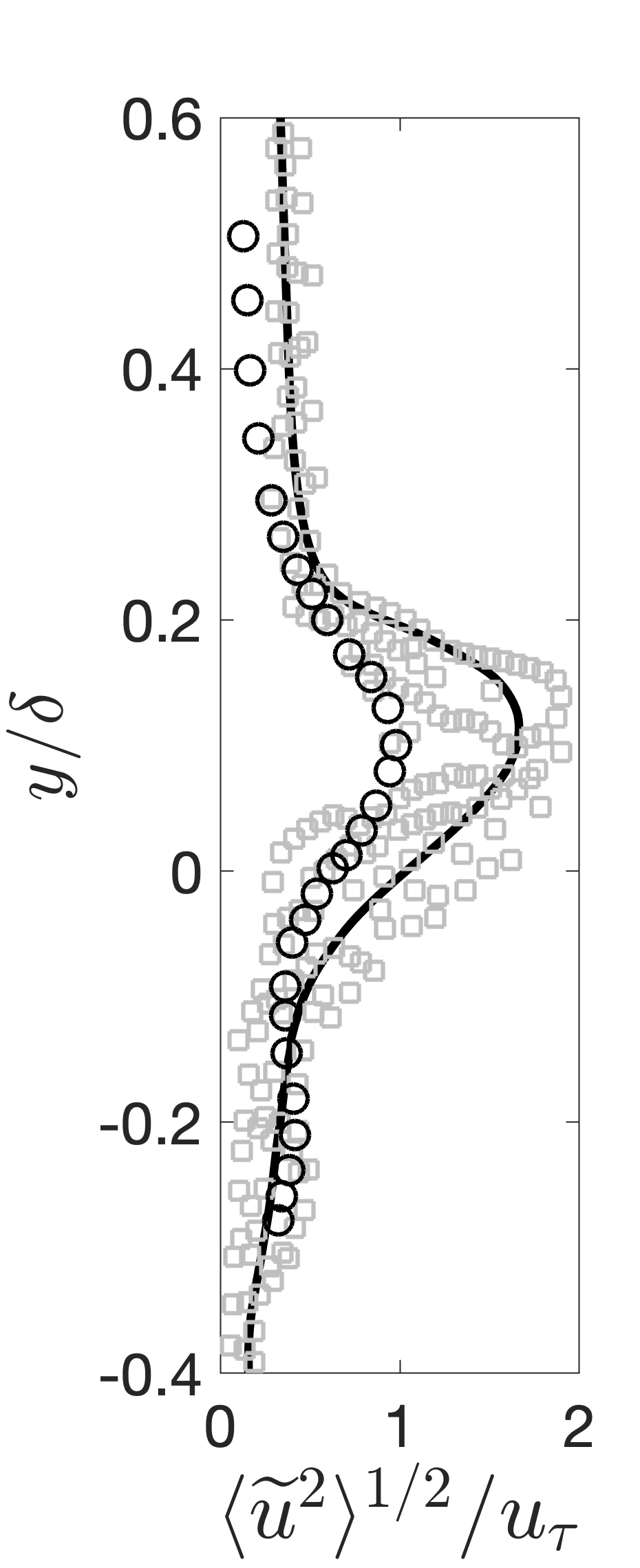}}
    \subfigure[]{
   \includegraphics[width=3.1cm,height=7.6cm,keepaspectratio]{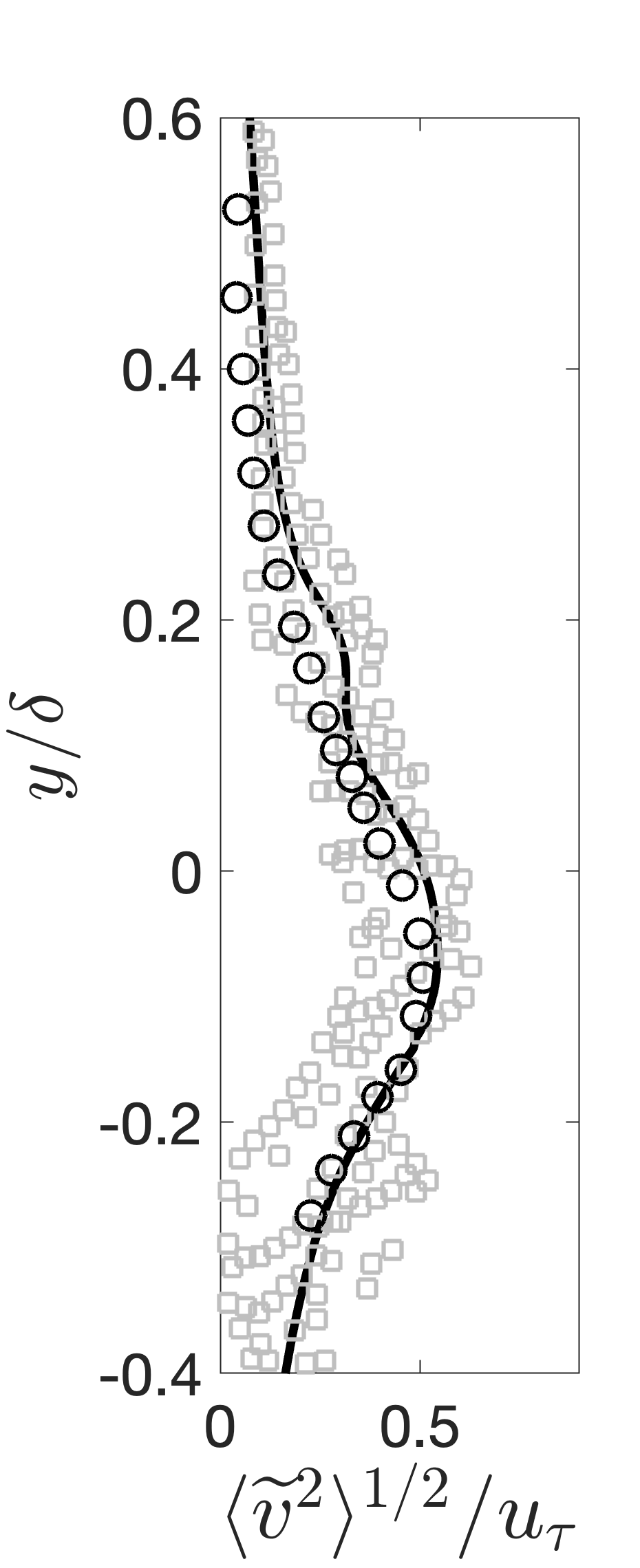}}
   \subfigure[]{
  \includegraphics[width=3.1cm,height=7.6cm,keepaspectratio]{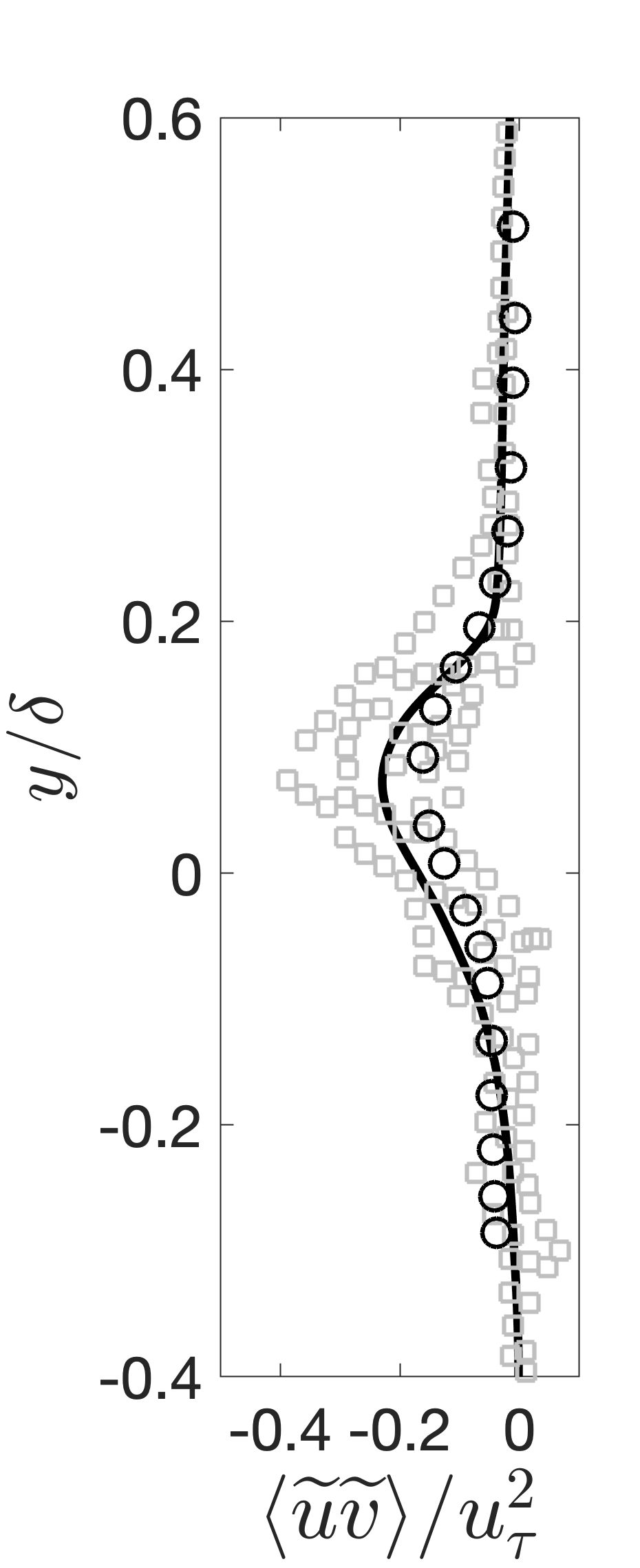}}
\caption{\small Comparison of (a) streamwise, (b) wall-normal, and (c) shear components of form induced stress tensor (\blkcircle), emulating experimental sampling (\graysquare), DNS (\blkline). }
\label{fig:vald2}
\end{figure}

To be consistent with the way results are presented in~\citet{voermans2017variation}, the location for $y=0$, is taken at the location where $\partial^2_{yy} \theta=0$ in figures~\ref{fig:vald1},\ref{fig:vald2}. However, for rest of the statistics shown in this paper $y=0$ is chosen to be the sediment crest location, while the virtual origin is chosen to be the zero-displacement plane, $y=-d$, instead. The double-averaged (DA) mean velocity profile normalized by free-surface velocity $U_{\delta}$ is shown in figure~\ref{fig:vald1}a. Excellent agreement is seen between the DNS data and experimental measurements. Figures~\ref{fig:vald1}b,~\ref{fig:vald1}c, and~\ref{fig:vald1}d show a comparison of DA turbulence intensities, namely stream-wise, wall normal and shear stresses. Again very good agreement between DNS and experiment is observed. The slight deviation in Reynolds stress in the free stream region can be attributed to the fact that~\citet{voermans2017variation} took measurements at three lateral positions (at one streamwise location), while the statistics were collected through the whole domain in the current study. Similar behavior was also noted in DNS predictions of~\citet{shen2020direct}.


The form-induced intensities or dispersive stresses normalized by $u_{\tau}$ are shown in figures~\ref{fig:vald2}a,~\ref{fig:vald2}b, and~\ref{fig:vald2}c. Upon initial observation, noticeable differences are observed between current DNS and experimental results. However, these differences can be explained as follows. Firstly, as mentioned in the previous section, spatial averaging is carried out over an entire $x-z$ plane at a given $y$ location for DNS results. While for the experimental data, spatial averaging was performed over three different spanwise locations over six different measurements. To quantify the differences in the sampling procedures between the experiments and DNS, the experimental sampling process is replicated in the DNS data whereby spatial averaging is carried out at a few finite uncorrelated spanwise locations and repeated over different streamwise locations. A family of curves, shown by grey squares, indicates the associated uncertainty in the processing of the experimental data. The averaged experimental and DNS data are within this scatter for all axial locations. Secondly, it is has been reported in literature~\citep{nikora2002zero, fang2018influence} that the spanwise averaging is highly sensitive to the geometry at the sediment-water interface. For the present DNS, only the mean porosity of the randomly distributed arrangement of mono-dispersed spherical particles is matched with the experimental geometry. However, the exact sediment-grain distribution in the experiments is unknown and is likely different compared to that used in DNS. This difference, especially near the top of the bed can also contribute to differences in the form-induced or dispersive stresses.

In spite of the potential differences in the sediment bed distribution between DNS and experimental work, the present results reproduce the mean flow and turbulence stresses observed in the experiment. The form-induced stresses fall within the experimental uncertainty and match qualitatively. In addition, turbulence statistics from the current work were compared with DNS predictions from~\citet{shen2020direct}. Good agreement between the two sets of DNS results was observed. The consistency with both experimental and numerical studies persuasively validates the numerical approach used in this work.

\subsection{The log-law and zero-displacement thickness}\label{sec:zero_disp}
In turbulent flows over rough walls and permeable beds the log-law has the following form
\begin{align}
     \dfrac{U(y)}{u_{\tau}} &= \dfrac{1}{\kappa} \log \left (\dfrac{y+d}{y_0}\right), 
    \label{eq:log_eq}
\end{align}
\begin{table}
\begin{center}
\def~{\hphantom{0}}
\begin{tabular}{@{}lc c c c c }
Case &  $\kappa$ & $d/{\delta}$ & $d^{+}$ & $y_{0}/\delta$  & $y_0^{+}$\\ 
PB & 0.32 & 0.175 & 47.0 & 0.0248 & 6.65 \\ 
IWF & 0.32 & 0.175  & 47.0 & 0.0248 & 6.65\\ 
IWH  & 0.35 & 0.137 & 37.0 & 0.0098 & 2.65\\ 
SW  & 0.41 & - & - & - & - \\ 
\end{tabular}
\caption{The von-K\'{a}rm\'{a}n constant ($\kappa$), zero-displacement thickness ($d$), and equivalent roughness height $y_0$ normalized by $\delta$ and $\nu/u_{\tau}$ for PB, IWF, IWH and SW cases.}
\label{tab:zdp}
\end{center}
\end{table}
where $\kappa$ is the von-K\'{a}rm\'{a}n constant, $d$ is distance between the zero-displacement plane and the top of the sediment crest, and $y_0$ is the equivalent roughness height. Several definitions of the zero-displacement thickness, $d$, have been used in the literature.~\citet{jackson1981displacement} postulated that $d$ is the level where the mean drag on the bed appears to act.~\citet{nikora2002zero} define $d$ as the level that large-scale turbulent eddies feel as the bed origin and thus, their sizes linearly scale with distance from this level. The equivalent roughness height, $y_0$, is related to a measure of the size of the roughness elements. 

\begin{figure}
   \centering
    \subfigure[]{
   \includegraphics[width=8cm,height=6cm,keepaspectratio]{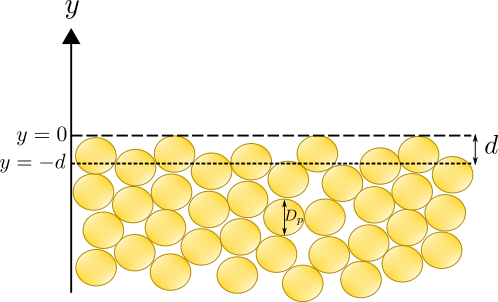}}
   \subfigure[]{
   \includegraphics[width=6cm,height=8cm,keepaspectratio]{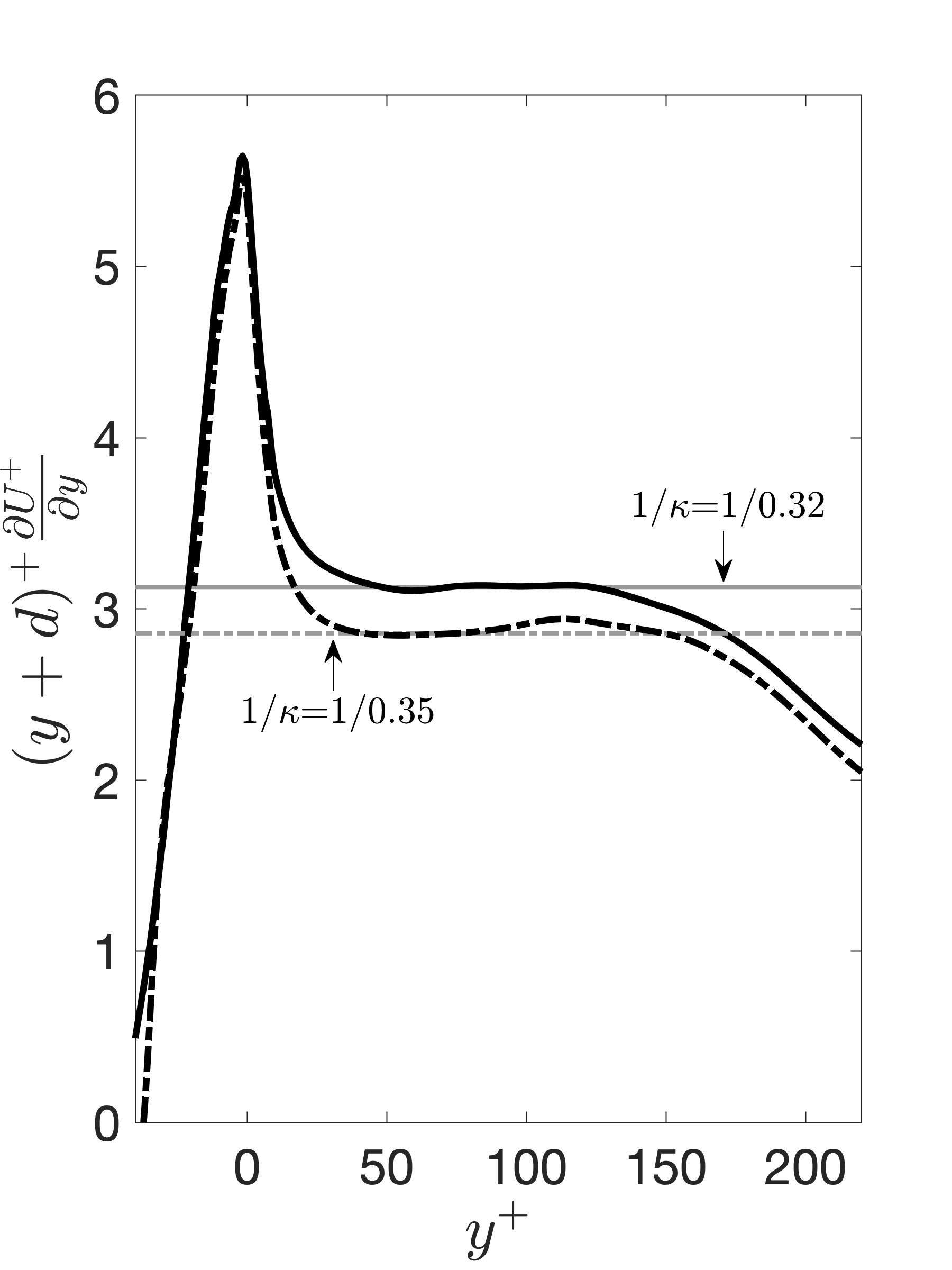}}
      \subfigure[]{
   \includegraphics[width=6cm,height=8cm,keepaspectratio]{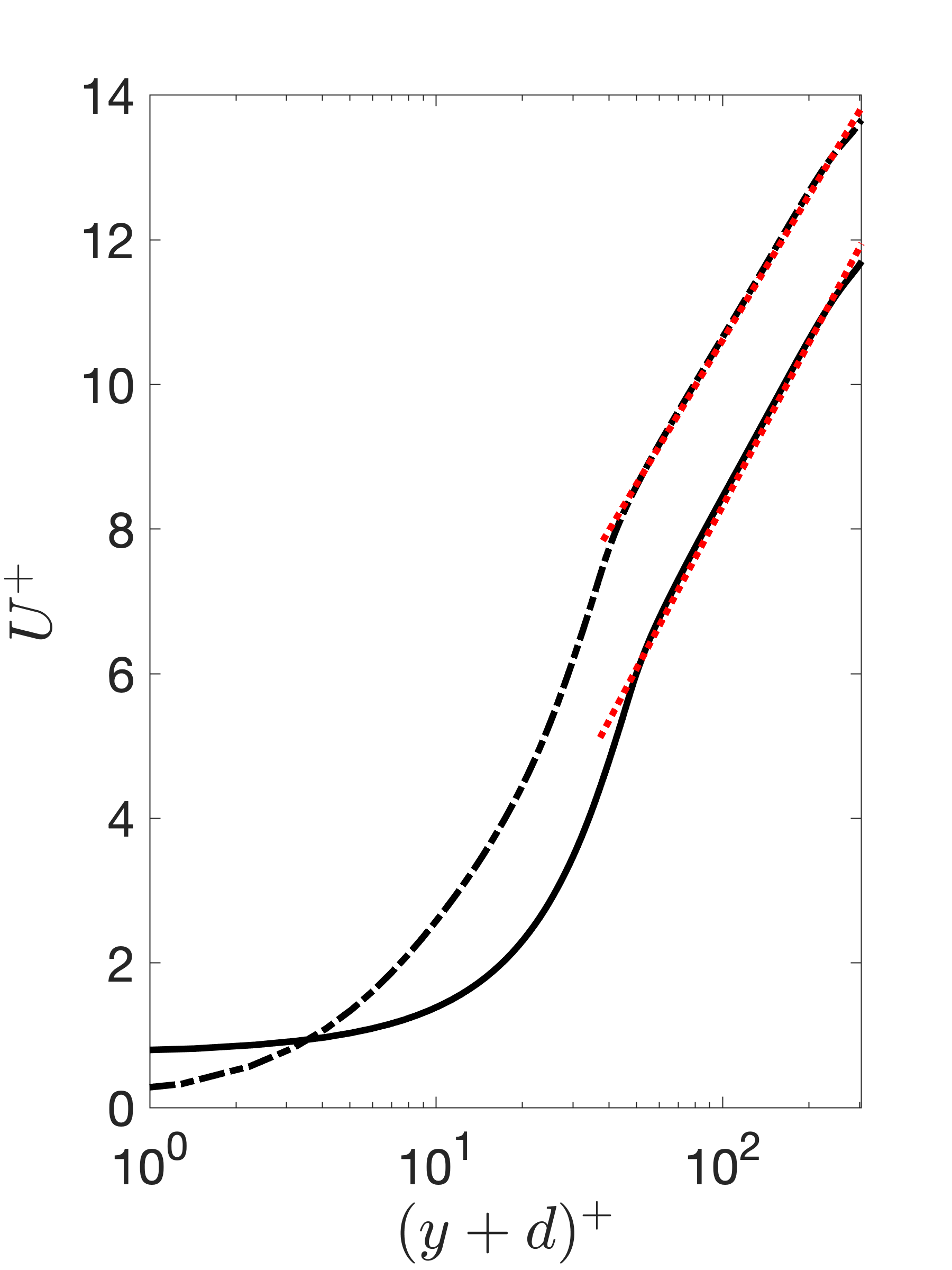}}
\caption{\small (a) Schematic showing positions of sediment crest ($y = 0$), zero-displacement plane ($y = -d$), and particle diameter ($D_{p}$), (b) log-law fitting, and (c) mean velocity profiles in wall units denoted by superscript ($+$). Legend: PB case (\blkline), IWH case (\blkdashdotline), and fitted logarithmic profiles (\reddottedline).
}
\label{fig:log}
\end{figure}

A schematic of the domain with the location of the sediment crest and the zero-displacement plane is shown in figure \ref{fig:log}a where $y = 0$ is chosen at the sediment crest in the domain. However, for comparison of flow statistics the virtual origin is chosen at the zero-displacement plane, $y = -d$ (except while comparing with the experimental data. please see section $3.1$).
Although several techniques have been used to determine these parameters in literature~\citep{raupach1991rough}, the procedure described by~\citet{breugem2006influence} is followed here. First, the extent of the logarithmic layer is determined by plotting $(y+d)^{+}{\partial_{y^{+}} U^{+}}$ against $y^{+}$ for several values of $d$. From equation~\ref{eq:log_eq} it is easy to see that the value of $(y+d)^{+}{\partial_{y^{+}} U^{+}}$ is a constant equal to $1/\kappa$ in the logarithmic layer. Therefore, the value of $d$ is the one that gives a horizontal profile in the logarithmic layer. Figure~\ref{fig:log}b shows the plot of $(y+d)^{+}\partial_{y^+} U^{+}$ against $y^{+}$ for the permeable bed and impermeable half-layer cases. Similar fitting procedure is carried out for impermeable full-layer case (not shown). The values of $d$, $\kappa$, and $y_0$ determined from a least squares fit to equation~\ref{eq:log_eq} to the velocity profile in the logarithmic layer are given in table~\ref{tab:zdp}. Mean velocity profiles for the permeable bed and impermeable half layer cases fitted with logarithmic profiles are shown in figure~\ref{fig:log}c. 

\subsection{Turbulence structure} \label{sec:comp_turbstruc}

\begin{figure}
   \centering
   \subfigure[]{
   \includegraphics[width=6cm,height=6cm,keepaspectratio]{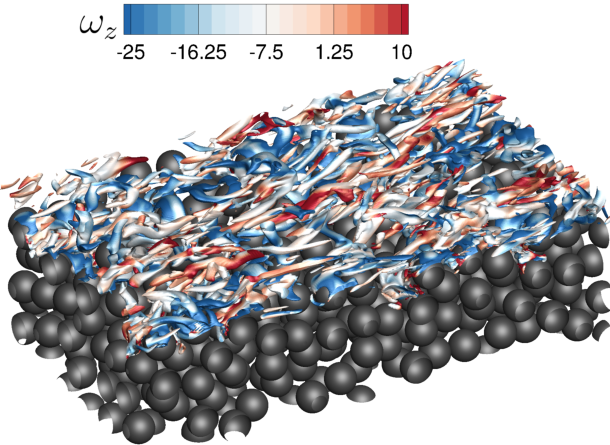}}
    \subfigure[]{
   \includegraphics[width=6cm,height=6cm,keepaspectratio]{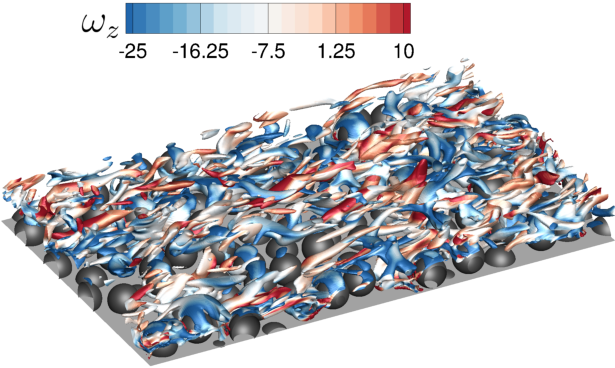}}
   \subfigure[]{
   \includegraphics[width=6cm,height=6cm,keepaspectratio]{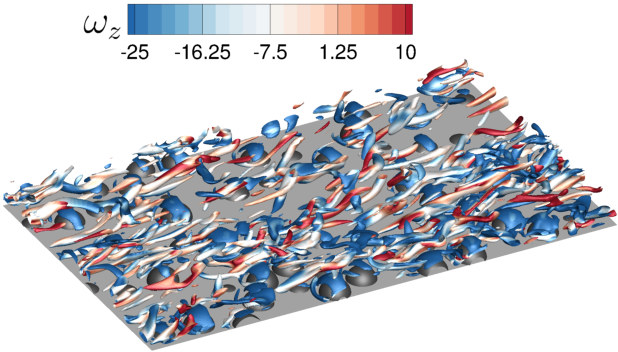}}
   \subfigure[]{
   \includegraphics[width=6cm,height=6cm,keepaspectratio]{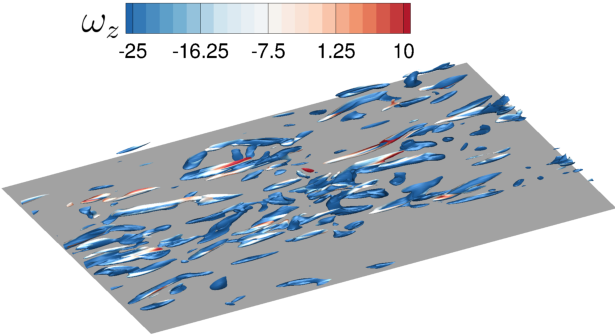}}
\caption{\small Iso-surface of swirling strength at $\lambda_{ci}$ = 0.1$\lambda_{ci,max}$, colored by z-component of vorticity, ${\omega}_{z}$, shown between $8.57 < x/\delta < 12.56$, $2.28 < z/\delta < 4.57$ and very close to the bed, $y/\delta < 0.057$ for (a) PB (b) IWF (c) IWH and (d) SW cases.}
\label{fig:turb_struc}
\end{figure}

Distinct characteristics in the primary turbulence statistics between permeable bed and non-permeable rough walls are presented in this section. Iso-surfaces of local swirling strength, $\lambda_{ci}$, defined as the imaginary part of the complex conjugate eigenvalue of the velocity gradient tensor, $\del_{x_j}u_i$~\citep{zhou1999mechanisms}, are shown in figure~\ref{fig:turb_struc} to gain a qualitative insight into the differences in turbulent structures between the permeable bed, impermeable rough walls, and smooth wall cases. 
The isosurfaces in figure~\ref{fig:turb_struc} are set at 10\% of the maximum value of $\lambda_{ci}$ and contours are colored by the z-component of vorticity, ${\omega}_{z}$. The PB and IWF cases show an abundance of vortical structures with a wide range of sizes. 
The IWH case on the other hand has much fewer structures with a small range of size distribution. Finally, the smooth wall case shows long elongated structures indicative of strong anisotropy. The porous bed and roughness elements break down these elongated flow structures and reduce the level of  anisotropy.



\begin{figure}
   \centering
   \subfigure[]{
   \includegraphics[width=3.6cm,height=7.6cm,keepaspectratio]{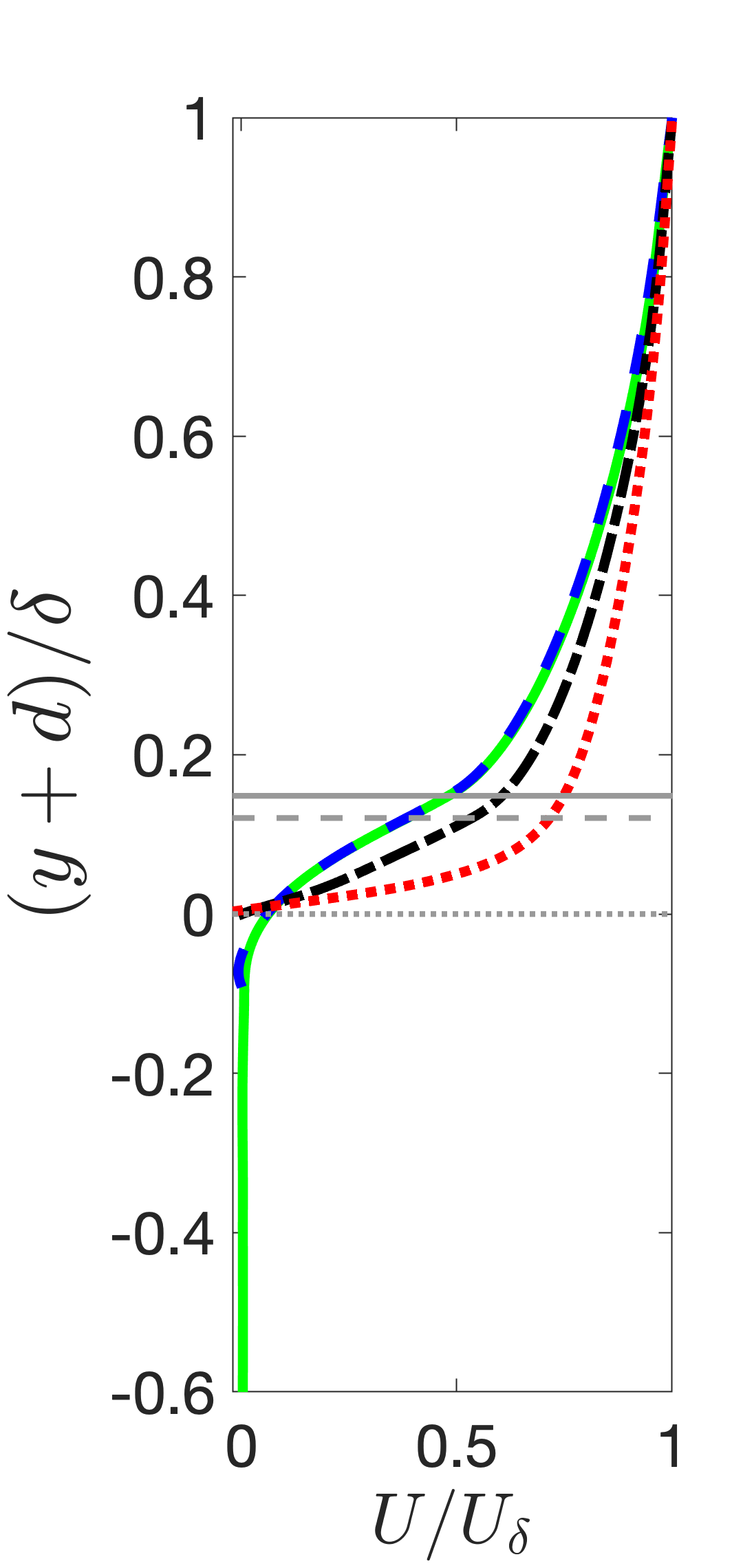}}
   \subfigure[]{
   \includegraphics[width=3.0cm,height=7.6cm,keepaspectratio]{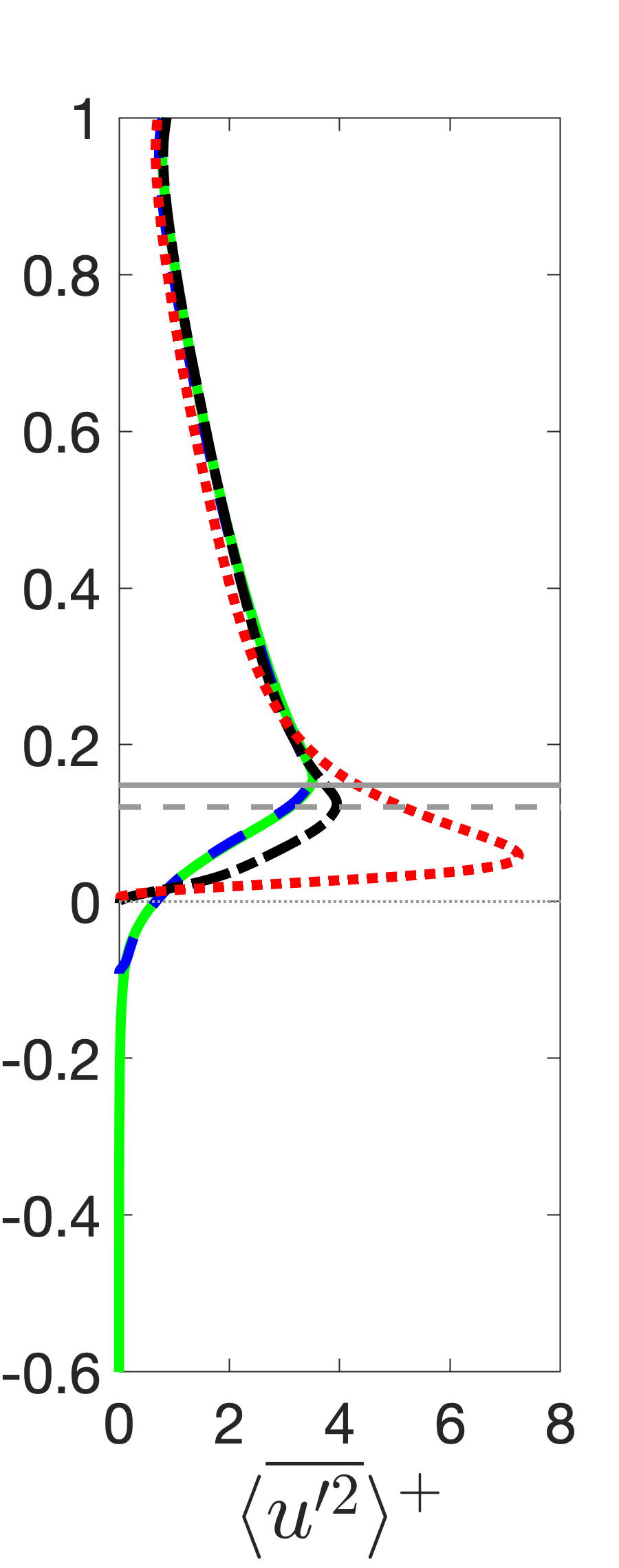}}
   \subfigure[]{
   \includegraphics[width=3.0cm,height=7.6cm,keepaspectratio]{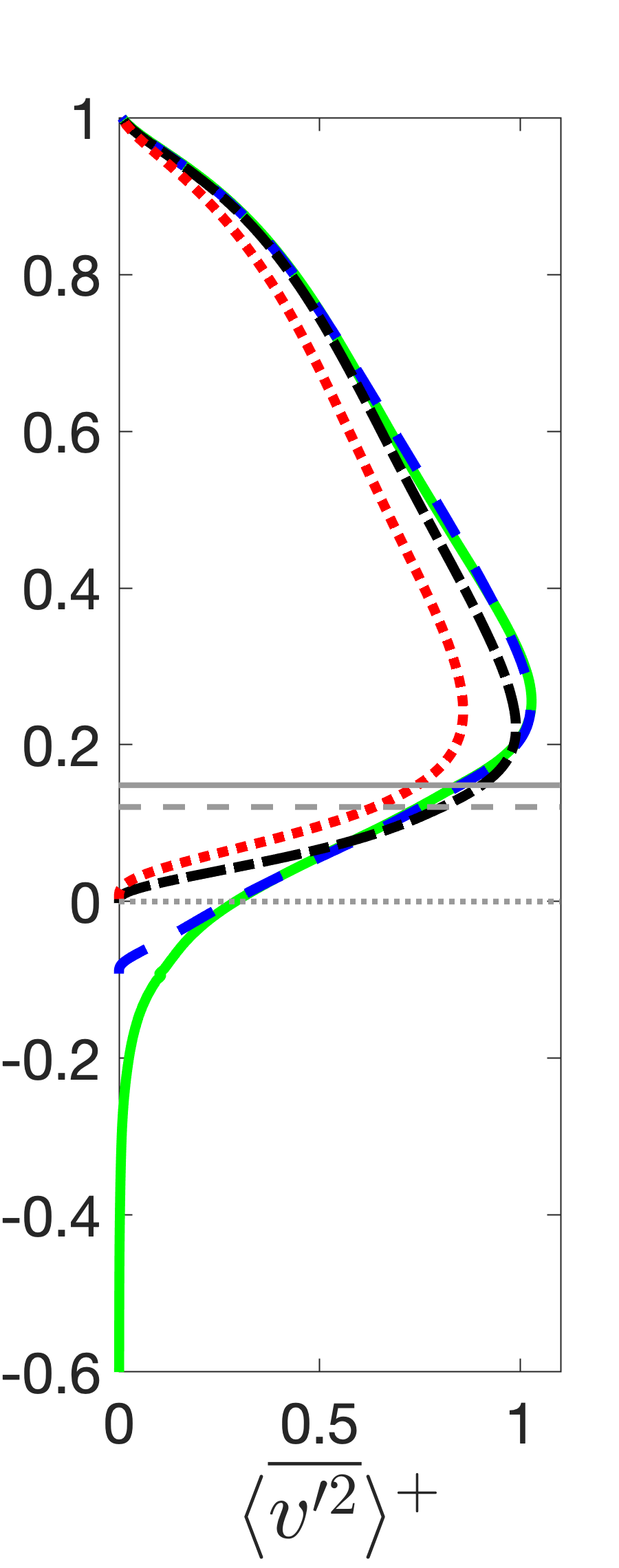}}
    \subfigure[]{
   \includegraphics[width=3.0cm,height=7.6cm,keepaspectratio]{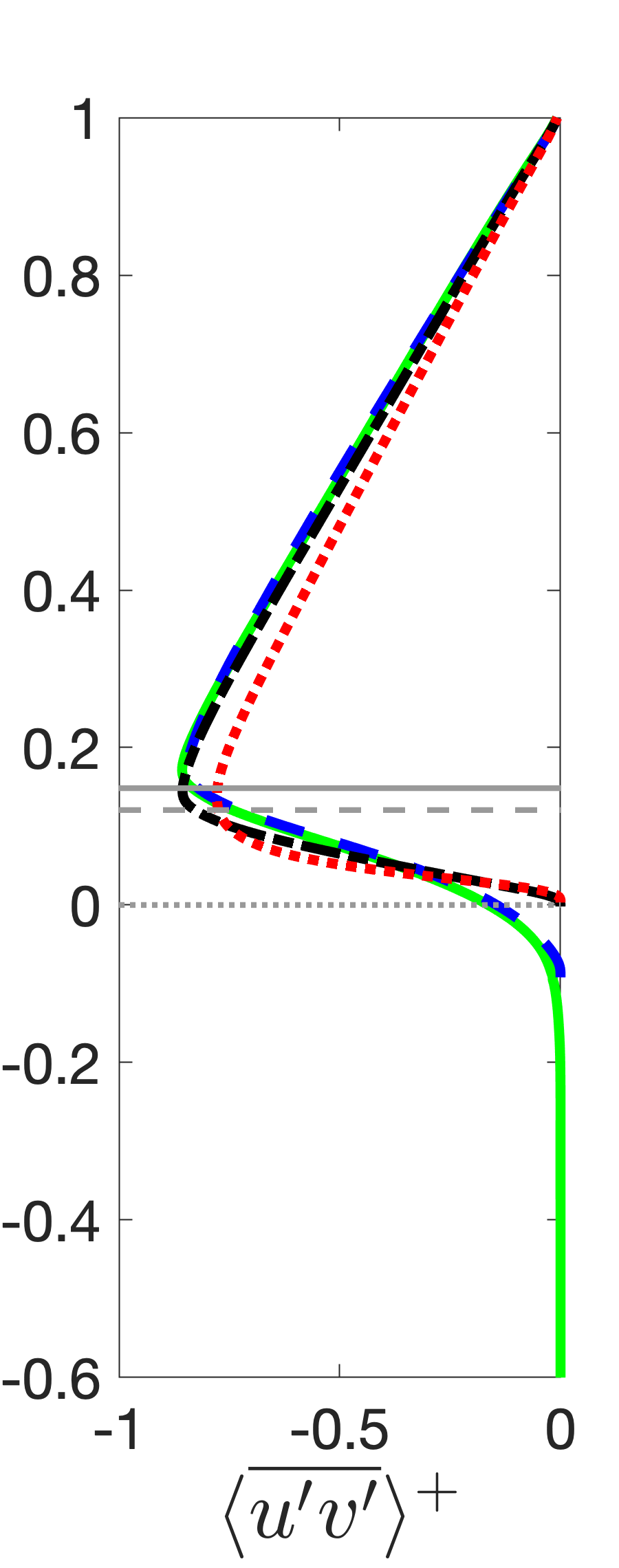}}
\caption{\small Comparison of the mean velocity and turbulent stress profiles for PB (\greenline), IWF (\bluedashline), IWH (\blkdashdotline), and SW cases (\reddottedline): (a) mean velocity and (b-d) streamwise, wall-normal, and shear components of Reynolds stress tensor. Horizontal lines show the crest of sediment bed for PB (\grayline), IWF (\graydashdotline), IWH (\graydashline), and SW cases (\graydottedline). (Note that PB and IWF sediment crest lines overlap).}
\label{fig:reys_prof}
\end{figure}

Figure \ref{fig:reys_prof} shows the double averaged mean velocity and Reynolds stresses for the four cases (PB, IWF, IWH and SW). The double-averaged variables are normalized by the friction velocity, $u_{\tau}$, and $y$ is shifted by $d$ and then normalized by $\delta$, effectively making virtual origin the same for all the cases. The discussion below compares the magnitudes for each case at their respective crest levels.
Compared to the smooth wall case, the permeable bed and both impermeable rough wall cases show a mean velocity deficit due to fluid momentum loss. The deficit is largest for the PB and IWF cases, followed by the IWH case showing that the loss in momentum due to wall roughness is enhanced by permeability. The axial component of turbulent velocity fluctuation, $\langle\overline{u^{\prime2}}\rangle^{+}$, is highest for the SW case followed by the IWH case, and is lowest for the PB and IWF cases with little difference between the two. The profiles of the $\langle\overline{u^{\prime2}}\rangle^{+}$ for all the cases overlap for $\left ({y+d}\right)/{\delta} >0.4$. This observation is consistent with the wall similarity hypothesis reported by \citet{raupach1991rough} and \citet{breugem2006influence}. For the smooth wall and impermeable half-layer cases, turbulent fluctuations into the wall (i.e., sweeps) are redirected into wall parallel components due to the wall blocking effect. The bed-normal velocity fluctuations, $\langle\overline{v^{\prime2}}\rangle^{+}$, are highest for the PB and IWF cases with little difference between them, suggesting that majority of the turbulent penetration is predominately observed in the top layer. 
As expected, the values of Reynolds stress $\langle\overline{u^{\prime}v^{\prime}}\rangle^{+}$ are higher for the permeable bed and impermeable rough wall cases compared to the smooth wall.
The results are quantified by comparing the mean and Reynolds stress values for each case at their respective crest locations. Compared to PB case, mean velocity and $\langle\overline{u^{\prime2}}\rangle^{+}$ for IWH case are 11.29\% and 13.22\% greater. Whereas $\langle\overline{v^{\prime2}}\rangle^{+}$ and $\langle\overline{u^{\prime}v^{\prime}}\rangle^{+}$ for IWH case are 4.70\% and 0.66\% lower than the PB case.

\begin{figure}
   \centering
   \subfigure[]{
   \includegraphics[width=3.6cm,height=7.6cm,keepaspectratio]{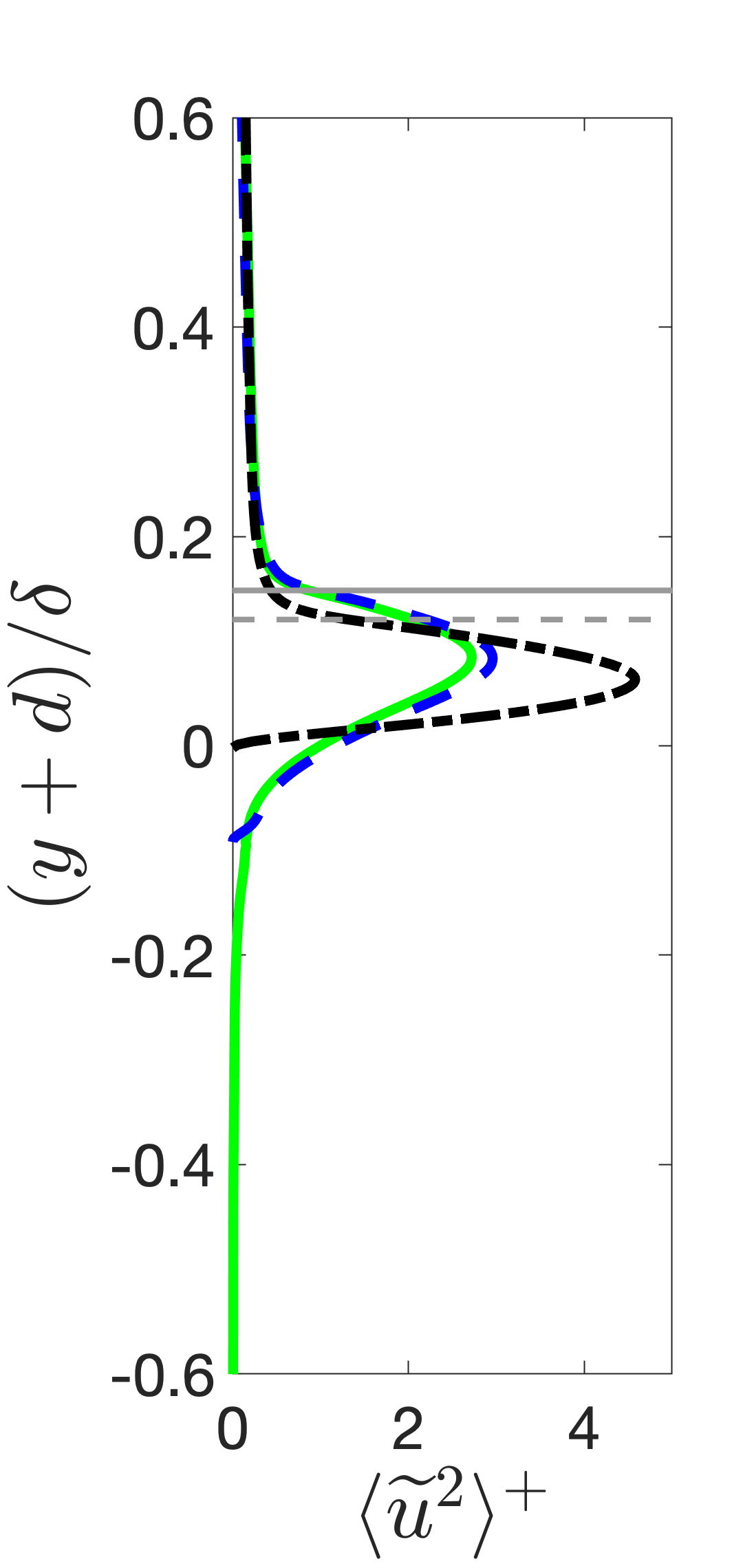}}
   \subfigure[]{
   \includegraphics[width=3.0cm,height=7.6cm,keepaspectratio]{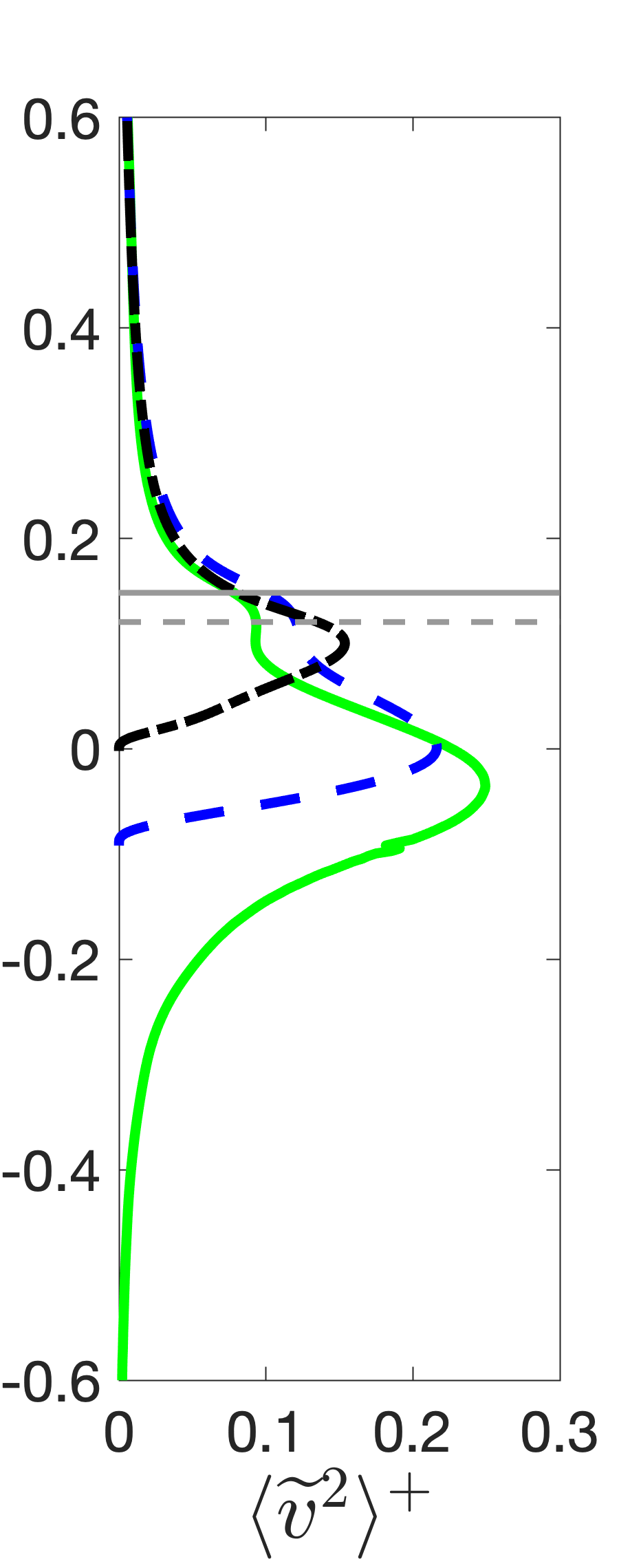}}
   \subfigure[]{
   \includegraphics[width=3.0cm,height=7.6cm,keepaspectratio]{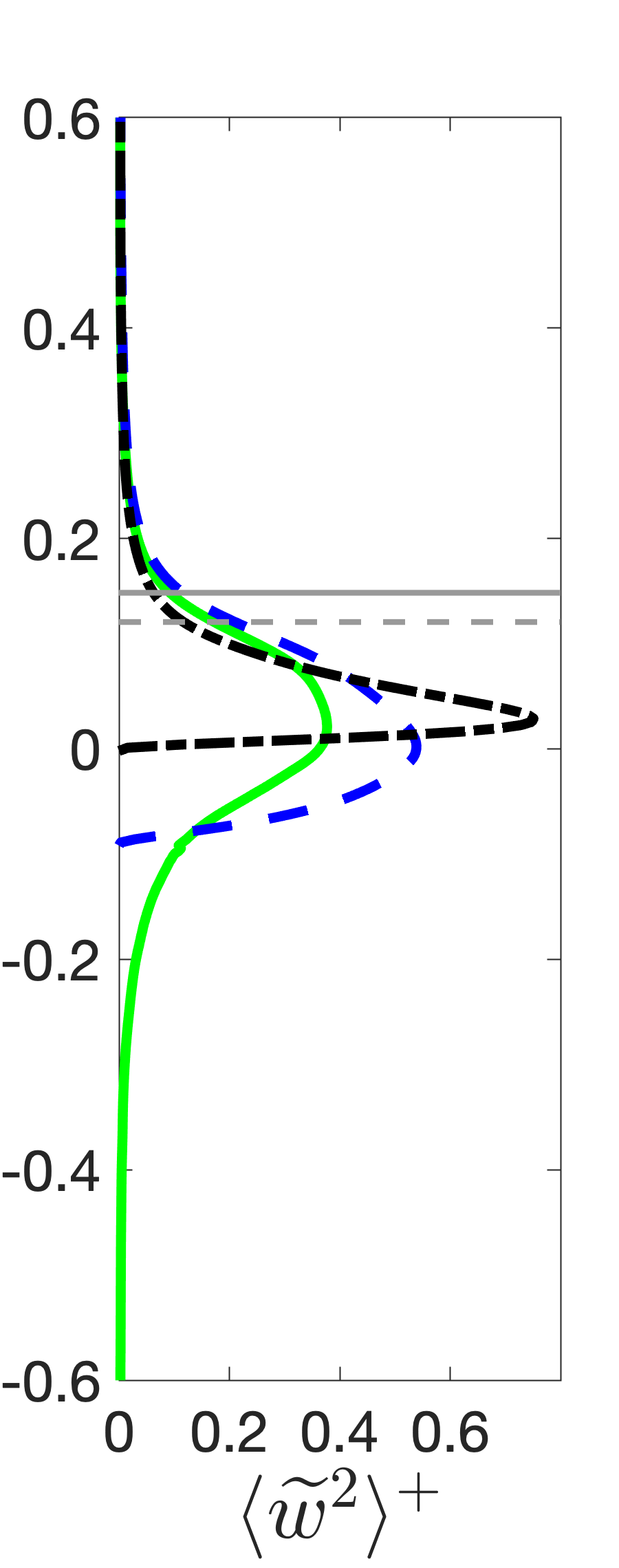}}
   \subfigure[]{
   \includegraphics[width=3.0cm,height=7.6cm,keepaspectratio]{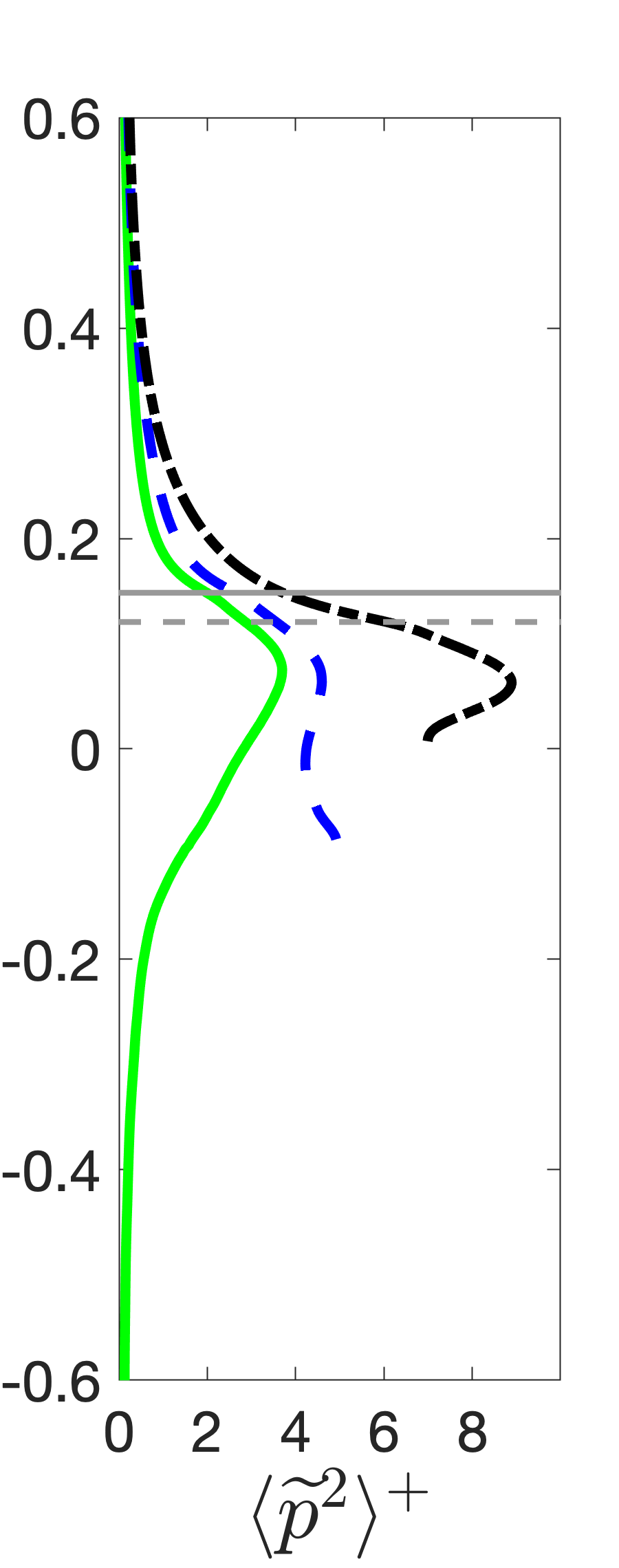}}
\caption{\small Comparison of form-induced stress profiles for PB (\greenline), IWF (\bluedashline), and IWH (\blkdashdotline) cases. (a) streamwise and (b) wall-normal, (c) spanwise, and (d) form-induced pressure fluctuations. Horizontal lines show the crest of sediment bed for PB (\grayline), IWF (\graydashdotline), and IWH case (\graydashline). (Note that PB and IWF sediment crest lines overlap).}
\label{fig:fis_prof}
\end{figure}

The permeable bed and impermeable full-layer cases show almost the same magnitudes of mean velocity and Reynolds stresses at the sediment crest. This shows that the presence of a solid wall underneath the full layer of spherical roughness elements has minimal influence on both mean flow field and turbulent fluctuations. The penetration of mean and turbulent fluctuations is thus restricted to the top layer of the bed for the permeability Reynolds number investigated. The full layer of roughness elements creates pockets underneath where the flow can penetrate and since the turbulent kinetic energy within this layer is still small, the flow characteristics and momentum transport mechanisms resemble that of a permeable bed. However, for the impermeable half-layer case the influence of the solid wall is observed on both mean and Reynolds stresses. Compared to the permeable bed and impermeable full-layer cases, longer streaky structures are observed for the half-layer case (see figure~\ref{fig:turb_struc}).  Wall blocking effect is more prominent for the IWH case, resulting in reduced penetration of turbulence. Lastly, the shear stress distribution indicates that while bed roughness increases Reynolds shear stress, permeability has minimal influence.
Figure \ref{fig:fis_prof} shows the double averaged form-induced stresses for the PB, IWF and IWH cases. Compared to the Reynolds stresses, the corresponding form-induced stresses are lower in magnitude. Moreover, the peak value of form-induced stresses occurs significantly below the sediment crest. It is seen from figure~\ref{fig:fis_prof}a that the peak value for axial component, $\langle\widetilde{u}^2\rangle^{+}$, is largest for the IWH case followed by the IWF and PB cases. On the other hand, the bed-normal component peak value, $\langle\widetilde{v}^2\rangle^{+}$, is largest for the PB case, followed by the IWF and IWH cases. The peak value and its location for  $\langle\widetilde{u}^2\rangle^{+}$,  $\langle\widetilde{v}^2\rangle^{+}$, $\langle\widetilde{w}^2\rangle^{+}$, $\langle\widetilde{p}^2\rangle^{+}$, and $\langle\widetilde{u}\widetilde{v}\rangle^{+}$ are given in table \ref{tab:fispk}. 

PB and IWF cases show significant differences in $\langle\widetilde{v}^2\rangle^{+}$ as can be seen from both the peak value and its location from table~\ref{tab:fispk}. This is in contrast to $\langle\overline{v^{\prime2}}\rangle^{+}$ where little difference was observed in both cases suggesting that majority of the Reynolds stress penetration is predominately observed in the top layer. However, the form induced stresses indicate that based on local variations in porosity between the PB and IWF cases, turbulence penetration differs. The peak value of $\langle\widetilde{v}^2\rangle^{+}$ for the PB case is $\sim$ 38.5\% greater than the IWH case as the solid impermeable wall prohibits penetration of fluid flow. This value is also $\sim$ 13.2\% greater than the IWF case, which shows that although the full layer of sediment captures most of the effects of the sediment bed, there are still some differences. The PB and IWH cases show a plateau in $\langle\widetilde{v}^2\rangle^{+}$ profiles near the sediment crest which is not observed for the IWH case. This may be attributed to the fact that, for the porous bed and the IWF cases, turbulence penetrates the top layer of sediments resulting in flow and pressure fluctuations underneath the sediment particles, which is not observed for the half-layer case.

The peak value for the spanwise component, $\langle\widetilde{w}^2\rangle^{+}$, is largest for the IWH case followed by IWF and PB cases. The peak value for the IWH case is $\sim$ 49.8\% greater than the PB case and $\sim$ 28.3\% greater than the IWF case. It is interesting to observe that the peak values of $\langle\widetilde{w}^2\rangle^{+}$ for all three cases are greater than their respective $\langle\widetilde{v}^2\rangle^{+}$ peak values, similar to Reynolds stress behavior observed in smooth wall boundary layer flow. The differences in the peak values between the $\langle\widetilde{w}^2\rangle^{+}$ and the $\langle\widetilde{v}^2\rangle^{+}$ form-induced stresses are $\sim$ 33.9\% for the PB case, $\sim$ 59.8\% for the IWF case and $\sim$ 79.6\% for the IWH case, showing that the permeability of the bed reduces the differences between spanwise and bed-normal components. 

The peak value for form-induced pressure fluctuations, $\langle\widetilde{p}^2\rangle^{+}$, is largest in the IWH case followed by the IWF and PB cases. From Figure~\ref{fig:fis_prof}d it can be seen that the peak value occurs below the crest for permeable bed and non-permeable rough wall cases. The IWF cases have two comparable peaks, one near the crest and the other near the underlying no-slip wall. Numerical values are reported for the peak value occurring near the crest. The peak value for the IWH case is $\sim$ 58.5\% greater than the PB case and $\sim$ 49.1\% greater than the IWF case. Similar to $\langle\widetilde{v}^2\rangle^{+}$ a deeper penetration of form-induced pressure fluctuations into the permeable bed is observed. 
The peak in pressure fluctuation for all three cases seems to occur at similar locations  as compared to the other stresses, suggesting that pressure fluctuations respond slowly to the changes in local permeability indicative of larger time scale than for the fluctuating velocities.

The peak value for the form induced shear stress (not shown), $\langle\widetilde{u}\widetilde{v}\rangle^{+}$, 
are similar for the PB and IWF cases, both being greater than the IWH case, suggesting that the top layer of the sediment bed particles is responsible for the majority of form induced shear production. The peak value of $\langle\widetilde{u}\widetilde{v}\rangle^{+}$ for the PB and IWF cases is approximately 40.4\% greater than the IWH case. In contrast to Reynolds stresses, bed permeability has more influence on the form-induced shear stress.

The skin friction coefficient for wall bounded turbulent channel flows is defined as $C_f = 2(u_{\tau}/U_{\delta})^2$ (where $U_{\delta}$ = free stream velocity). The mean values of $C_f$ for PB, IWF, IWH and SW cases are 0.0144, 0.0141, 0.0107 and 0.0054, respectively. The mean $C_f$ value in a permeable bed is about 2.66 times that of the SW case and 1.35 times that of the IWH case, and is associated with greater penetration of turbulence shear into the bed thereby increasing flow resistance. These results are consistent with results reported in literature~\citep{manes2009turbulence}. Similar enhancement is also observed with the IWF case.

\begin{table}
  \begin{center}
\def~{\hphantom{0}}
\caption{The normalized location, $(y+d)/\delta$, and the peak value of normalized form-induced stresses (shown in brackets) for the PB, IWF, and IWH cases. Location is normalized by $\delta$, form-induced axial, bed-normal, and spanwise stresses are normalized by $u_{\tau}^2$ and form-induced pressure fluctuations by $\rho u_{\tau}^2$.}
\begin{tabular}{@{}lc c c c c }
Case &  $\langle\widetilde{u}^2\rangle^{+}$ & $\langle\widetilde{v}^2\rangle^{+}$ & $\langle\widetilde{w}^2\rangle^{+}$ &  $\langle\widetilde{p}^2\rangle^{+}$ & $\langle\widetilde{u}\widetilde{v}\rangle^{+}$  \\ 
PB &0.0838 (2.71) &-0.0351 (0.249) &0.0251 (0.377)  &0.0753 (3.685) &0.0555 (-0.21)  \\ 
IWF &0.0835 (2.95)&0.0032 (0.216) &0.0032 (0.538) &0.0627 (4.52) &0.0597 (-0.22)  \\ 
IWH  &0.0627 (4.58) &0.0996 (0.153) &0.0288 (0.751) &0.0627 (8.884) &0.0596 (-0.131)   \\ 
\end{tabular}
\label{tab:fispk}
 \end{center}
\end{table}

\subsection{Form-induced fluctuations and joint PDFs}\label{sec:fis_qual}

\begin{figure}
   \centering
   \subfigure[]{
   \includegraphics[width=4.3cm,height=4.3cm,keepaspectratio]{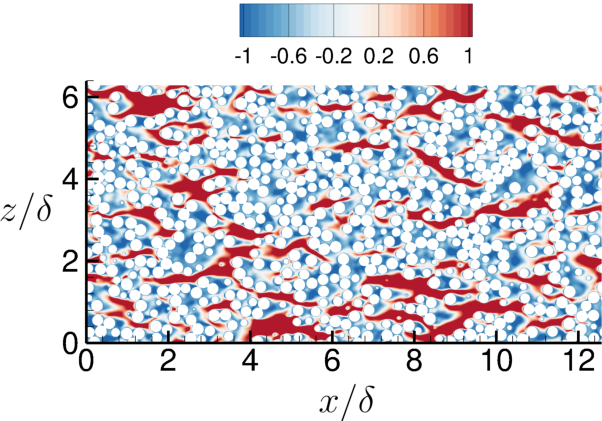}}
   \subfigure[]{
   \includegraphics[width=4.3cm,height=4.3cm,keepaspectratio]{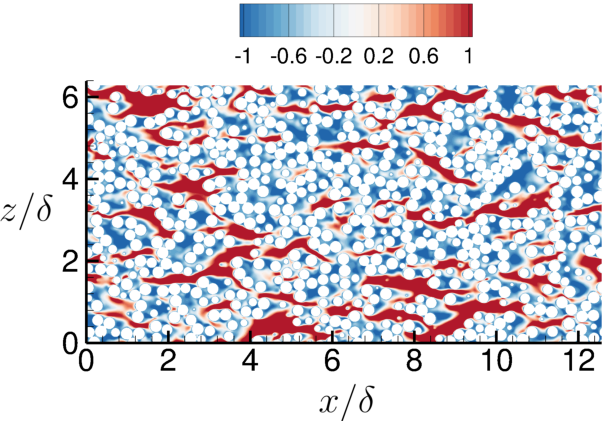}}
   \subfigure[]{
   \includegraphics[width=4.3cm,height=4.3cm,keepaspectratio]{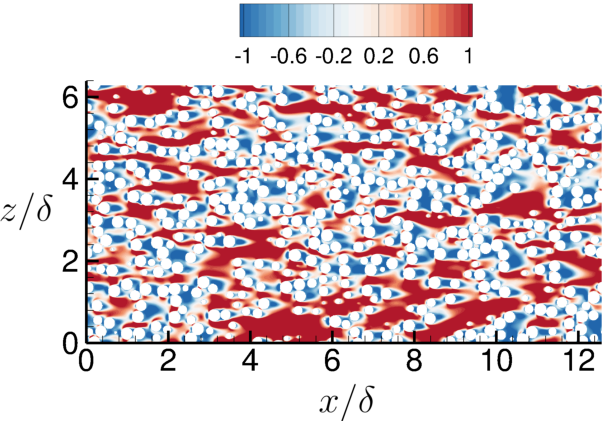}}
   \subfigure[]{
   \includegraphics[width=4.3cm,height=4.3cm,keepaspectratio]{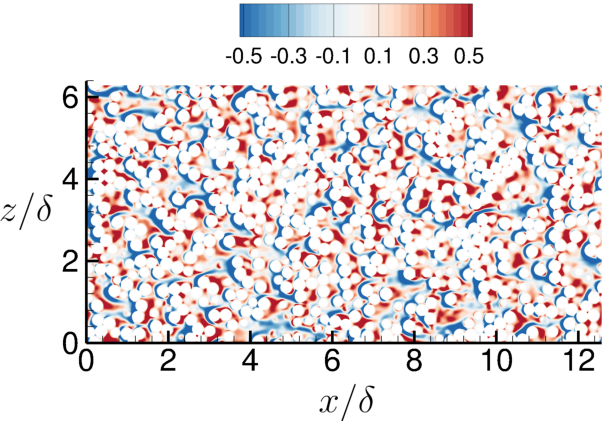}}
   \subfigure[]{
   \includegraphics[width=4.3cm,height=4.3cm,keepaspectratio]{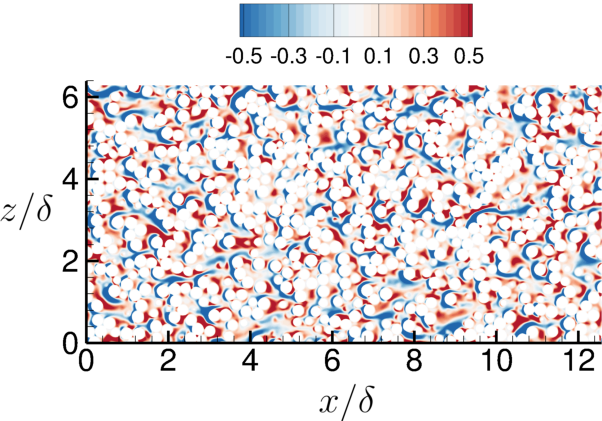}}
   \subfigure[]{
   \includegraphics[width=4.3cm,height=4.3cm,keepaspectratio]{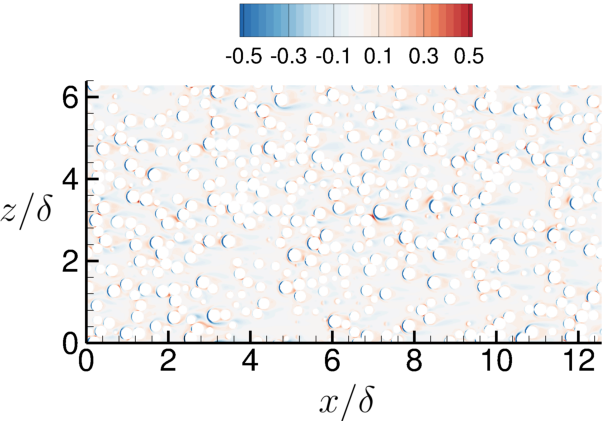}}
   \subfigure[]{
   \includegraphics[width=4.3cm,height=4.3cm,keepaspectratio]{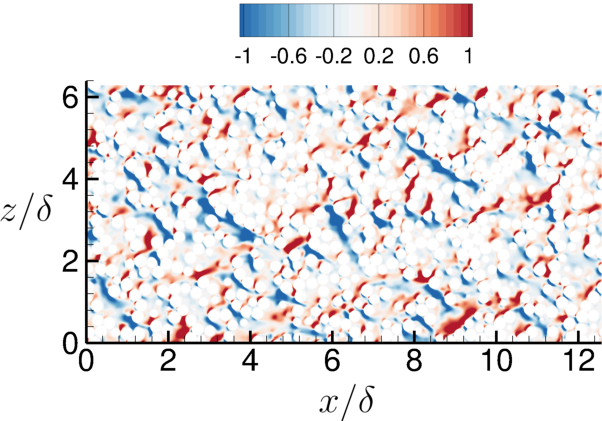}}
   \subfigure[]{
   \includegraphics[width=4.3cm,height=4.3cm,keepaspectratio]{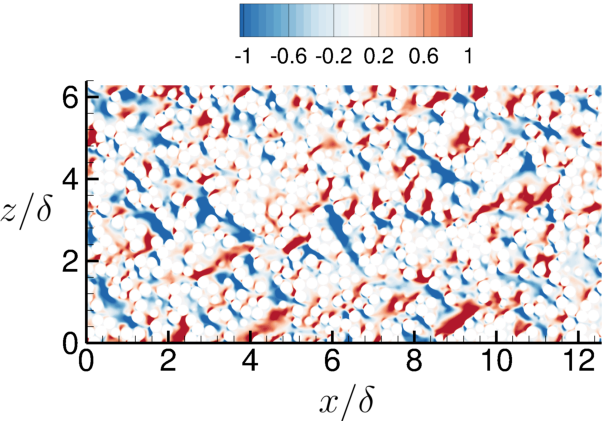}}
   \subfigure[]{
   \includegraphics[width=4.3cm,height=4.3cm,keepaspectratio]{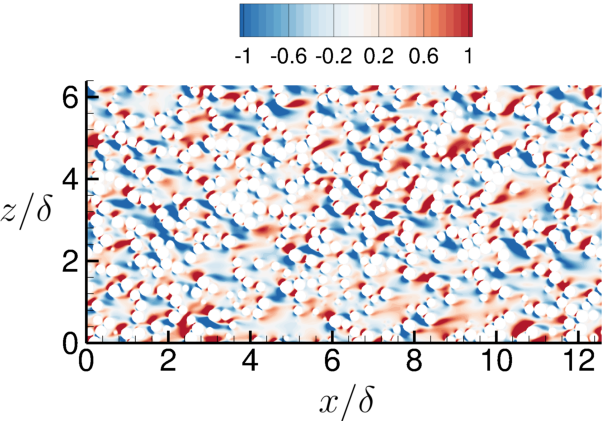}}
   \subfigure[]{
   \includegraphics[width=4.3cm,height=4.3cm,keepaspectratio]{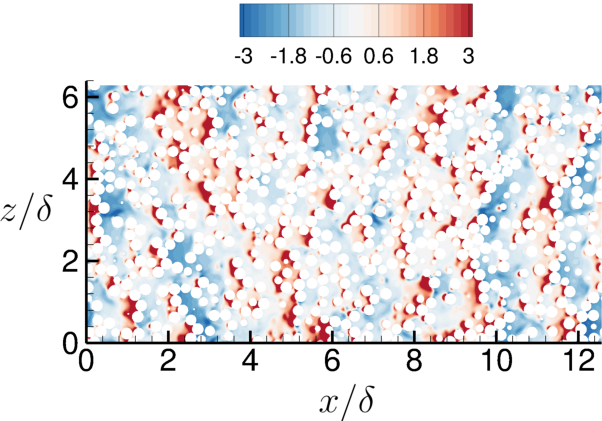}}
   \subfigure[]{
   \includegraphics[width=4.3cm,height=4.3cm,keepaspectratio]{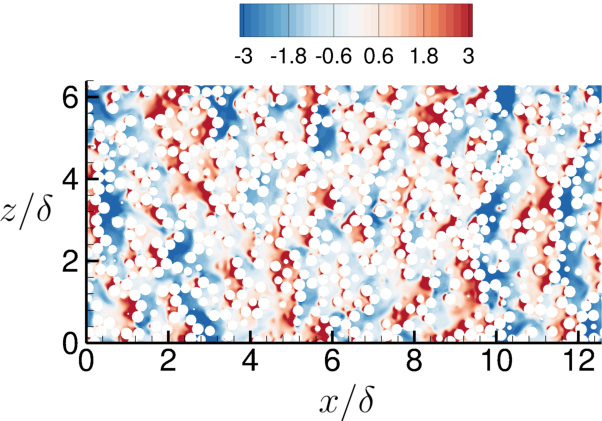}}
   \subfigure[]{
   \includegraphics[width=4.3cm,height=4.3cm,keepaspectratio]{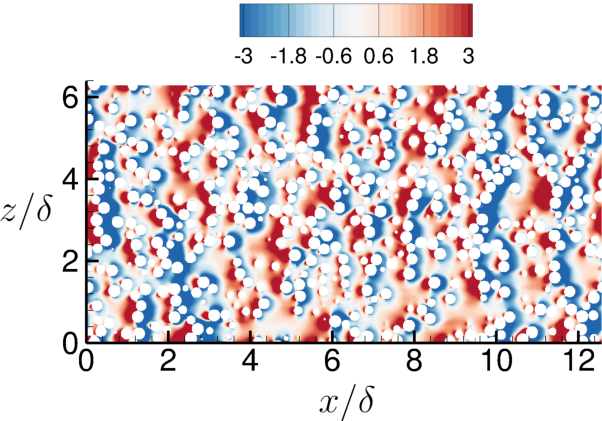}}
\caption{\small Form-induced fluctuations in the $y$--$z$ plane at $y=-d$ showing  (a--c) $\langle\widetilde{u}\rangle^{+}$, (d--f) $\langle\widetilde{v}\rangle^{+}$,  (g--i) $\langle\widetilde{w}\rangle^{+}$, and  (j--l) $\langle\widetilde{p}\rangle^{+}$. Left panel (PB), middle panel (IWF), and right panel (IWH). The velocities are normalized by $u_{\tau}$ and pressure by $\rho u_{\tau}^2$.} 
\label{fig:zerodis_form}
\end{figure}

\begin{figure}
   \centering
   \subfigure[]{
   \includegraphics[width=4.3cm,height=4.3cm,keepaspectratio]{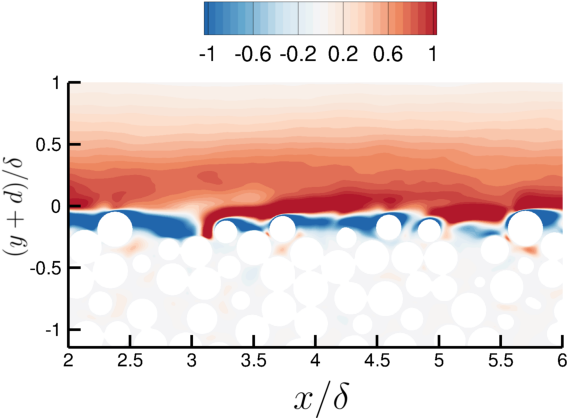}}
   \subfigure[]{
   \includegraphics[width=4.3cm,height=4.3cm,keepaspectratio]{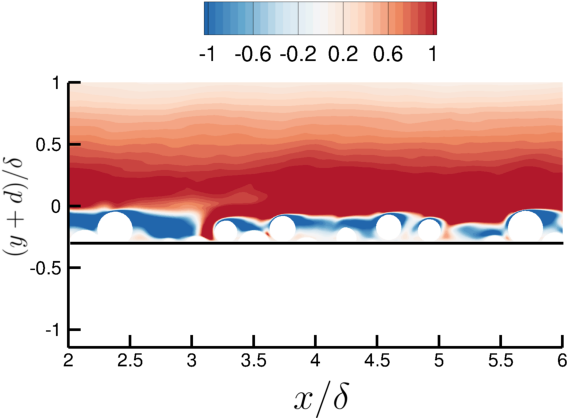}}
   \subfigure[]{
   \includegraphics[width=4.3cm,height=4.3cm,keepaspectratio]{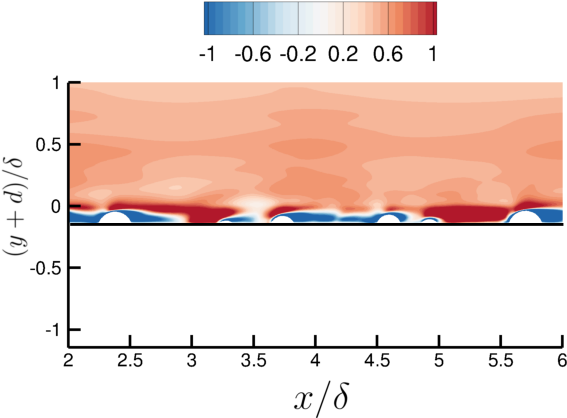}}
    \subfigure[]{
   \includegraphics[width=4.3cm,height=4.3cm,keepaspectratio]{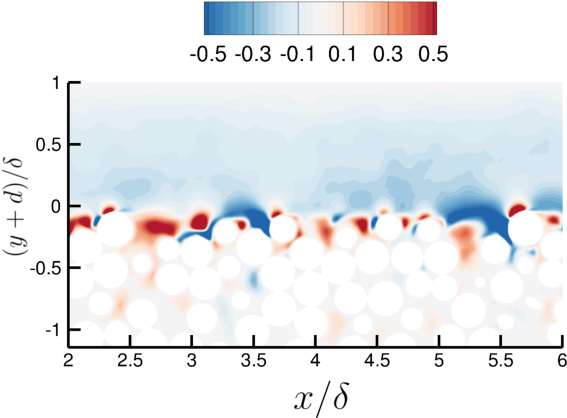}}
   \subfigure[]{
   \includegraphics[width=4.3cm,height=4.3cm,keepaspectratio]{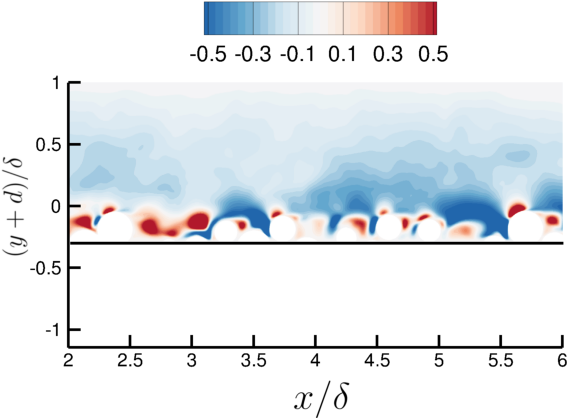}}
   \subfigure[]{
   \includegraphics[width=4.3cm,height=4.3cm,keepaspectratio]{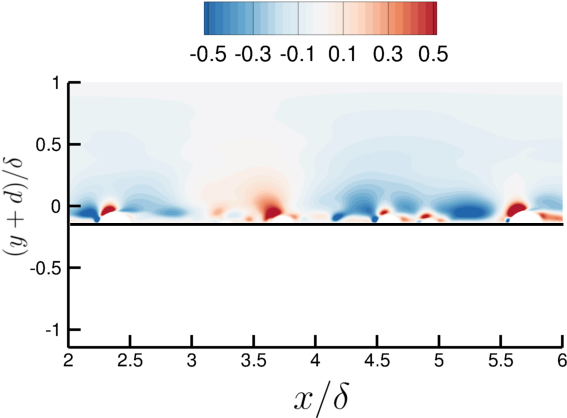}}
   \subfigure[]{
   \includegraphics[width=4.3cm,height=4.3cm,keepaspectratio]{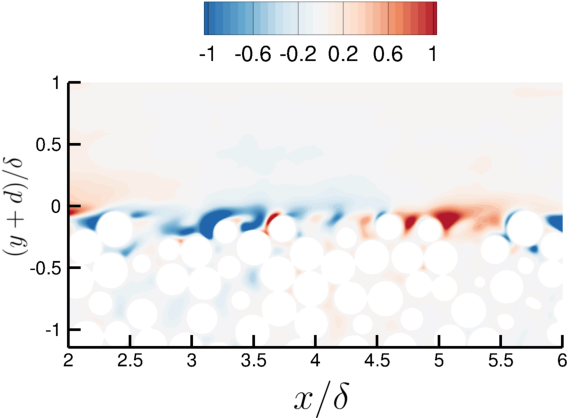}}
   \subfigure[]{
   \includegraphics[width=4.3cm,height=4.3cm,keepaspectratio]{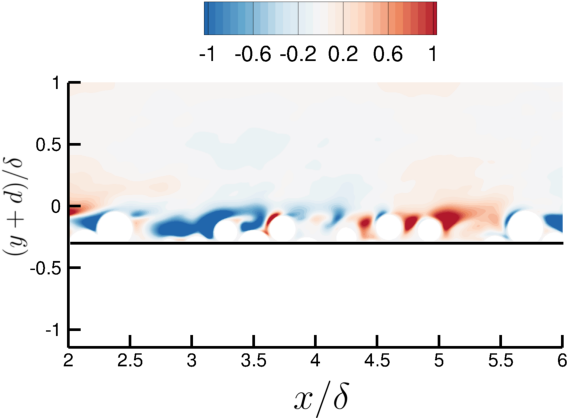}}
   \subfigure[]{
   \includegraphics[width=4.3cm,height=4.3cm,keepaspectratio]{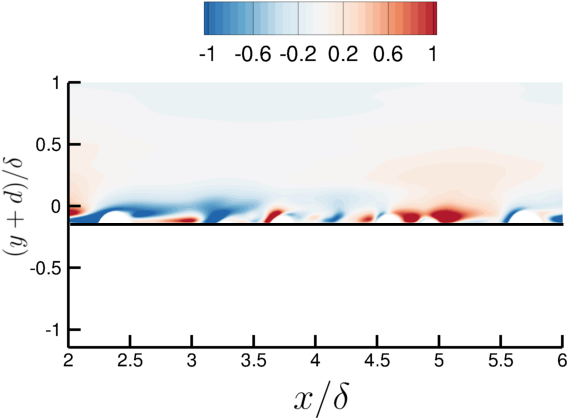}}
   \subfigure[]{
   \includegraphics[width=4.3cm,height=4.3cm,keepaspectratio]{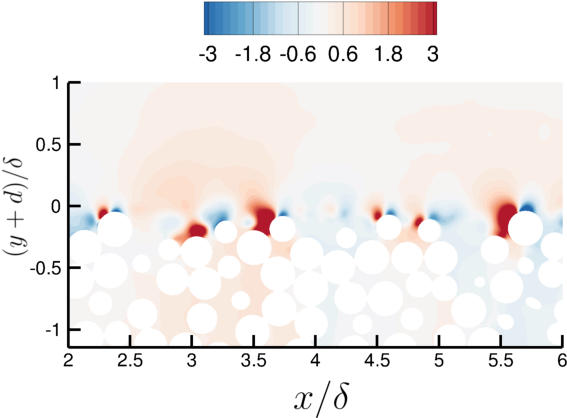}}
   \subfigure[]{
   \includegraphics[width=4.3cm,height=4.3cm,keepaspectratio]{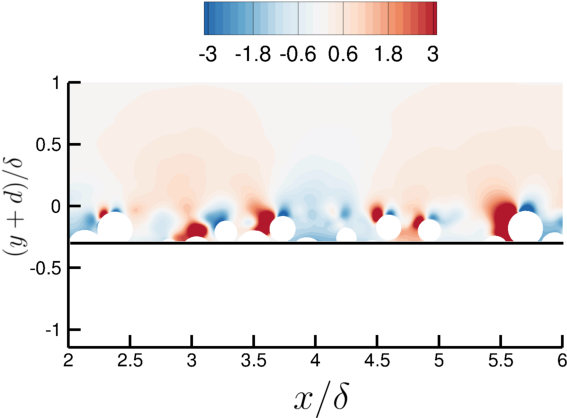}}
   \subfigure[]{
   \includegraphics[width=4.3cm,height=4.3cm,keepaspectratio]{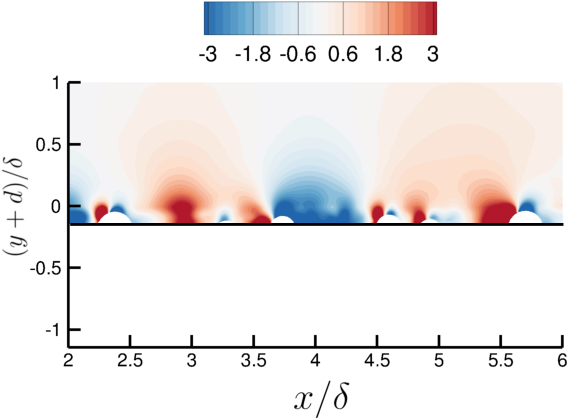}}
   \vspace{-3mm}
\caption{\small Form-induced fluctuations in the $x$--$z$ plane at at the $xy$-symmetry plane showing (a--c) $\langle\widetilde{u}\rangle^{+}$, (d--f) $\langle\widetilde{v}\rangle^{+}$, (g--h) $\langle\widetilde{w}\rangle^{+}$, and  (j--l) $\langle\widetilde{p}\rangle^{+}$. Left panel (PB), middle panel (IWF), and right panel (IWH). The solid black line (\blkline) underneath roughness elements for IWF and IWH cases represents the no-slip wall. The velocities are normalized by $u_{\tau}$ and pressure by $\rho u_{\tau}^2$. 
} 
\label{fig:symmetry_form}
\end{figure}
The double-averaged form-induced fluctuations for the PB, IWF and IWH cases are analyzed at their respective zero-displacement planes ($y=-d$) in figure~\ref{fig:zerodis_form} and at the $xy$-symmetry plane (normal to z) in figure \ref{fig:symmetry_form}, respectively. Qualitative influence of roughness topography and permeability on the turbulence structure and penetration are evident from these figures. 
Figures \ref{fig:zerodis_form}a--c show that large positive $\langle\widetilde{u}\rangle^{+}$ (red region) occurs in the trough regions between the roughness elements, while large negative $\langle\widetilde{u}\rangle^{+}$ (blue region) occurs in wake behind the roughness elements. Figures \ref{fig:symmetry_form}a--c show these negative wake regions coalescing  between adjacent roughness elements. Positive red regions in troughs seem to extend up to the front of the roughness elements. 
For the IWH case (figure~\ref{fig:zerodis_form}c), the streaks of red and blue corresponding to high and low momentum, respectively, are longer than those observed in both IWF (figure~\ref{fig:zerodis_form}b) and PB (figure~\ref{fig:zerodis_form}a) cases. For PB, the shortened streaks are a result of reduced wall blocking effect allowing the turbulence to penetrate into the bed, consistent with the behavior observed by~\citep{breugem2006influence}. The IWF case, with its single layer of particles, also has shorter streaks, showing penetration of turbulence underneath the top layer of the bed. However, the $\langle\widetilde{u}\rangle^{+}$ streaks for this case have less diffusive heads and tails compared to the PB case which can be seen in figures~\ref{fig:zerodis_form}a--c, as the flow is affected by the presence of the underlying impermeable wall.

Figures~\ref{fig:zerodis_form}d--f and \ref{fig:symmetry_form}d--f show the bed-normal double-averaged form-induced fluctuations in the $y=-d$ and $y$-$z$ symmetry plane for the PB, IWF, and IWH cases, respectively. Large positive vertical velocities $\langle\widetilde{v}\rangle^{+}$ (red regions), occur mostly in the wake of the roughness elements (figure~\ref{fig:zerodis_form}d--f)  and near the crests of individual roughness elements (figure~\ref{fig:symmetry_form}d--f). On the other hand, large negative downward velocities $\langle\widetilde{v}\rangle^{+}$ (blue regions) are observed in front of the roughness elements (figure~\ref{fig:symmetry_form}d--f) and in the trough regions (figure~\ref{fig:zerodis_form}d--f). In the PB case (figures~\ref{fig:symmetry_form}d), penetration of $\langle\widetilde{v}\rangle^{+}$ (shades of negative blue and positive red contours) are observed deep inside the bed. Figure~\ref{fig:symmetry_form}d qualitatively demonstrates an important distinction in turbulent flow structures between the permeable bed and impermeable full layer rough wall. Since turbulent fluctuations due to spatial in-homogeneity can penetrate deeper in a permeable bed, they can help in mass transport of solutes deeper compared to impermeable full layer case with an impenetrable wall underneath. Therefore this indicates that even though the PB and IWF case show similar momentum transport characteristics, mass transport behavior could be very different and needs further investigation. 

Figures~\ref{fig:zerodis_form}g--i and \ref{fig:symmetry_form}g--i show the spanwise double-averaged fluctuations in the $y=-d$ and $y$-$z$ symmetry planes for the PB, IWF, and IWH cases, respectively. Pairs of large positive $\langle\widetilde{w}\rangle^{+}$  (red regions) and large negative  $\langle\widetilde{w}\rangle^{+}$ (blue regions) are observed in front of the roughness crests (figures~\ref{fig:zerodis_form}g--i). This suggests that the interaction of turbulence with the spherical roughness elements produces a pair of clockwise and anti-clockwise vortices in front of them. For the IWH case, the size of these vortices is bigger followed by the IWF and the PB cases. Due to the wall blocking effect of the underlying no-slip wall in the impermeable cases, the pair of counter rotating vortices are stretched tangential to the wall, whereas in the permeable bed cases the vortices break down sooner due to flow penetration into the bed (figure~\ref{fig:symmetry_form}g--i).

Contours of form-induced pressure fluctuations are shown in figures \ref{fig:zerodis_form}j--l and \ref{fig:symmetry_form}j--l in the two different planes. High $\langle\widetilde{p}\rangle^{+}$ (red regions) is observed in front of the roughness elements, where the flow stagnates, whereas low $\langle\widetilde{p}\rangle^{+}$ (blue regions) is observed behind the roughness elements~(figure \ref{fig:zerodis_form}j--l), and at roughness elements crests ~(figure\ref{fig:symmetry_form}j--l) in all the three cases. This shows that the protruding spherical roughness elements experience form drag. In the PB case, the regions of high and low pressures (red and blue, respectively) start to diffuse a little further away from the bed as permeability allows for penetration of pressure fluctuations (figure~\ref{fig:zerodis_form}j). However, in the IWF case, figure~\ref{fig:zerodis_form}k, the underlying impermeable wall blocks the penetration of pressure fluctuations, resulting in both higher magnitude (dark red and blue regions) of pressures stretched further away from the roughness elements. The IWH case,  figure~\ref{fig:zerodis_form}l, shows same qualitative behavior as the IWF case, but with even larger magnitude (darker red and blue regions) of pressures stretched much farther away from the roughness elements.

\begin{figure}
   \centering
   \subfigure[]{
   \includegraphics[width=4.0cm,height=4.0cm,keepaspectratio]{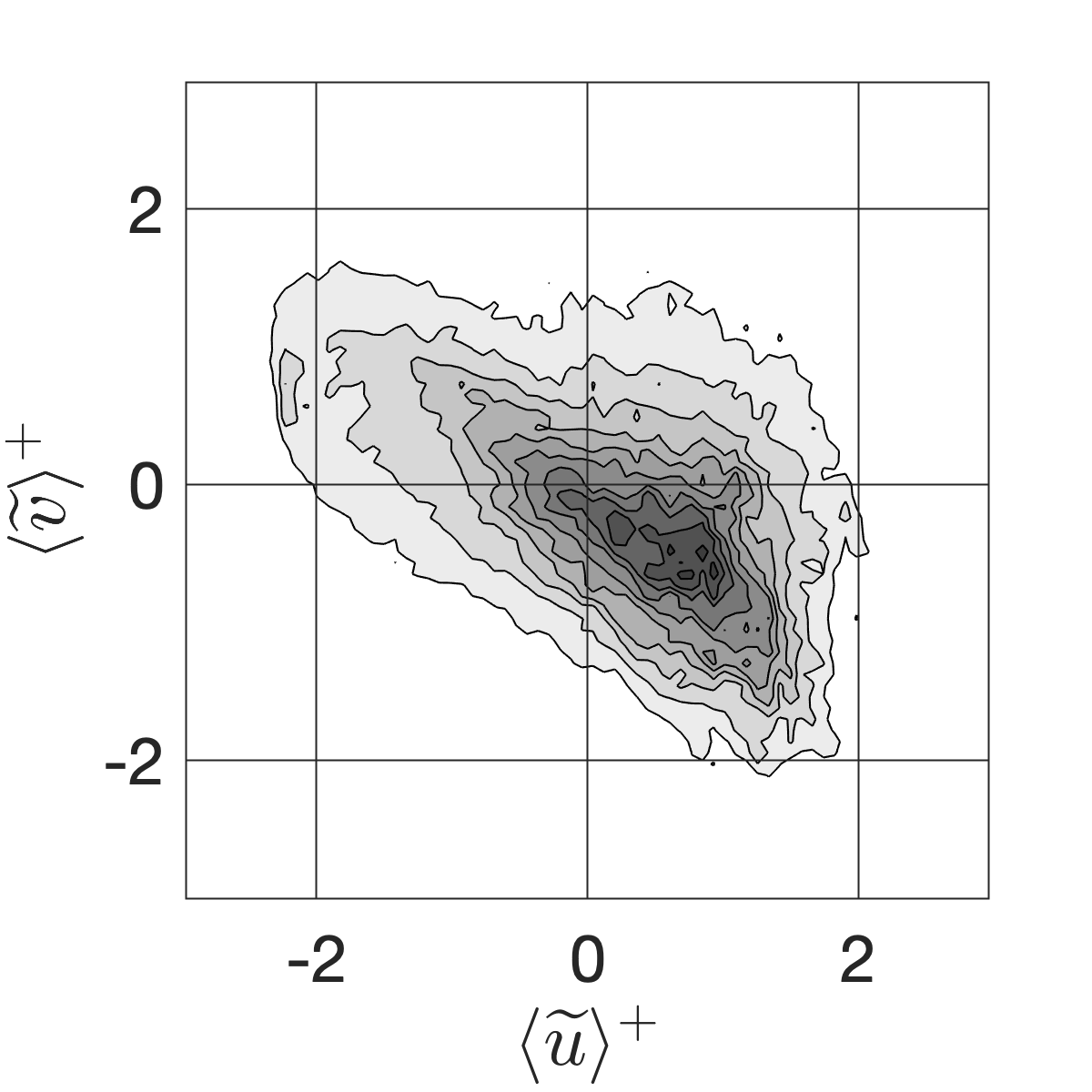}}
   \subfigure[]{
   \includegraphics[width=4.0cm,height=4.0cm,keepaspectratio]{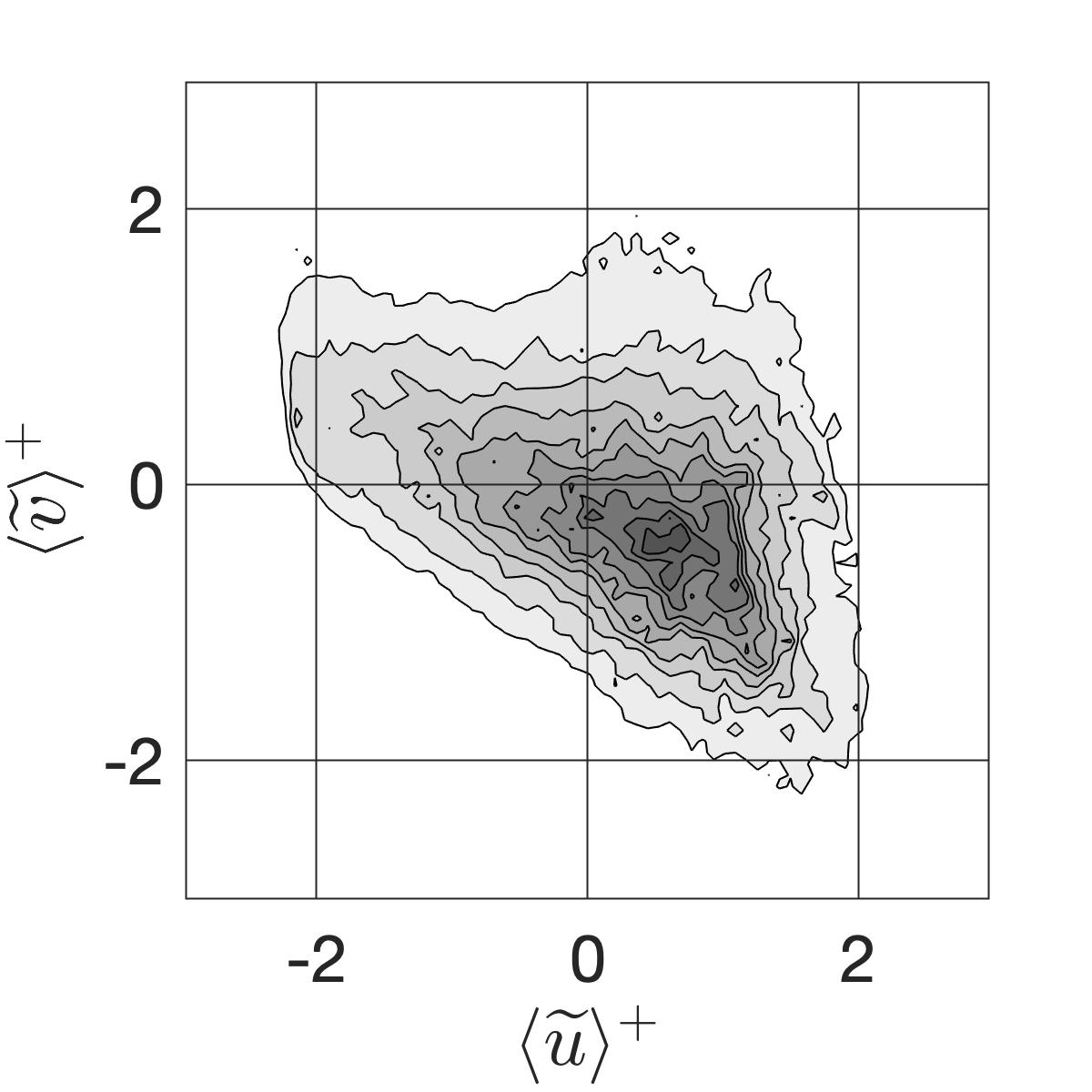}}
   \subfigure[]{
   \includegraphics[width=4.6cm,height=4.0cm,keepaspectratio]{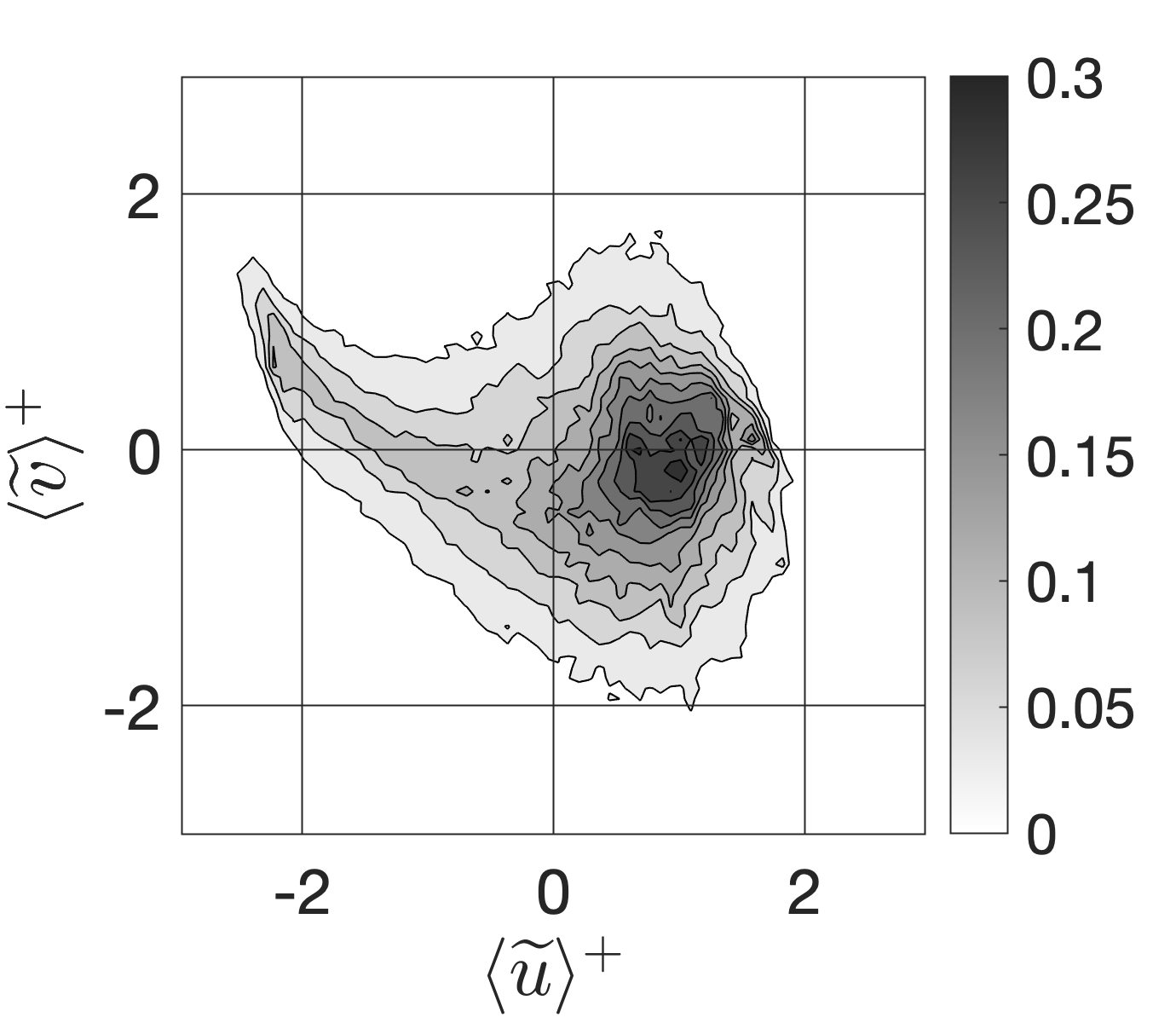}}
    \subfigure[]{
   \includegraphics[width=4.0cm,height=4.0cm,keepaspectratio]{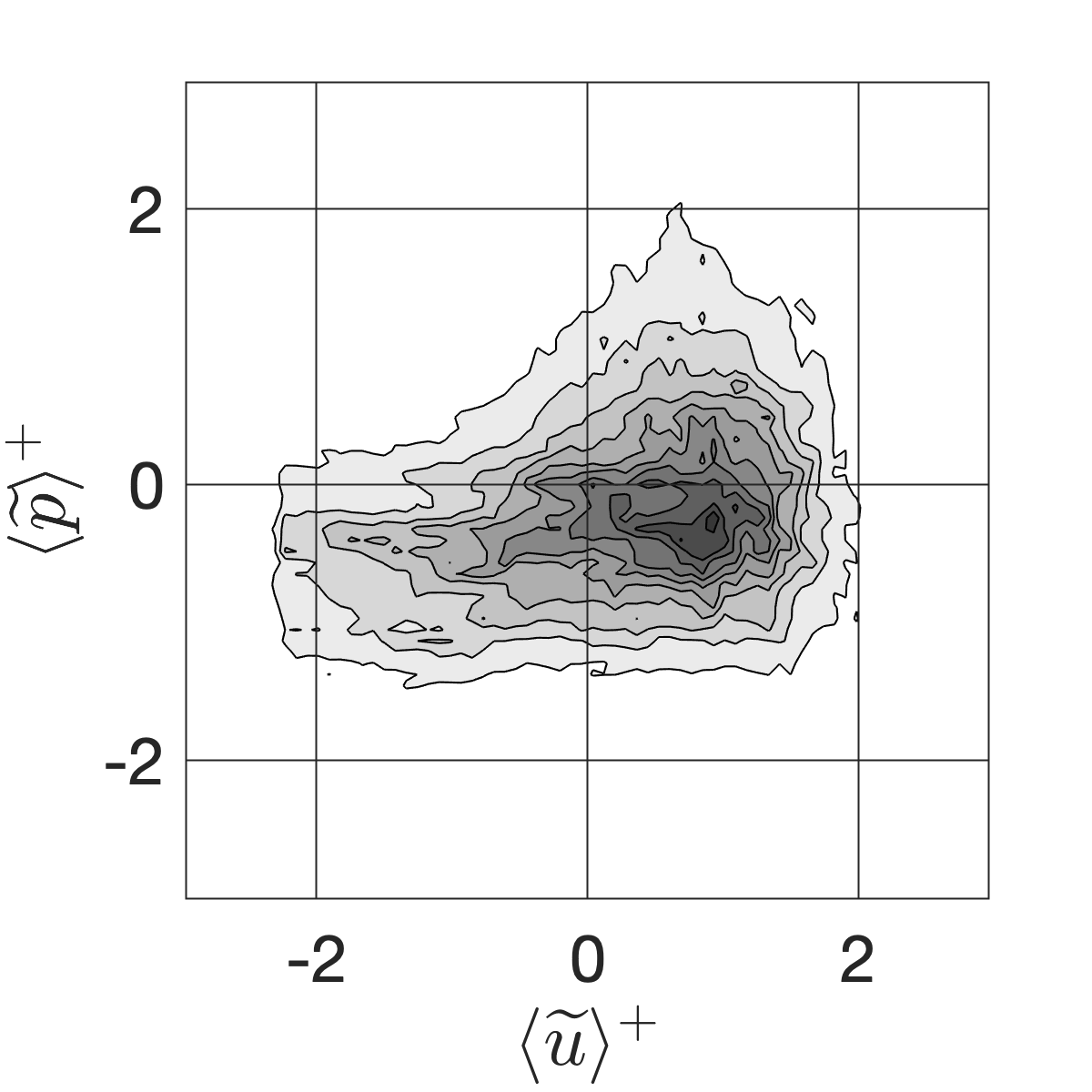}}
   \subfigure[]{
   \includegraphics[width=4.0cm,height=4.0cm,keepaspectratio]{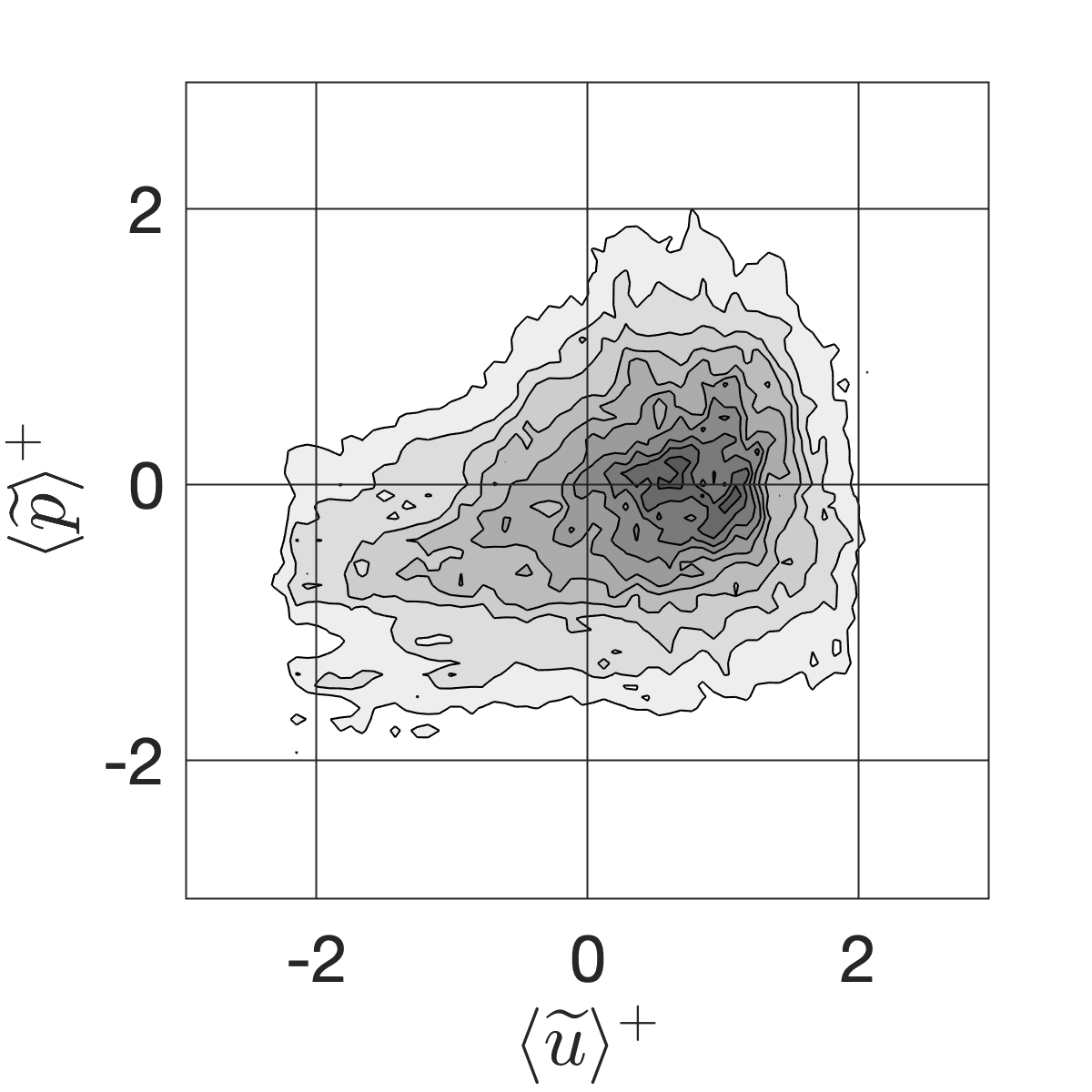}}
   \subfigure[]{
   \includegraphics[width=4.6cm,height=4.0cm,keepaspectratio]{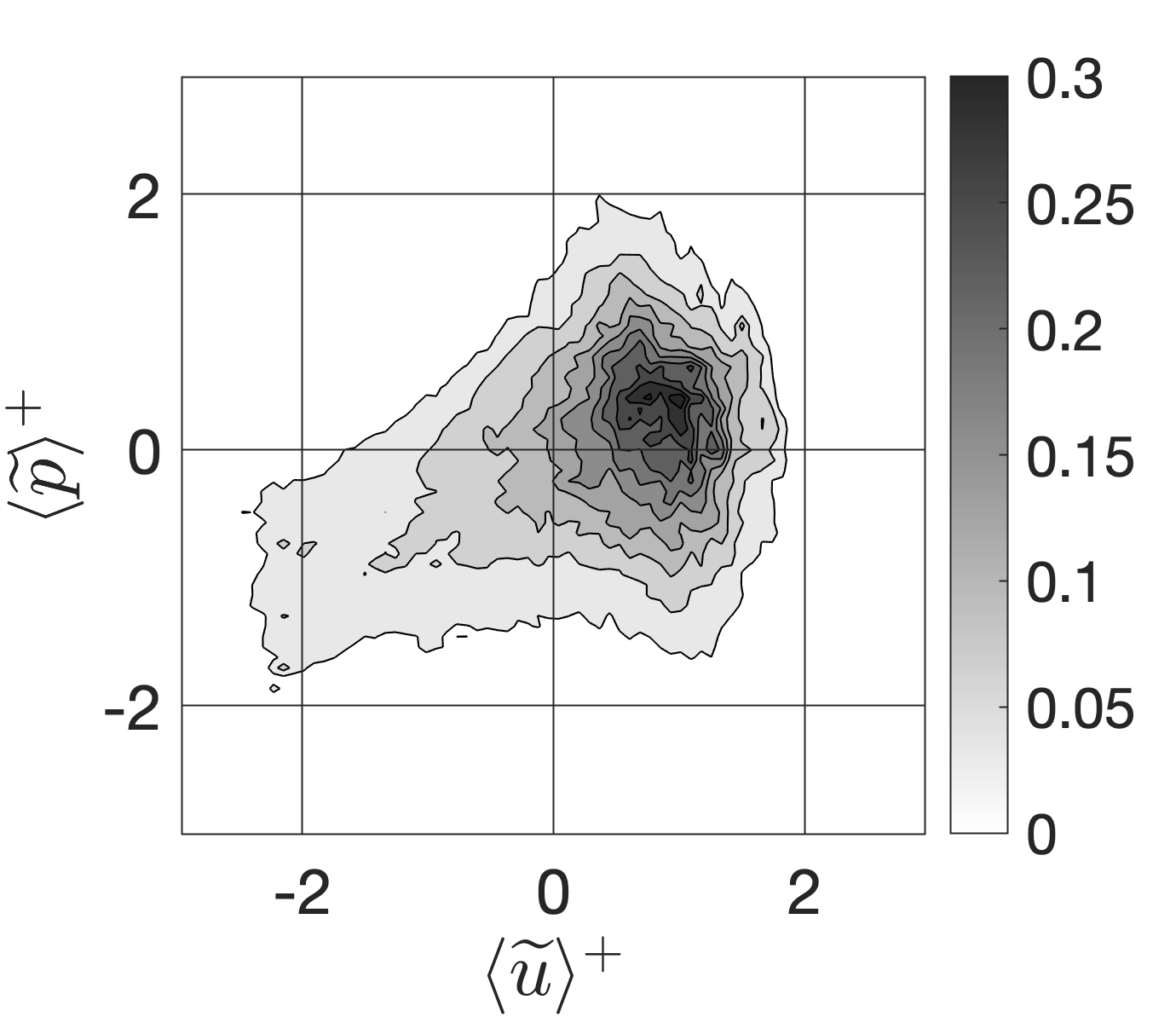}}
     \subfigure[]{
   \includegraphics[width=4.0cm,height=4.0cm,keepaspectratio]{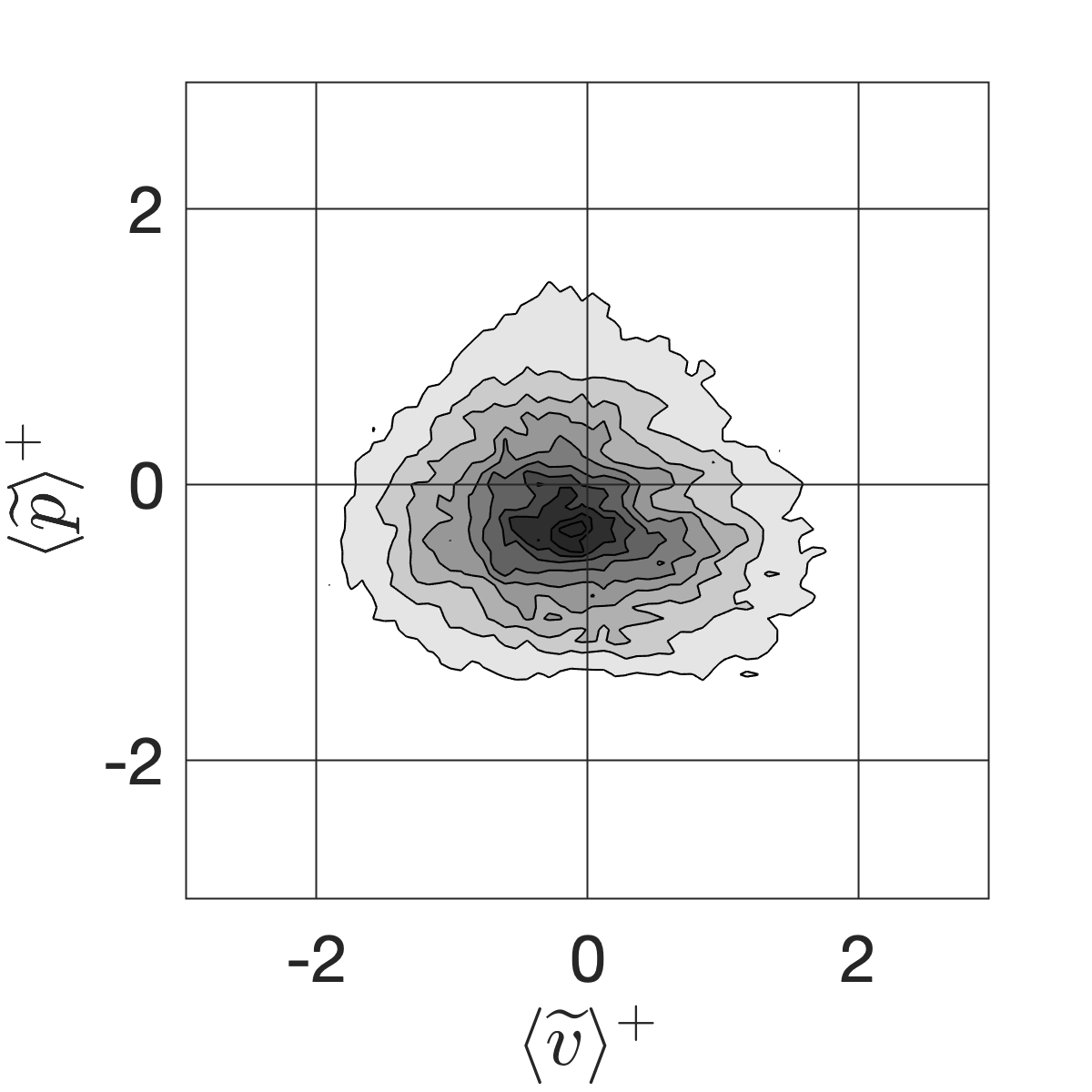}}
   \subfigure[]{
   \includegraphics[width=4.0cm,height=4.0cm,keepaspectratio]{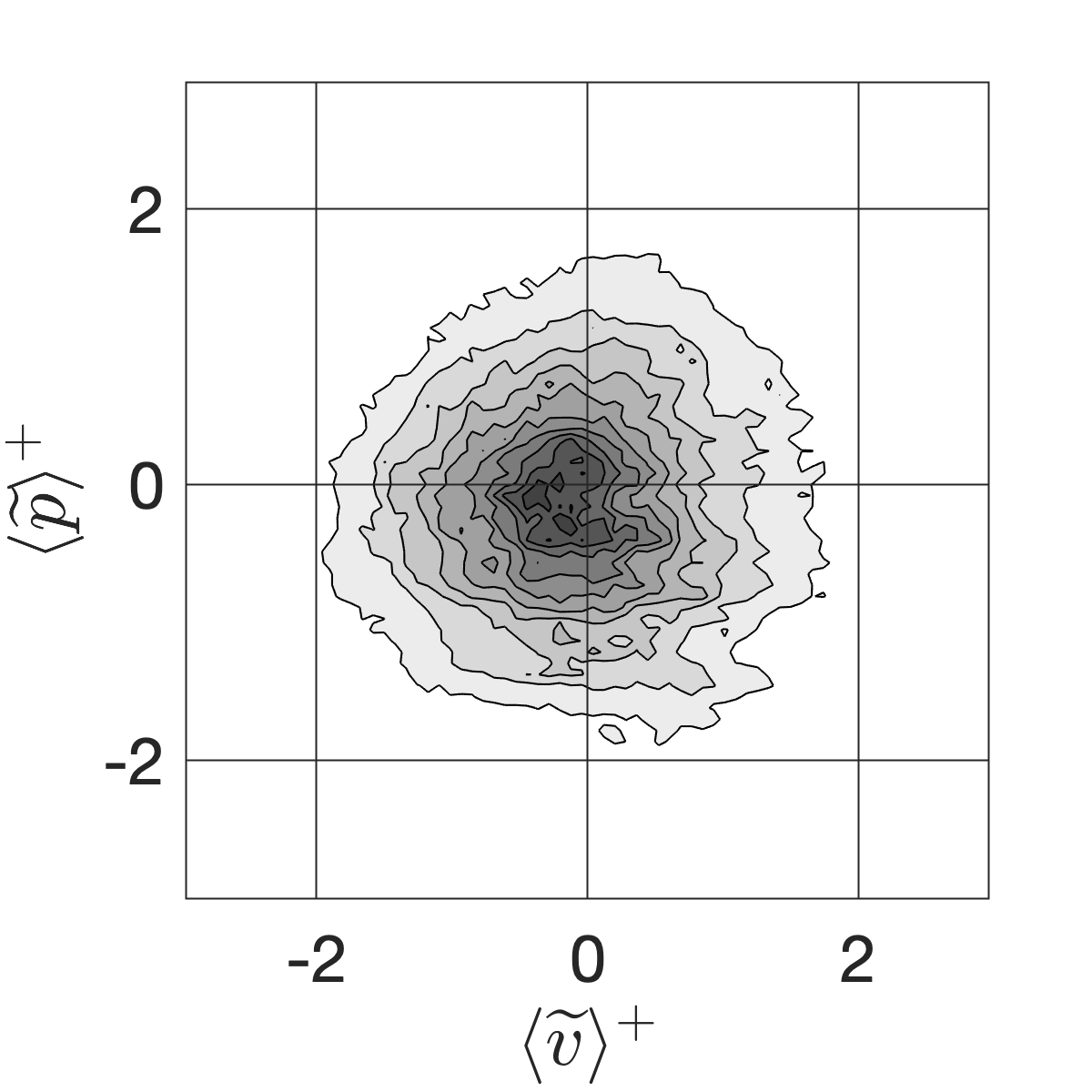}}
   \subfigure[]{
   \includegraphics[width=4.6cm,height=4.0cm,keepaspectratio]{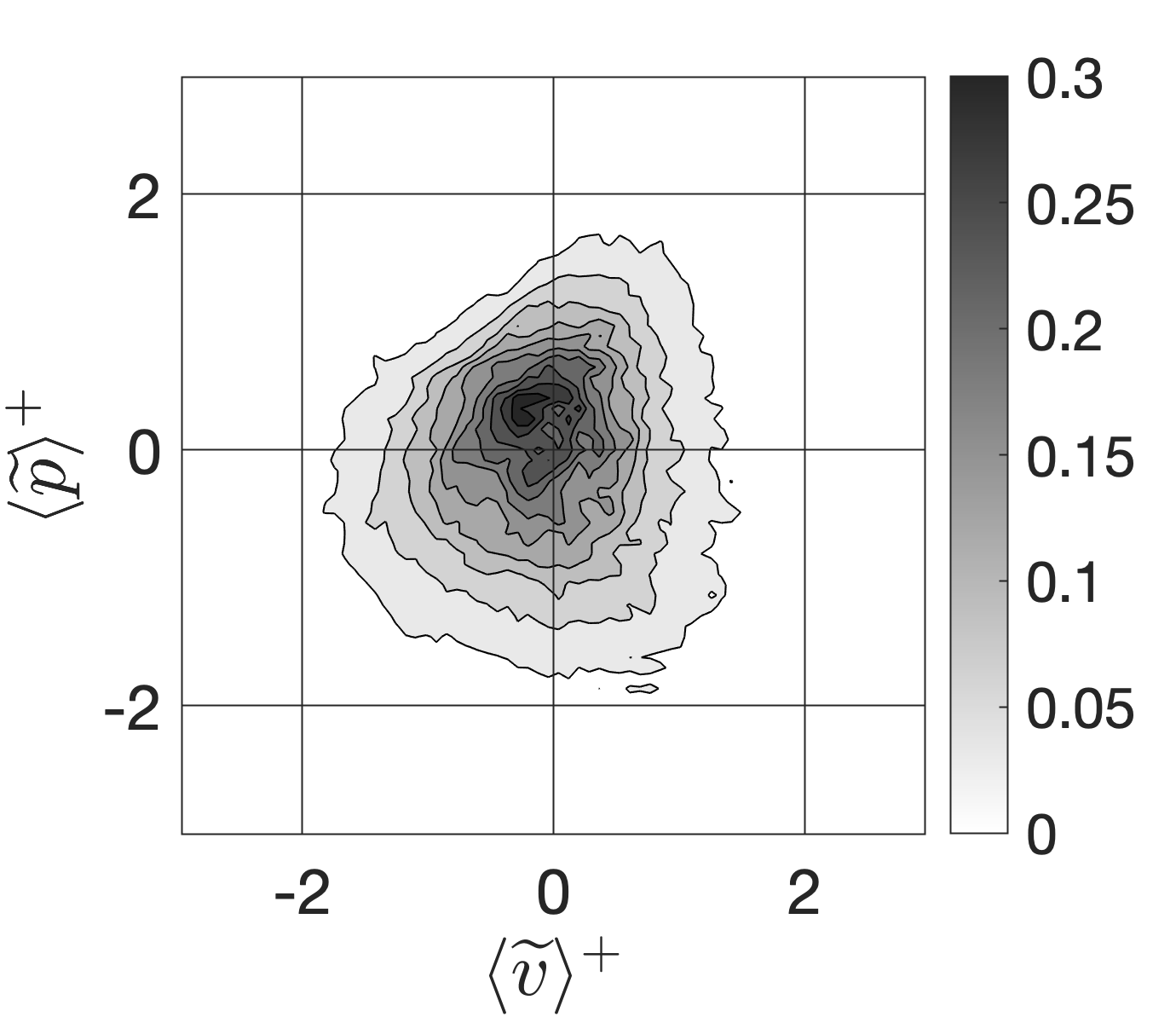}}
\caption{\small Joint PDFs of dispersive stresses for the PB (left panel), IWF (middle panel), and IWH (right panel) cases at their respective zero-displacement planes, $y=-d$: (a--c) ($\widetilde{u}^{+}$, $\widetilde{v}^{+}$), (d--f) ($\widetilde{u}^{+}$, $\widetilde{p}^{+}$), (g--i) ($\widetilde{v}^{+}$, $\widetilde{p}^{+}$). The velocities have been normalized by $u_{\tau}$ and pressure by $\rho u_{\tau}^2$.}
\label{fig:jpdf}
\end{figure}

Upwelling (ejection) and downwelling (sweep) events are analyzed through joint PDFs of ($\widetilde{u}^{+}$, $\widetilde{v}^{+}$), ($\widetilde{u}^{+}$, $\widetilde{p}^{+}$), ($\widetilde{v}^{+}$, $\widetilde{p}^{+}$) at their respective zero-displacement planes. 
Figures~\ref{fig:jpdf}a,\ref{fig:jpdf}b,\ref{fig:jpdf}c show the joint PDFs of ($\widetilde{u}^{+}$, $\widetilde{v}^{+}$) for the PB, IWF and IWH cases. A strong correlation is observed in the second and fourth quadrants. A positive $\widetilde{u}^{+}$ is correlated with a negative $\widetilde{v}^{+}$, in the fourth quadrant, results in a downwelling motion and negative $\widetilde{u}^{+}$ and positive $\widetilde{v}^{+}$, in the second quadrant, results in an upwelling motion. For the IWH case, negative $\widetilde{v}^{+}$ events are less likely to happen due to underlying impermeable-wall blocking effect. Long tail is observed where negative $\widetilde{u}^{+}$ corresponds to positive $\widetilde{v}^{+}$ representative of packets of fluid being pushed away from the wall.

Figures~\ref{fig:jpdf}d,\ref{fig:jpdf}e,\ref{fig:jpdf}f show the joint PDFs of ($\widetilde{u}^{+}$, $\widetilde{p}^{+}$) for the PB, IWF and IWH cases. A strong correlation is observed in the third and fourth quadrants. A negative $\widetilde{p}^{+}$ correlated with a negative $\widetilde{u}^{+}$ in the third quadrant is representative of an upwelling motion, whereas negative $\widetilde{p}^{+}$ and positive $\widetilde{u}^{+}$ in the fourth quadrant results in an downwelling motion. For the IWH case, stronger correlation is observed in the first quadrant compared to the PB and IWF cases. This happens in trough regions surrounded by particles which are clustered together resulting in a stronger positive $\widetilde{p}^{+}$ corresponding to a positive $\widetilde{u}^{+}$ and a downwelling motion. Negative $\widetilde{p}^{+}$ events are less likely to happen in the IWH case due to the presence of wall-blocking effect due to an impermeable solid wall.

Figures~\ref{fig:jpdf}g,\ref{fig:jpdf}h,\ref{fig:jpdf}i show the joint PDFs of ($\widetilde{v}^{+}$, $\widetilde{p}^{+}$). A strong correlation is observed in the third and fourth quadrants for PB and IWF cases. A negative $\widetilde{p}^{+}$ is correlated with a positive $\widetilde{v}^{+}$ in the third quadrant, resulting in a upwelling motion. Similarly, a negative $\widetilde{p}^{+}$ correlated with a negative $\widetilde{v}^{+}$ in the fourth quadrant corresponds to a downwelling motion. For the IWH case, fewer negative $\widetilde{p}^{+}$ events are likely to happen, due to the underlying impermeable wall. 

\subsection{Turbulent pressure fluctuations}\label{sec:pres_fluc}
Pressure fluctuations at SWI play a critical role in hyporheic transport. Specifically, pressure fluctuations due to turbulence are conjectured to have significant impact on mass transport within the hyporheic zone as it can directly influence the residence times through turbulent advection. Pressure fluctuation statistics for the different cases are compared first.

\begin{figure}
   \centering
   \subfigure[]{
   \includegraphics[width=6cm,height=6cm,keepaspectratio]{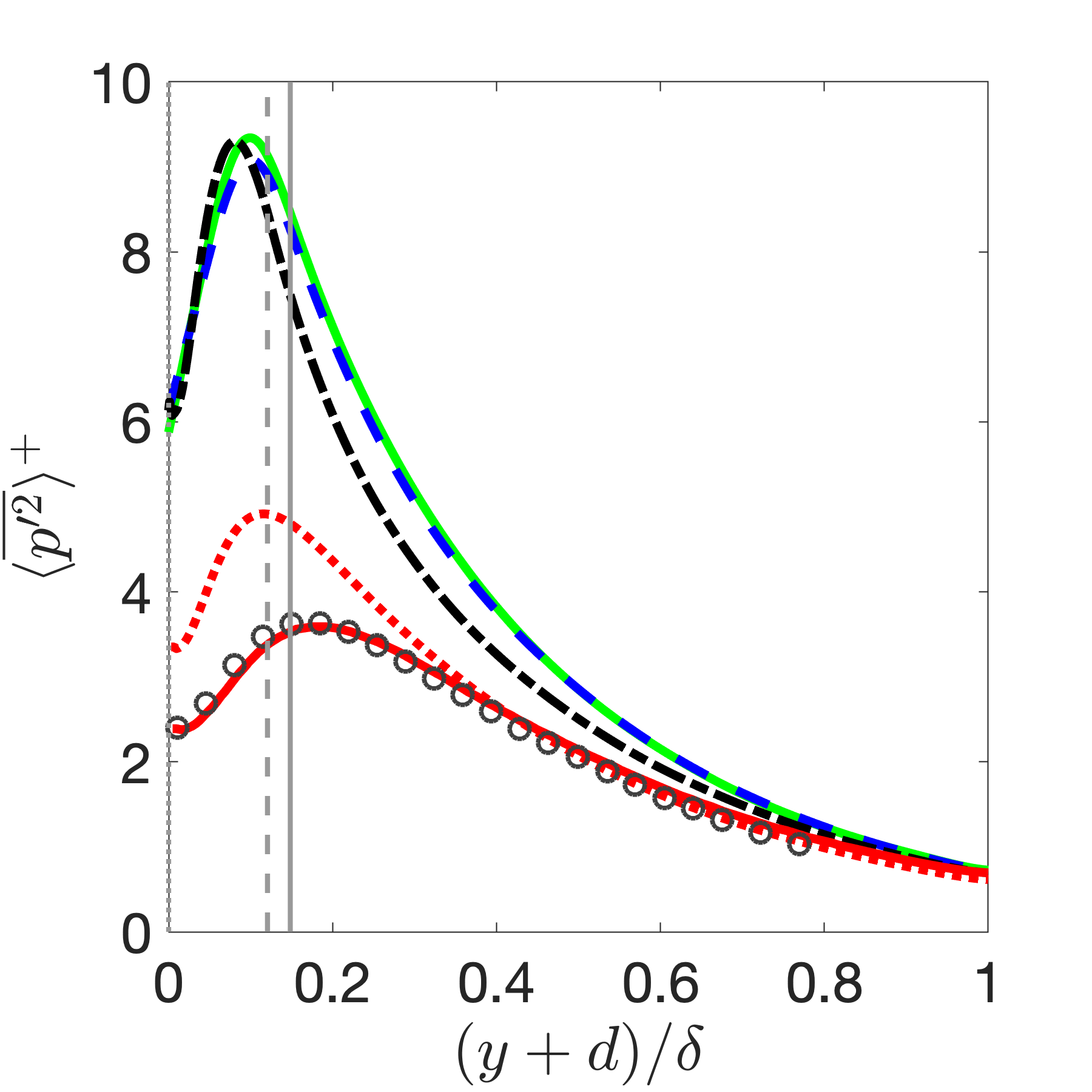}}
    \subfigure[]{
   \includegraphics[width=6cm,height=6cm,keepaspectratio]{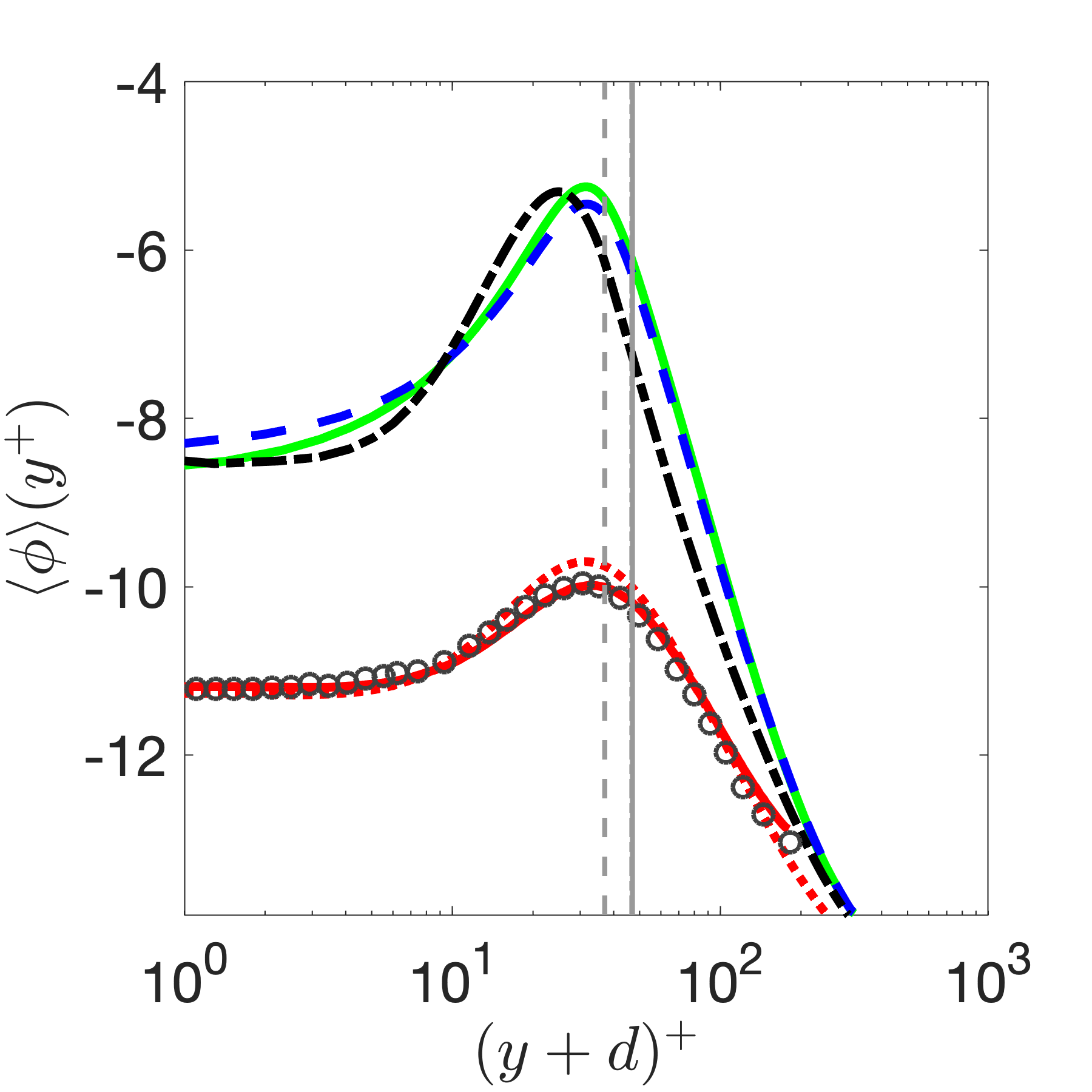}}
\caption{\small Profiles of (a) mean-square pressure fluctuations $\langle\overline{p^{\prime2}}\rangle^{+}$ (pressure is normalized by $\rho u_{\tau}^2$) and (b) inner correlation $\langle{\phi}\rangle(y^+)$ for PB (\greenline), IWF (\bluedashline), IWH (\blkdashdotline), SW (\reddottedline), and SW180 (\redline) cases. The symbols (\darkgraycircle) show data for $Re_{\tau}$ = 182 from \citet{panton2017correlation}. Vertical lines show the crest of sediment bed for each case with the following legend: PB (\grayline), IWF (\graydashdotline), and IWH (\graydashline). (Note that PB and IWF sediment crest lines overlap). }
\label{fig:pres_fluc}
\end{figure}

Figure~\ref{fig:pres_fluc}a shows the mean square pressure fluctuations, $\langle\overline{p^{\prime2}}\rangle^{+}$, for the permeable bed, impermeable full and half layers, and the smooth wall cases. A smooth wall case (SW180) at $Re_{\tau} = 180$ was simulated and compared with smooth wall data for $Re_{\tau} = 182$ from \citet{panton2017correlation}. The results  compare well with the data collapsing on to the logarithmic matching correlation given by \citet{panton2017correlation}
\begin{align}
     \langle\overline{p^{\prime2}}\rangle_{cp}^{+}(y/\delta) &= -2.5625\ln(y/\delta) + 0.2703.
    \label{eq:ppsq}
\end{align}
For the rough wall and permeable bed cases, $\langle\overline{p^{\prime2}}\rangle^{+}$ values collapse on the correlation given by~\ref{eq:ppsq} for $(y+d)/\delta > 0.95$.
For the smooth wall case, it is seen that the peak value for $\langle\overline{p^{\prime2}}\rangle^{+}$ occurs at $y^+ \sim 30$. In contrast, for the rough wall and permeable bed cases, the peak value for $\langle\overline{p^{\prime2}}\rangle^{+}$ happens underneath the sediment crest level. The roughness elements significantly increase the pressure fluctuations compared to the smooth wall case for $y^+<30$. The maximum value of $\langle\overline{p^{\prime2}}\rangle^{+}$ for PB case is $\sim 47.5\%$ greater than than the SW case. On the other hand, permeability has a smaller effect on increase in magnitude of $\langle\overline{p^{\prime2}}\rangle^{+}$ with the difference in peak values of $\sim 2.33\%$ between PB and IWF cases and $\sim 0.57\%$ between the PB and IWH cases. Small differences are observed in the free-stream for the pressure fluctuations in the IWH case as compared to the PB and IWF cases.

\citet{panton2017correlation} also derived a inner function correlation to absorb the $Re_{\tau}$ dependence in $\langle\overline{p^{\prime2}}\rangle^{+}$,
\begin{align}
    \langle\overline{\phi}\rangle(y^+)  &= \langle\overline{p^{\prime2}}\rangle^{+}(y^+,Re_{\tau})-2.5625\ln(Re_{\tau}) - 0.2703.
    \label{eq:phi}
\end{align}
Figure~\ref{fig:pres_fluc}b shows the $\langle\overline{\phi}\rangle(y^+)$ computed for PB, IWF, IWH and SW cases. The data for $\langle\overline{\phi}\rangle(y^+)$, at $Re_{\tau} = 182$,  from \citep{panton2017correlation} is also shown, comparing well with SW180  $\langle\overline{\phi}\rangle(y^+)$ profile. Both the rough wall cases and the permeable bed do not collapse onto the correlation given by~\ref{eq:phi} but come close to it for $(y+d)^{+}>200$.

\subsection{Double averaged TKE Budget}\label{sec:tke_budget}
The double-averaged TKE budget for all the four cases is analyzed to understand the role of different terms in energy production, transport, and dissipation. Following the literature on flow over macro-roughness beds~\citep{mignot2009double} and canopies~\citep{raupach1982averaging,raupach1981turbulence} the equations become, 
\begin{multline}
\label{eq:TKE_budget_equation}
{\partial \langle\overline{u_i^{\prime}u_i^{\prime}}\rangle/2 \over \partial t} = 
\underbrace{-\langle\overline{u_i^{\prime}v^{\prime}}\rangle  {\partial \langle\overline{u_i}\rangle \over \partial y}}_{P_s} + 
\left[-\underbrace{\Bigg \langle\widetilde{u_i^{\prime}u_j^{\prime}}  {\partial \tilde{u_i} \over \partial x_j}\Bigg \rangle}_{P_w}  -\underbrace{\langle\overline{u_i^{\prime}u_j^{\prime}}\rangle           \Bigg \langle{\partial \tilde{u_i} \over \partial x_j}\Bigg \rangle}_{P_m}\right] \\
- {\partial \over \partial y} \left[ \underbrace{\langle\overline{u_i^{\prime}u_i^{\prime}v^{\prime}}\rangle/2}_{T_t} + \underbrace{\langle\widetilde{u_i^{\prime}u_i^{\prime}}\tilde{v}\rangle/2}_{T_w}\right] \underbrace{-{1 \over \rho} {\partial \over \partial y}     {\langle\overline{p^{\prime}v^{\prime}}\rangle}}_{T_p}  \underbrace{+{\nu} {\del^2 \over \del y^2}     {\langle\overline{u_i^{\prime}u_i^{\prime}}\rangle/2}}_{T_{\nu}}  - \langle{\overline{\epsilon}}\rangle.
\end{multline}

Following the description given by~\citet{ghodke2016dns}, the eight terms on the right hand side of the equation~\ref{eq:TKE_budget_equation} are defined as  follows: the shear production term, $P_s$, represents the work of the double-averaged velocity against the double averaged shear; the wake production term, $P_w$, is the work of wake-induced velocity fluctuations against the bed-induced shear; form induced production, $P_m$, is the work of the bed-induced velocity fluctuations against double-averaged shear; $\del{T_t}/\del{y}$, is turbulent transport; $\del{T_w}/\del{y}$, is the bed-induced turbulent transport; $T_p$ is pressure transport; $T_{\nu}$ is viscous transport of TKE and the last term on
the right represents viscous dissipation, $\epsilon$. Here the terms $P_m$, $P_w$ and $T_w$ arise as a result of spatial heterogeneity at the roughness element length scale. 

\begin{figure}
   \centering
   \subfigure[]{
   \includegraphics[width=4.3cm,height=4.3cm,keepaspectratio]{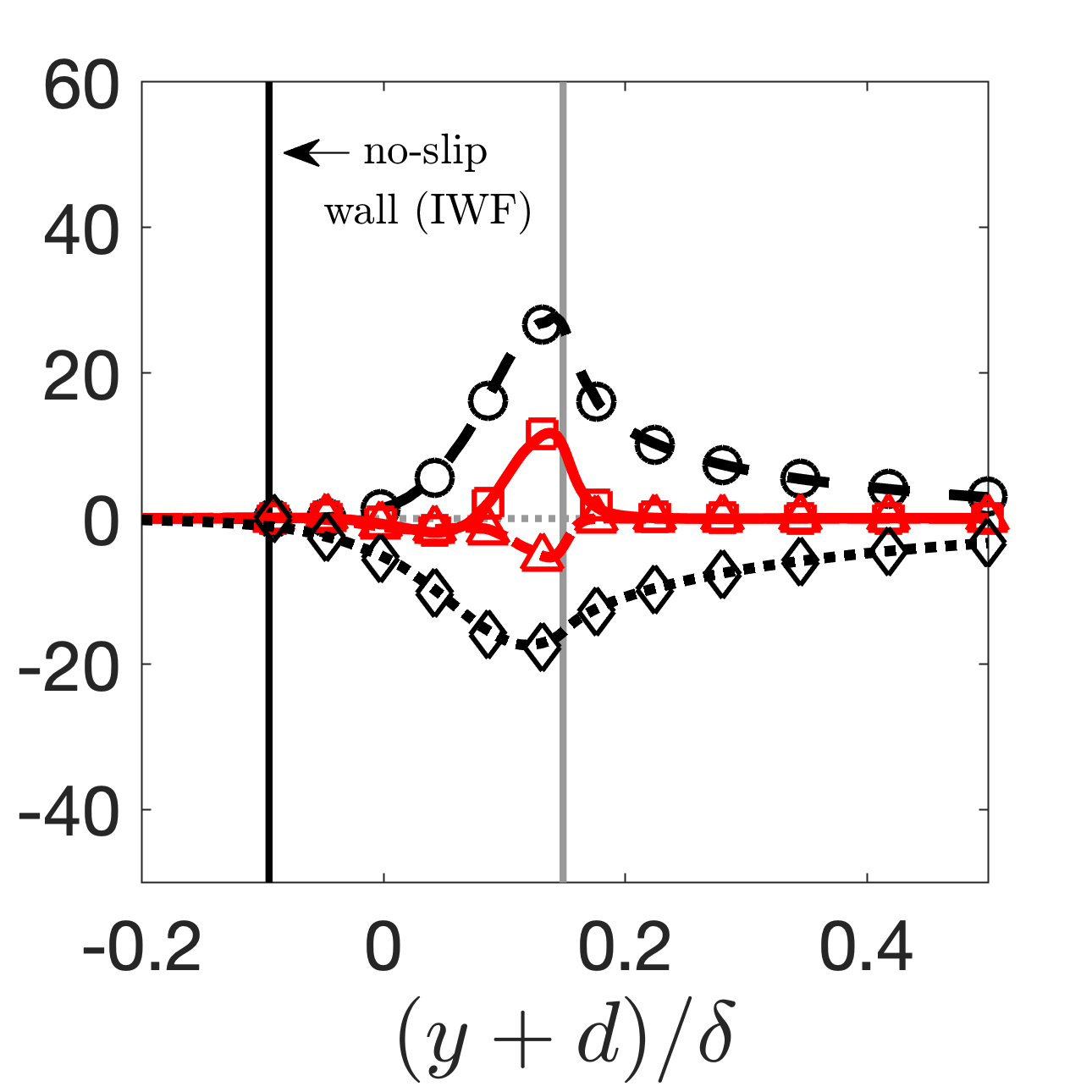}}
    \subfigure[]{
   \includegraphics[width=4.3cm,height=4.3cm,keepaspectratio]{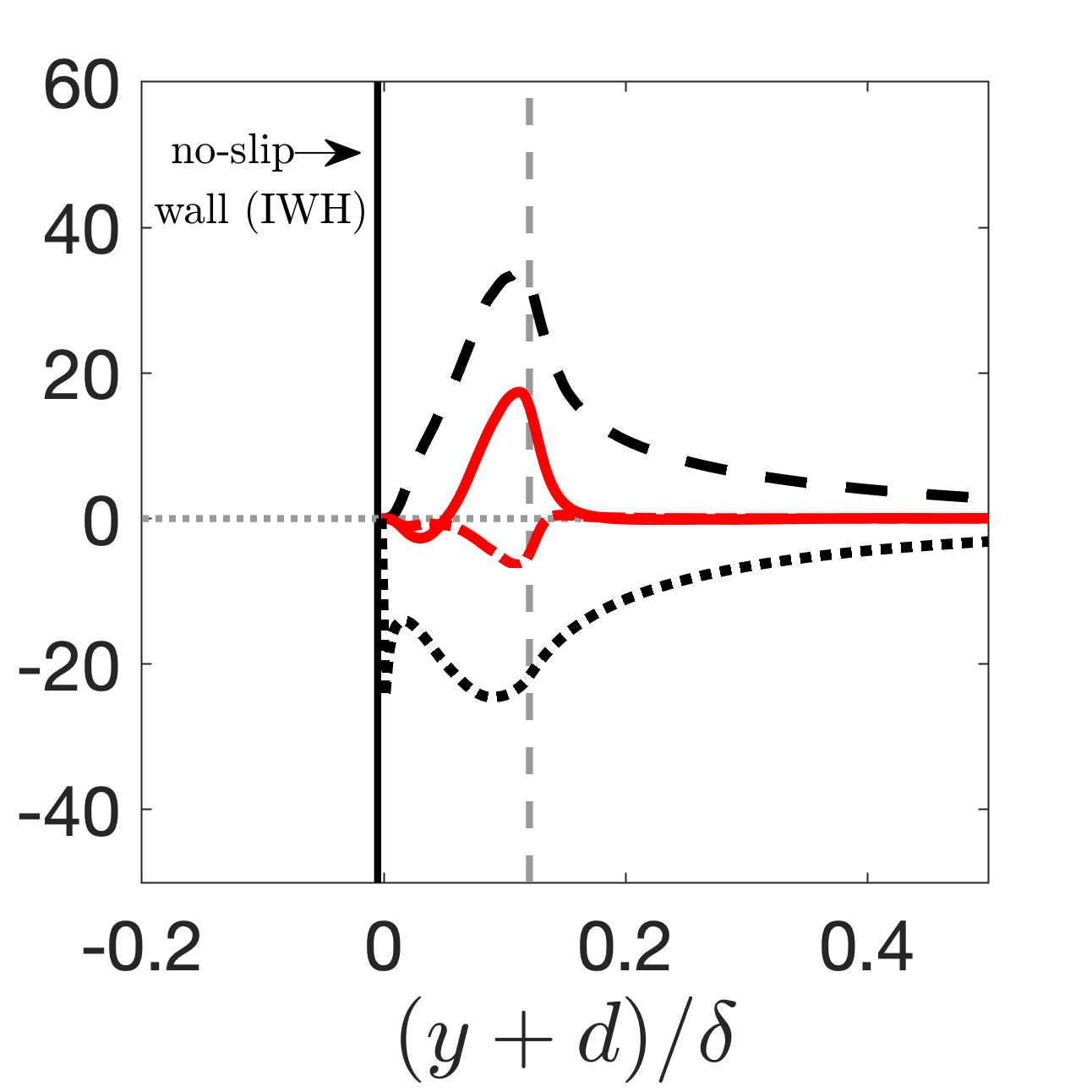}}
   \subfigure[]{
  \includegraphics[width=4.3cm,height=4.3cm,keepaspectratio]{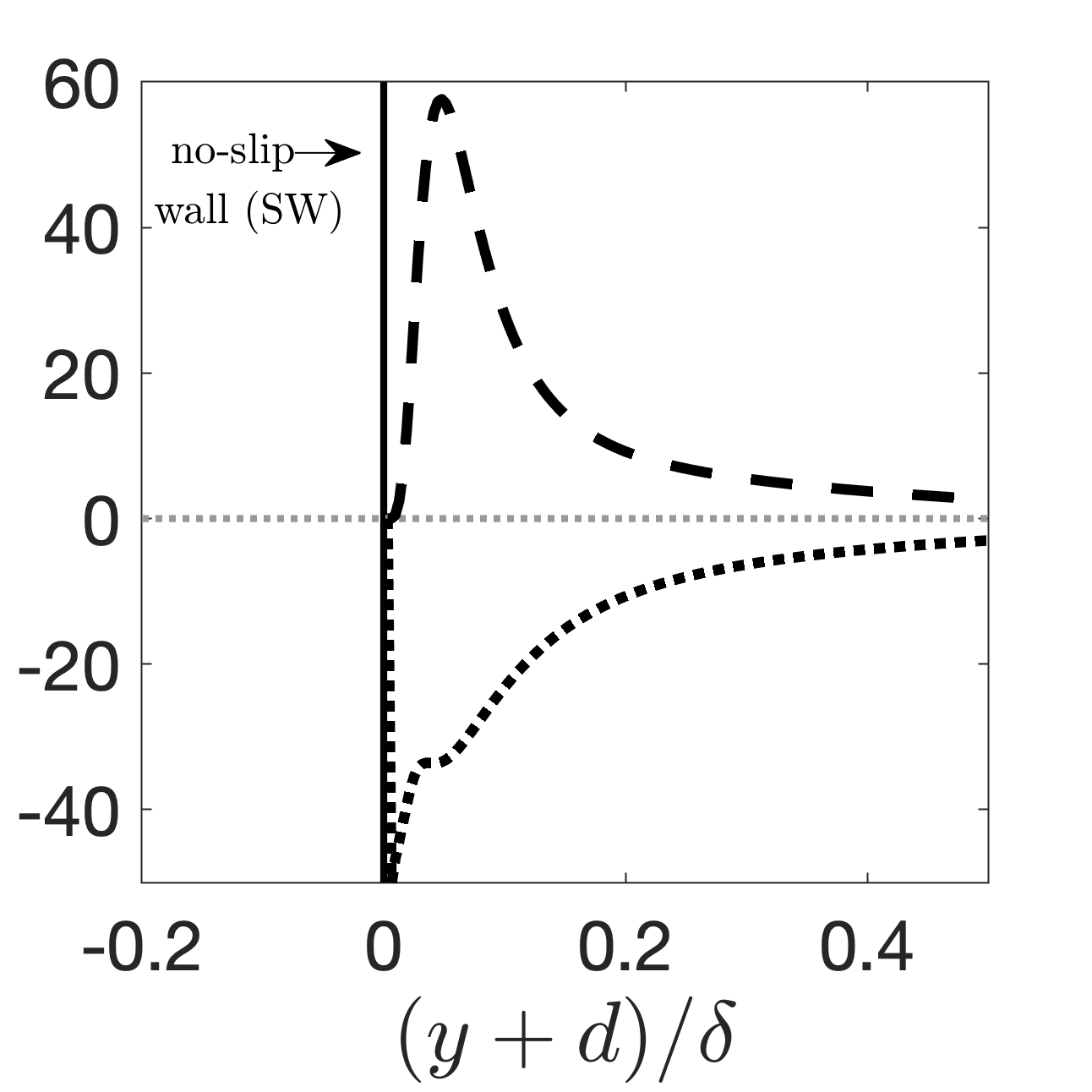}}
\caption{\small Bed-normal variation of TKE production and dissipation terms for (a) PB (thick lines) and IWF cases (symbols), (b) IWH case, and (c) SW case. ($P_s$, \blkdashline, \blkcircle ), ($P_m$, \redline, \redsquare), ($P_w$, \reddashdotline, \redtriangle), ($\epsilon$, \blackdottedline, \blkdiamond). Vertical lines show bed crest and no-slip wall for PB (\grayline), IWF (\graydashdotline), IWH (\graydashline) and no-slip wall (\blkline). (Note that PB and IWF sediment crest lines overlap). }
\label{fig:tke_pd_1}
\end{figure}

\begin{figure}
   \centering
   \subfigure[]{
   \includegraphics[width=4.3cm,height=4.3cm,keepaspectratio]{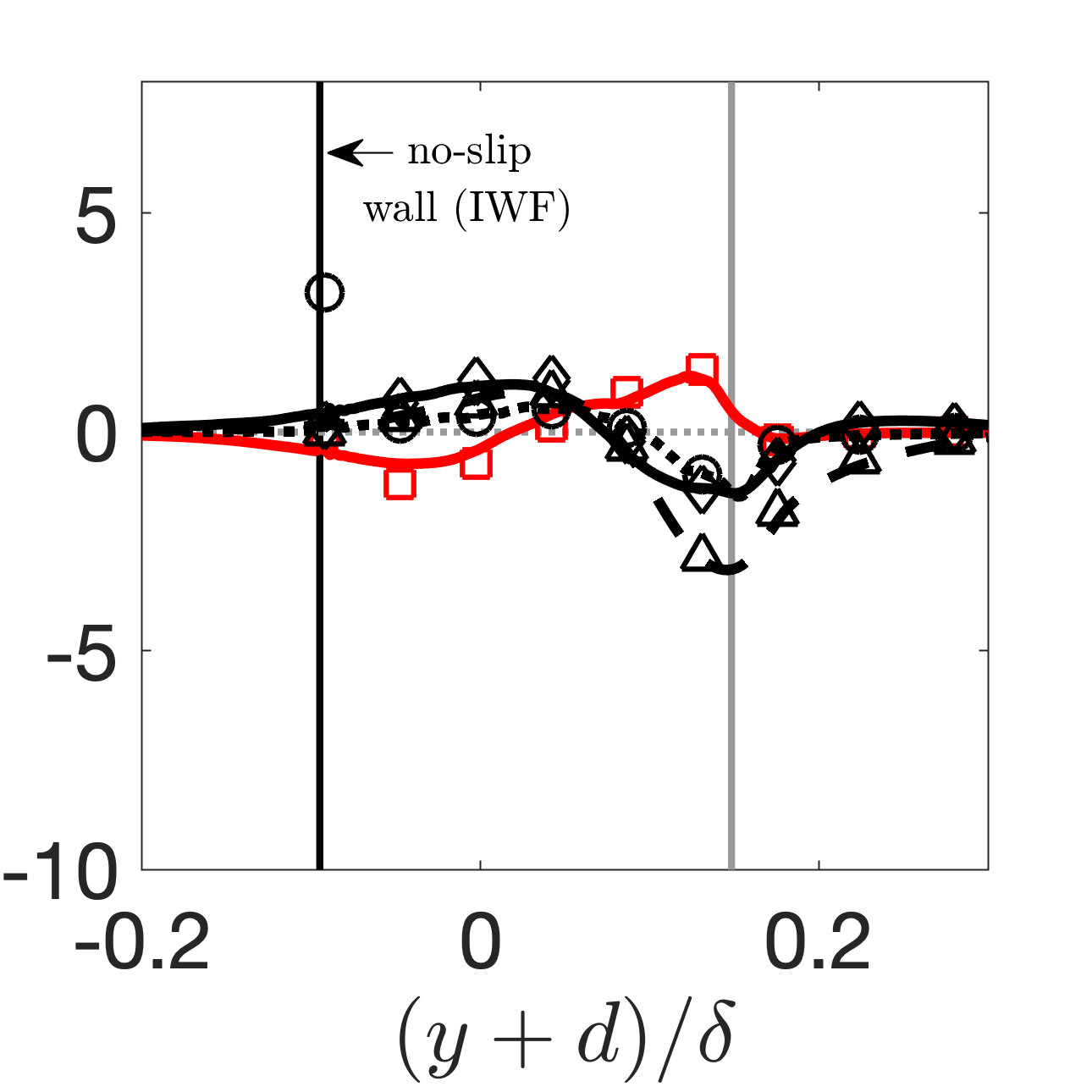}}
    \subfigure[]{
   \includegraphics[width=4.3cm,height=4.3cm,keepaspectratio]{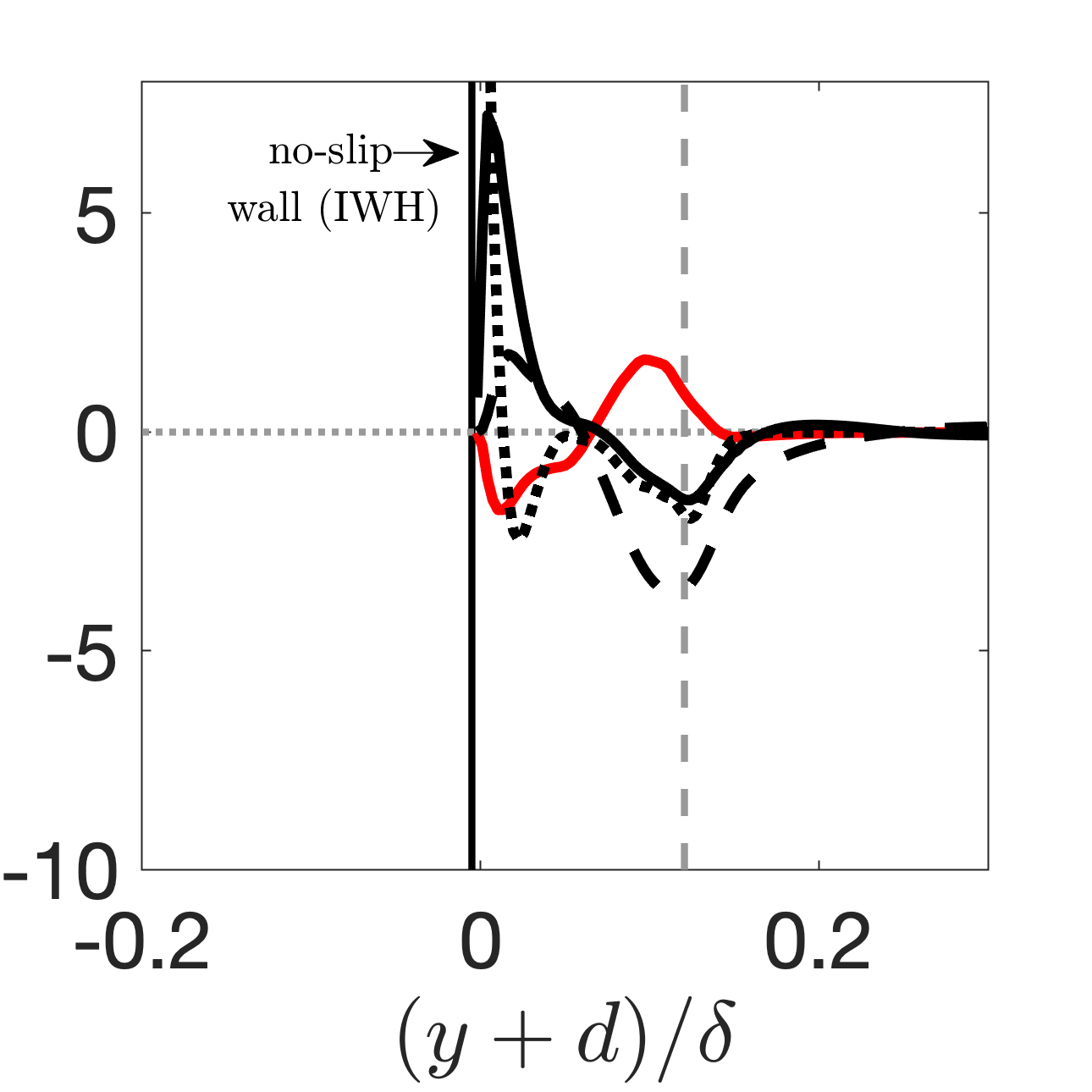}}
   \subfigure[]{
  \includegraphics[width=4.3cm,height=4.3cm,keepaspectratio]{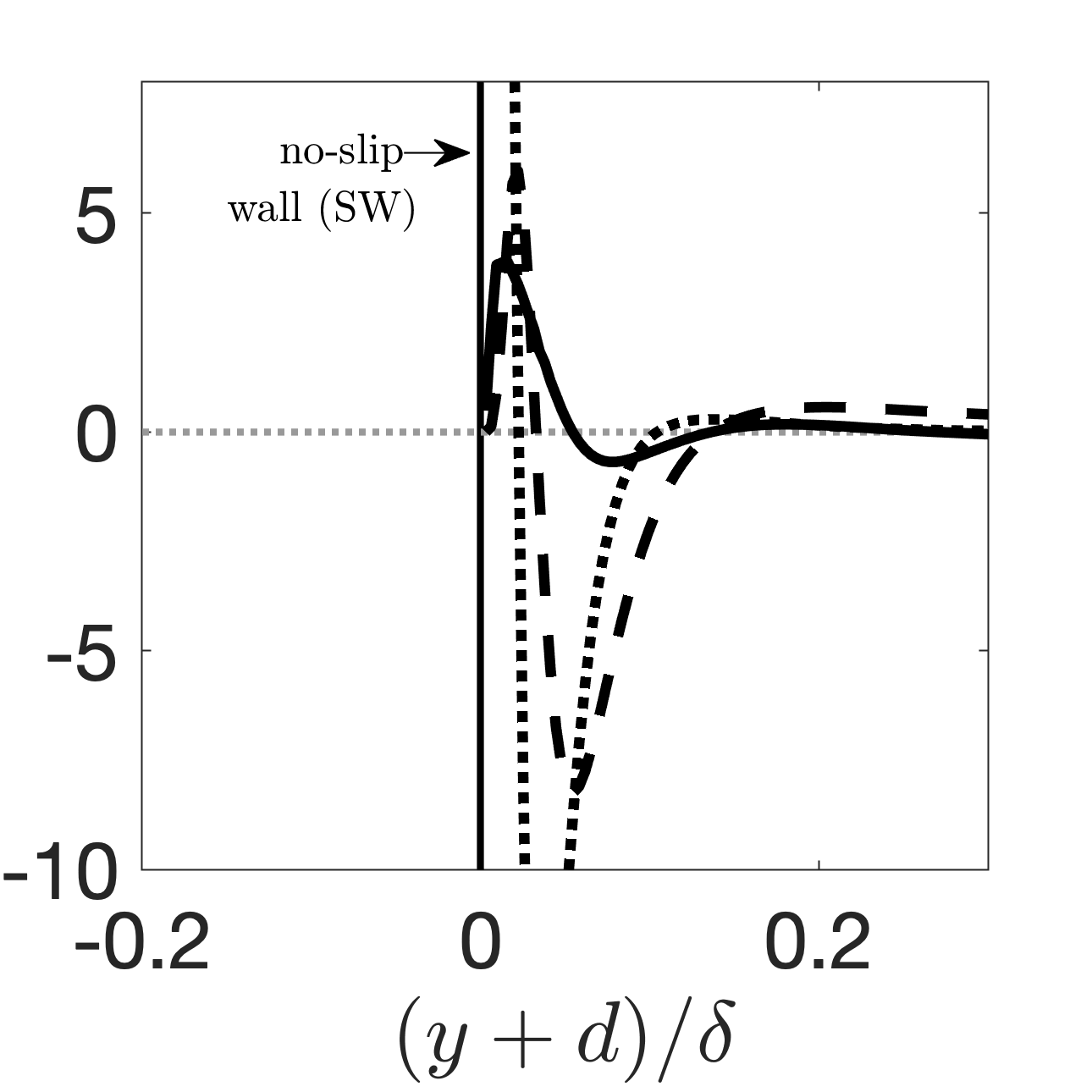}}
\caption{\small Bed-normal variation of TKE transport terms for (a) PB (thick lines) and IWF cases (symbols), (b) IWH case, and (c) SW case. ($T_t$, \blkdashline, \blktriangle), ($T_w$, \redline, \redsquare), ($T_{\nu}$, \blackdottedline, \blkcircle), ($T_p$, \blkline, \blkdiamond). Vertical lines show bed crest and no-slip wall with the following legend: PB (\grayline), IWF (\graydashdotline), IWH (\graydashline) and no-slip wall (\blkline). (Note that PB and IWF sediment crest lines overlap).}
\label{fig:tke_trans}
\end{figure}

Figure~\ref{fig:tke_pd_1} shows the variation of the production, dissipation terms, and figure~\ref{fig:tke_trans} includes the TKE transport terms. Shear production, $P_s$, and the wake productions terms $P_m, P_w$ (figure~\ref{fig:tke_pd_1}a,b) peak below the crest, inside the bed, for both the permeable bed and impermeable wall cases. The peak value for shear production, $P_s$, for the PB case is observed at $(y+d)/\delta\sim 0.1405$, for the IWF case at $(y+d)/\delta\sim 0.14$ and for the IWH case at  $(y+d)/\delta\sim 0.11$. The peak value for dissipation, $\epsilon$ , is observed slightly below at $(y+d)/\delta\sim 0.115$ for the PB case, $(y+d)/\delta\sim 0.116$ for the IWF case, and $(y+d)/\delta\sim 0.09$ for the IWH case. For the SW case the peak value for shear productions is observed at $(y+d)/\delta\sim 0.048$ ($y^+\sim13.0$) . Further away from the sediment crest, for $(y+d)/\delta > 0.5$, all of the terms except shear production $P_s$ and dissipation $\epsilon$ decay to zero, establishing equilibrium outer layer where the rate of production balances the rate of dissipation.

Bed-induced production terms, both $P_m$ and $P_w$ (figure~\ref{fig:tke_pd_1}a,b), are comparable to the shear production terms, $P_s$. Negative peak in $P_w$ at the roughness crest level is observed and could be attributed to the conversion of turbulent kinetic energy to wake kinetic energy as a result of work of large-scale structures (greater than roughness scale) associated with $\langle\overline{u^{\prime2}}\rangle^{+}$ at this location acting against  the pressure drag of roughness elements (figure~\ref{fig:tke_pd_1}).

Figure~\ref{fig:tke_trans} shows the transport terms. The form induced transport, $T_w$, is significant for PB, IWF and IWH cases and works against the other transport processes by moving the TKE upwards from a low-TKE region inside the bed to the crest region. The pressure transport term, $T_p$, is quite significant and works in moving high TKE from crest into lower levels of the bed.
 
\subsection{Wall shear stress fluctuations}\label{sec:pdf_wallshear}
\begin{figure}
   \centering
   \subfigure[]{
  \includegraphics[width=6cm,height=4cm,keepaspectratio]{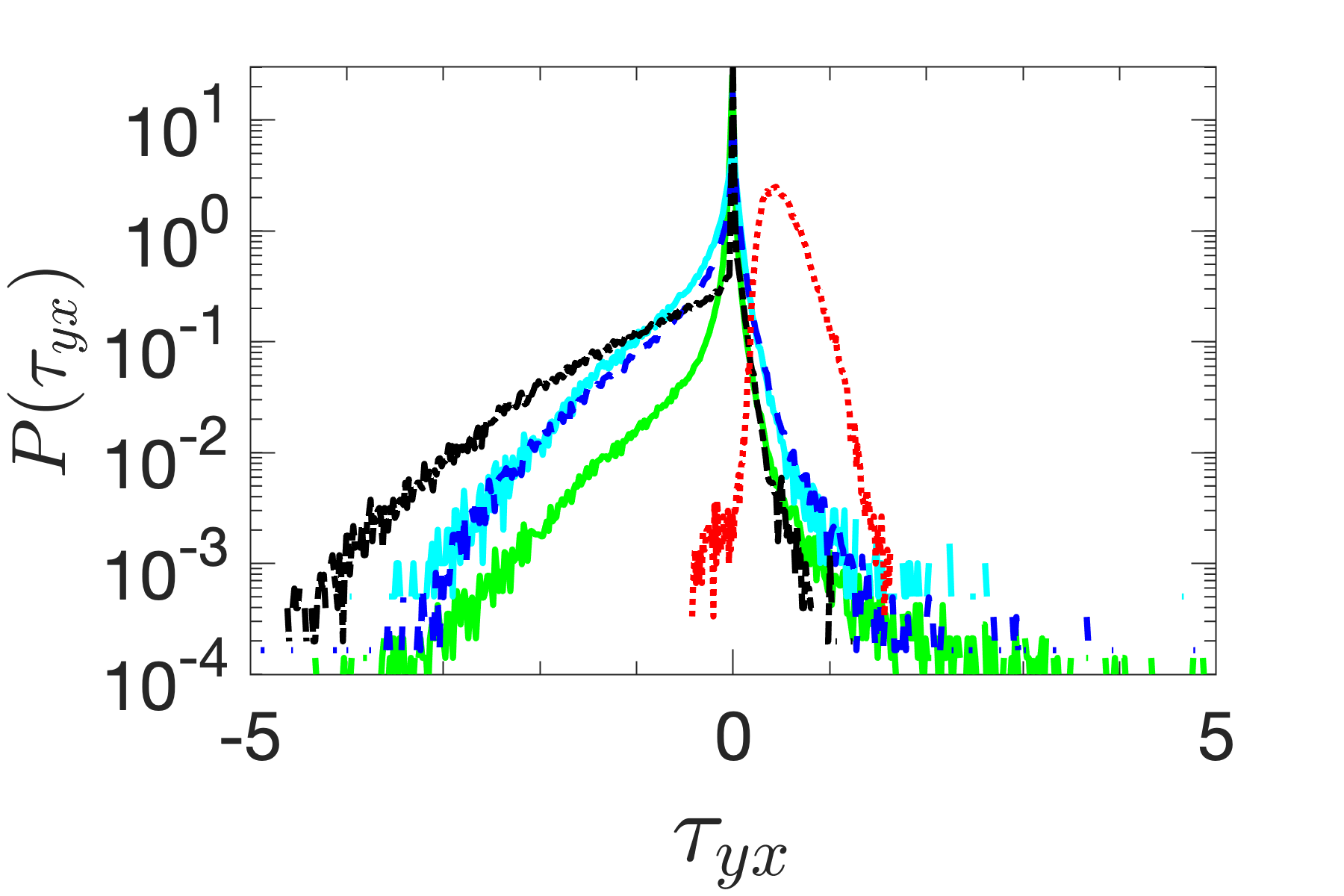}}
  \subfigure[]{
  \includegraphics[width=6cm,height=4cm,keepaspectratio]{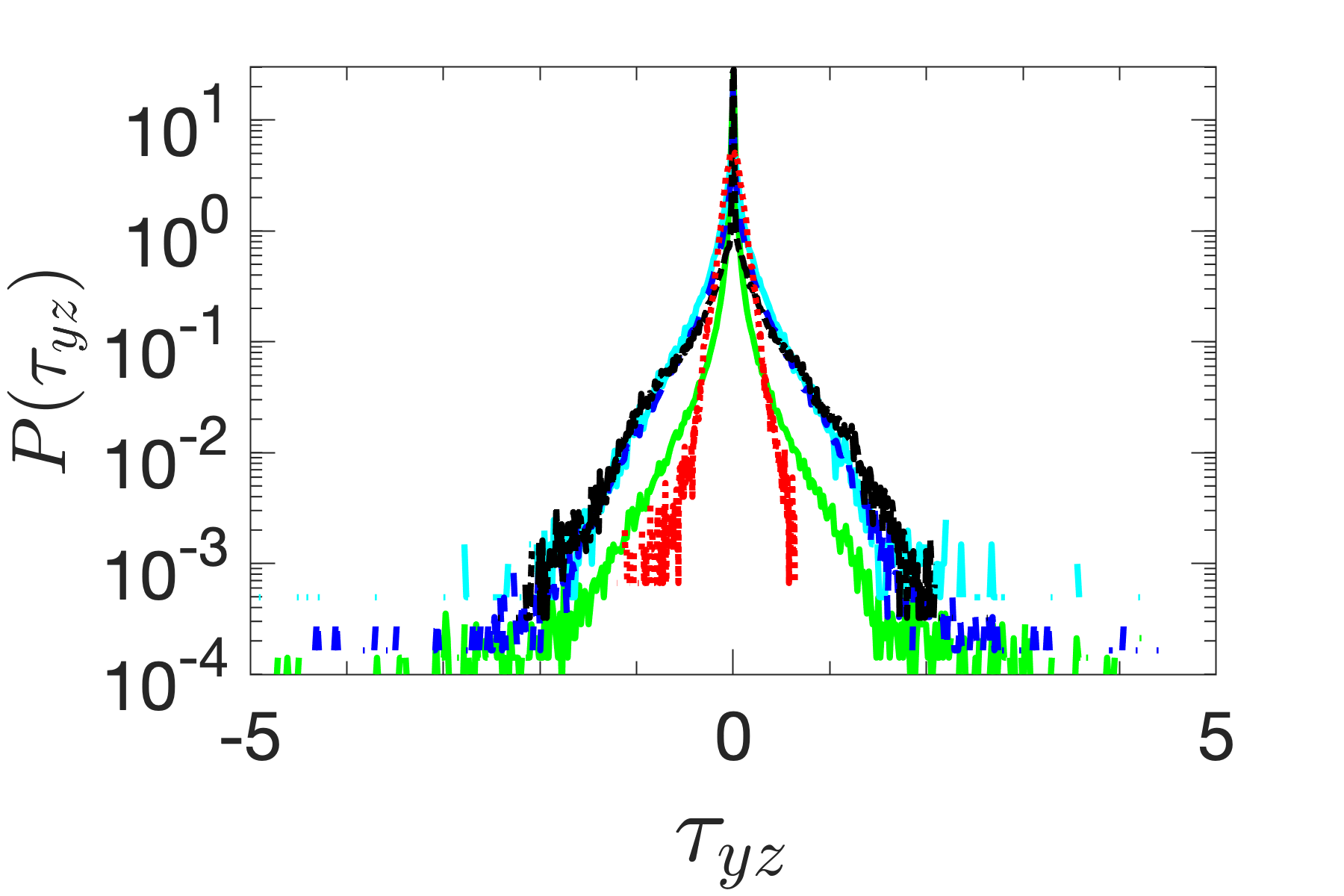}}
\caption{\small PDFs of the (a) stream-wise wall shear stress ($\tau_{yx}$), and (b) spanwise wall shear stress ($\tau_{yz}$) for PB (\greenline), PBTL (\cyanline), IWF (\bluedashline), IWH (\blkdashdotline), and SW cases (\reddottedline). }
\label{fig:pdf_wssa}
\end{figure}

\begin{figure}
   \centering
       \subfigure[]{
   \includegraphics[width=6cm,height=4cm,keepaspectratio]{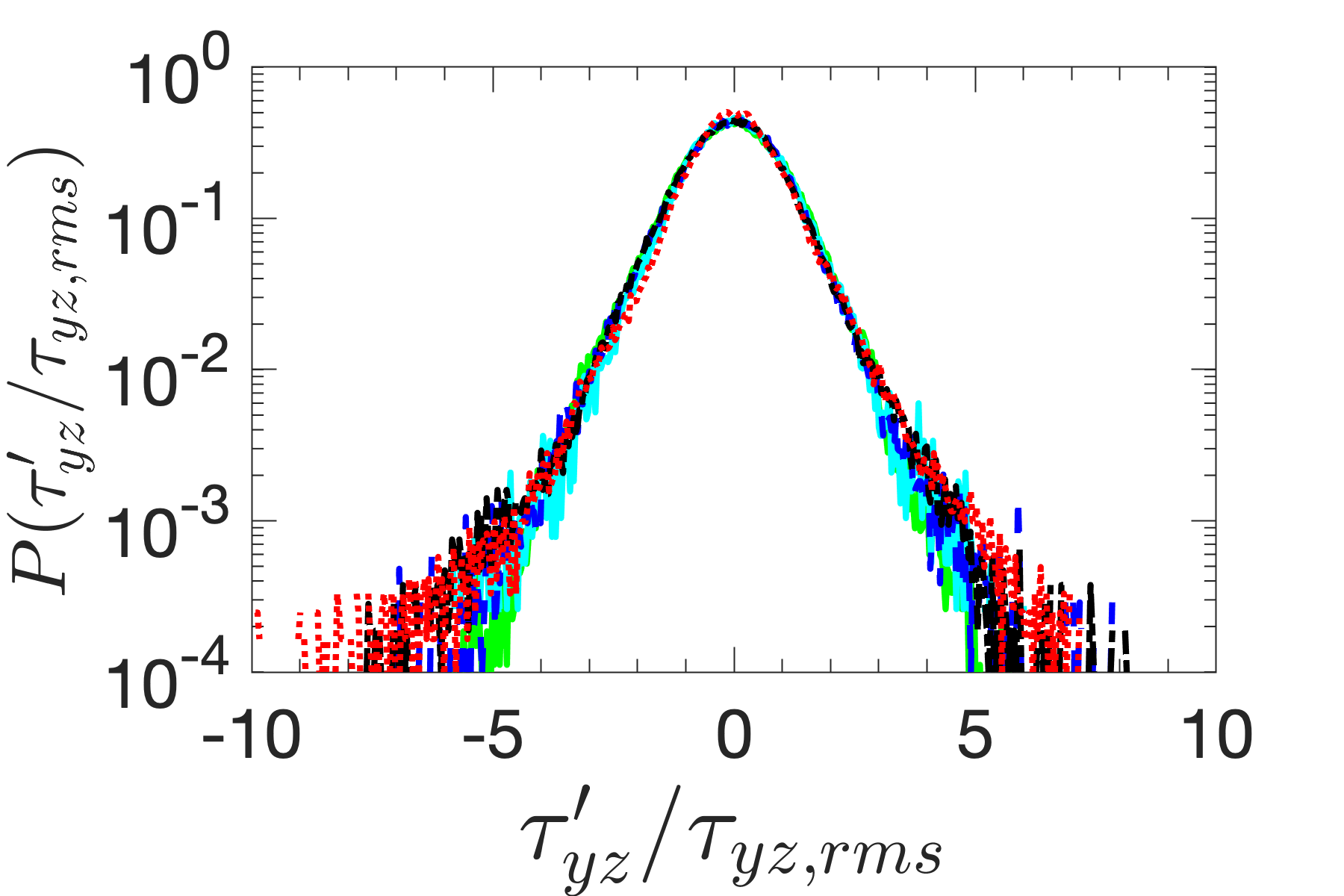}}
    \subfigure[]{
   \includegraphics[width=6cm,height=4cm,keepaspectratio]{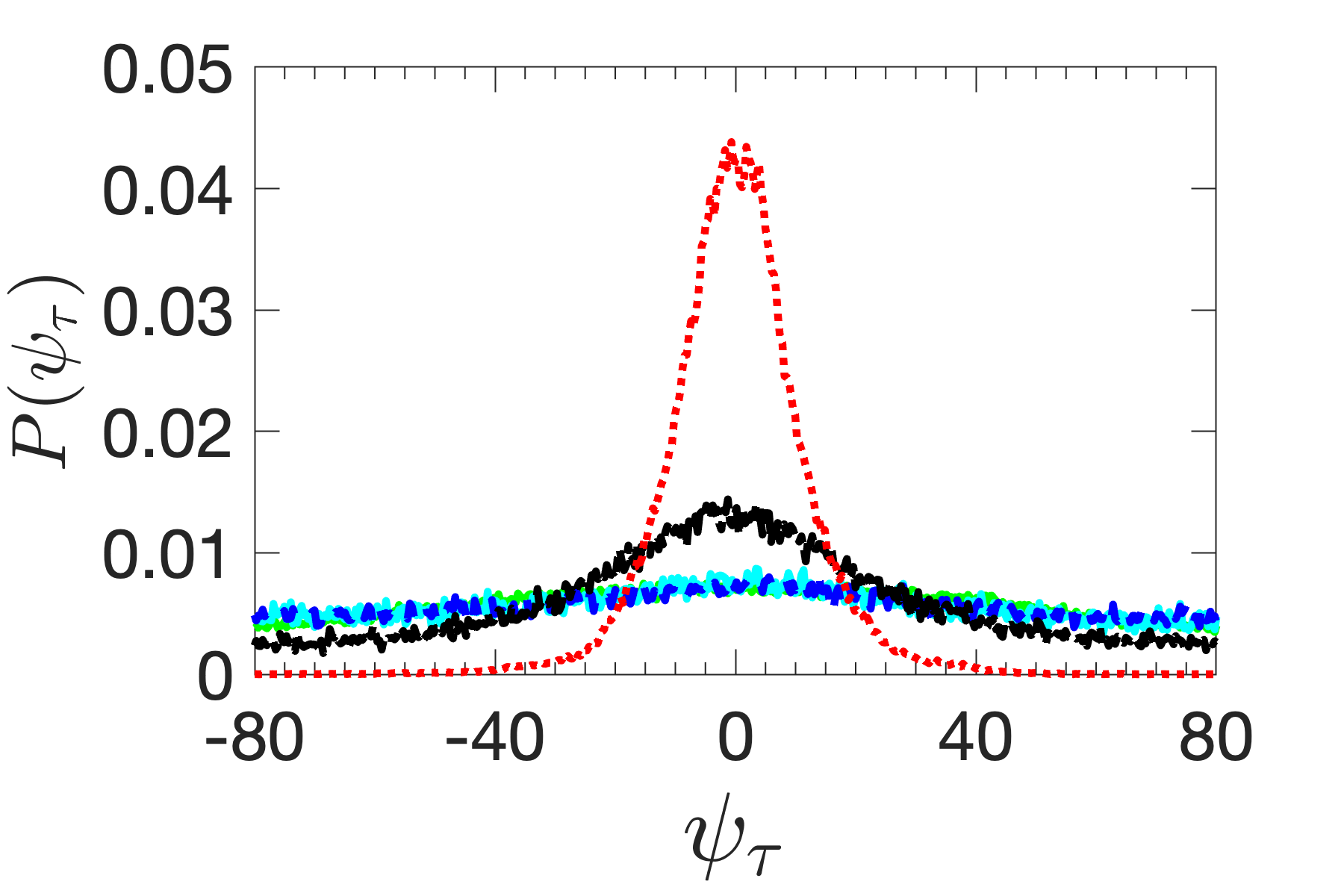}}
\caption{\small PDFs of (a) spanwise-wise wall shear stress fluctuations 
normalized by the rms values, and (b) the yaw angle ($\phi_{\tau}$) for PB (\greenline), PBTL (\cyanline), IWF (\bluedashline), IWH (\blkdashdotline), and SW cases (\reddottedline).}
\label{fig:pdf_wssb}
\end{figure}

The influence of permeability and roughness layers on the wall shear stress is studied in this section for the PB, IWF, IWH, and SW cases. For the permeable bed, the shear stresses are calculated on the surfaces of the spherical elements, whereas for the impermeable rough wall cases, IWF and IWH, the stress on underlying no-slip horizontal wall is also included. For the permeable bed case shear stress statistics for only the top layer are analyzed separately. This is referred to as the PBTL case in  table~\ref{tab:wss_mom} and figures~\ref{fig:pdf_wssa},~\ref{fig:pdf_wssb}. The root mean-squared fluctuations of the streamwise shear stress for the SW case is in good agreement with the correlation, $\tau^{+}_{yx} = \tau_{yx}/\tau_{w} = 0.298 + 0.018\ln{Re_{\tau}}$, proposed by~\citet{orlu2011fluctuating}. 
Table~\ref{tab:wss_mom} provides the mean, standard deviation, skewness, and kurtosis values for the streamwise ($\tau_{yx}$) and spanwise ($\tau_{yz}$) shear stresses and shear stress fluctuations. The standard deviation for the yaw angle ($\psi_{\tau} = {\rm atan}(\tau_{yz}/\tau_{yx})$) is also shown.

\begin{table}
\begin{center}
\def~{\hphantom{0}}
\caption{Higher order statistics for streamwise shear stress, $\tau_{yx}$, spanwise shear stress, $\tau_{yz}$, and yaw angle, $\psi_{\tau}$ showing the mean $\mu(\cdot)$,
standard deviation $\sigma(\cdot)$, skewness $Sk(\cdot)$, and kurtosis $Ku(\cdot)$. 
Cases PB and PBTL show statistics for all layers and only the top layer of the permeable sediment bed, respectively. 
}
\begin{tabular}{@{}l c c c c c c c c c }
Case & $\mu(\tau_{yx})$ & $\sigma(\tau_{yx})$& $Sk({\tau^{\prime}_{yx}})$ & $Ku({\tau^{\prime}_{yx}})$ & $\mu(\tau_{yz})$ & $\sigma(\tau_{yz})$& $Sk({\tau^{\prime}_{yz}})$ & $Ku({\tau^{\prime}_{yz}})$ & $\sigma(\psi_{\tau})$\\ 
PB &-0.038 &0.184 &-0.229 &3.79 &-1.43e-4 &0.122 &-0.1 &3.96 &46.98\\ 
PBTL &-0.191 &0.403 &-3.1e-3 &5.35 &-2.8e-3 &0.260 &3.60e-2 &5.02 &48.29 \\ 
IWF &-0.137 &0.372 &0.159 &5.36 &1.1e-3 &0.232 &-2.64e-2 &5.12 &48.90 \\ 
IWH  &-0.248 &0.555 &0.36 &5.09 &4.0e-3 &0.238 &2.96e-2 &5.94 &38.80 \\
SW  &0.505 &0.184 &0.951 &4.47 &1.7e-3 &0.098 &-0.316 &9.94 &11.40 \\ 
\end{tabular}
\label{tab:wss_mom}
\end{center}
\end{table}

The PDFs of the streamwise and the spanwise shear stresses are shown in figures~\ref{fig:pdf_wssa}a,b, respectively. Both shear stress PDFs show that the probability for extreme events increases with the permeable bed and roughness cases as compared to the smooth wall case. The tails of PDFs for PB, IWF, and IWH cases are negatively skewed in the streamwise stress, figure~\ref{fig:pdf_wssa}a. This is because as the flow goes past the spherical particles or roughness elements, it separates and creates a recirculation zone behind them. The skewness is more negative for the IWH case as the exposed underlying horizontal surface helps the separated flow to reattach. The full layer IWF case has less negative skewness compared to the IWH case, and is slightly different compared to the complete permeable bed (figure~\ref{fig:pdf_wssa}a). It is important to note that if only the top layer of the PB case is used for the data analysis (PBTL), the PDFs for both components of the shear stresses are very similar to the single layer IWF case (figures~\ref{fig:pdf_wssa}a,b,). This is  critical information for large scale hyporheic exchange models as it shows that almost all of the extreme events of shear stress happen within the top layer of the sediment bed. Therefore, including just the effect of the top layer of the sediment in large domain hyporheic modeling is potentially sufficient to capture realistic values of the shear stress distribution on the bed. 

The PDFs of the spanwise shear stress fluctuations (normalized by the corresponding rms values; $\tau^{\prime}_{yz}/\tau_{yz,rms}$) and the yaw angle ($\psi_{\tau} = {\rm atan}(\tau_{yz}/\tau_{yx})$) are shown in figures~\ref{fig:pdf_wssb}a,b, respectively. Normalizing the spanwise shear stress fluctuations with spanwise shear stress rms values collapses the PDFs (see figure~\ref{fig:pdf_wssb}a), and has almost a zero mean and zero skewness. More extreme events are observed in the permeable bed and impermeable rough wall cases compared to the smooth wall for the spanwise shear stress, figure~\ref{fig:pdf_wssb}a), similar to the streamwise component.

Figure~\ref{fig:pdf_wssb}b shows the PDF of the shear stress yaw angle, $\psi_{\tau}$. In the smooth wall case, the probability of events with shear stress yaw angle outside the range of $-45$ to $45$ degrees is found to be very small. This is consistent with behavior reported in literature by~\citet{jeon1999space}, ~\citet{diaz2017wall} and ~\citet{ma2021direct}. Roughness elements and permeability enhance the probability of events with yaw angles greater than 45 degrees with most increase in such events observed in the IWF and PB cases. This implies that the probability of simultaneously obtaining large scale fluctuations in the streamwise, $\tau^{\prime}_{yx}$, and spanwise, $\tau^{\prime}_{yz}$, shear stress increases for the IWF and PB cases. Table~\ref{tab:wss_mom} shows the standard deviation of $\psi_{\tau}$ for all four cases. The $\sigma$ of $\psi_{\tau}$ for the IWH case is higher than SW case. The $\sigma$ for $\psi_{\tau}$ for IWF, PB  cases is even higher than the IWH case, quantifying the increase in large scale shear stress fluctuations due to roughness elements and permeability. It is interesting to note that the $\sigma$ for PBTL case is is very similar to the IWF and PB cases confirming that these large scale fluctuations are confined to the top layer of the permeable bed. 

\section{Conclusions} 
\label{sec:conclusions}
Pore-resolved direct numerical simulations of turbulent flow over a sediment bed were performed to test the hypothesis that the structure and dynamics of turbulence over a porous sediment bed can be significantly different than that over an impermeable rough wall. A fully validated fictitious domain based numerical approach~\citep{apte2009frs,finn2013relative,ghodke2016dns,ghodke2018roughness,he2019characteristics} is used to conduct these simulations. First, the numerical predictions are validated against the experimental data by~\citet{voermans2017variation} for a randomly packed, monodispersed bed of spherical particles
at $Re_K \sim 2.56$, representative of natural aquatic systems. After a thorough validation, the numerical approach was used to investigate turbulent boundary flow at $Re_{\tau} = 270$ (and $Re_K \sim 2.56$)  for four different geometries; namely, (i) permeable bed  with randomly packed sediment grains, (ii) an impermeable-full layer rough wall matching the top layer of the permeable sediment bed, (iii) an impermeable-half layer rough wall matching top half layer of the sediment bed, (iv) a smooth wall. The double-averaging methodology is used to compute the mean velocity, Reynolds stresses and form-induced stresses for all four cases. Differences in near-bed turbulence structure, statistics of bed shear stress and pressure field at the sediment-water interface were quantified.
The {key findings} of this work are summarized below. 

(i) Mean velocity and Reynolds stresses are similar in magnitude for permeable bed and impermeable full layer rough wall cases. However, for the impermeable half layer rough wall case, mean velocity and axial Reynolds stress are $\sim$ 11.29\% and $\sim$ 13.22\% greater, and bed-normal and shear stresses are $\sim$ 4.70\% and $\sim$ 0.66\% lower than the permeable bed case.

(ii) Bed roughness increases Reynolds shear stress whereas permeability has minimal influence. Bed permeability, in contrast to Reynolds stresses, significantly influences the form-induced shear stress. Peak value for form-induced wall normal stress for both permeable-bed and impermeable full layer cases, happen deep inside the bed top layer, with peak value for the permeable bed case $\sim$ 38.5\% greater than the impermeable-half layer case and $\sim$ 13.2\% greater than the impermeable-full layer.

(iii) Permeable bed and impermeable full layer cases show similar behavior in upwelling and downwelling form-induced flux correlations. Due to wall blocking in impermeable half layer case, negative $\widetilde{v}$ and  $\widetilde{p}$ events are less likely to occur.

(iv) Roughness elements increase the mean square pressure fluctuations $\langle\overline{p^{\prime2}}\rangle^{+}$ in the inner layer compared to a smooth wall.

(v) The spatial inhomogeneity at the roughness length scale introduces additional production and transport terms in double-averaged TKE budget. Bed-induced production terms,
both $P_m$ and $P_w$, are comparable to the shear production terms, $P_s$. Negative peak in $P_w$ at the roughness crest level was observed and and is attributed to the conversion of turbulent kinetic energy to wake kinetic energy as a result of work of large-scale structures associated with $\langle\overline{u^{\prime2}}\rangle^{+}$ at this location against pressure drag of roughness elements. The form induced transport, $T_w$, is significant for PB, IWF, and IWH cases and works against the other transport processes by moving the TKE upwards from a low-TKE region inside the bed to the crest region. The pressure transport term, $T_p$, is quite significant and works in moving high TKE from crest into lower levels of the bed. 

(vi) The pdf of wall-shear stress fluctuations show that probability of extreme events increase with bed permeability and bed roughness compared to a smooth wall.

The above findings have important implications for {\it reach-scale modeling of hyporheic exchange}. Typical large-scale simulations are based on Reynolds-averaged Navier Stokes computations of turbulent flow over complex topology of bottom boundary wherein the bottom boundary is treated as impermeable. The pressure field at the impermeable boundary is then used as a boundary condition to conduct mass and momentum transport through the hyporheic zone of sediment in a sequential and separate computation. The present study indicates that such an assumption can significantly impact the transport of momentum (and by analogy, transport of mass) at the sediment-water interface. Modeling the roughness effect of at least the top layer of sediment bed can capture the pressure and shear-stress fluctuations at the sediment-water interface more accurately. Additionally, the standard, sequential reach scale models can be potentially improved by incorporating the pressure fluctuation data obtained from such detailed turbulent flow simulations of permeable beds as boundary conditions for hyporheic flow computation within the porous bed. A model that incorporates high-order statistics of the PDF of pressure and shear stress fluctuations at the SWI can be developed (similar to work by~\citet{ghodke2018roughness}) by collecting data for a range of Reynolds numbers, and will be the focus of future studies.

\begin{acknowledgments}
\section*{Acknowledgements} 
This work was initiated as part of SKK's internship at Pacific Northwest National Laboratory. Simulations were performed at the Texas Advanced Computing Center's (TACC) Frontera system. Computing resource from Pacific Northwest
National Laboratory’s EMSL (Environmental Molecular Sciences Laboratory) is also acknowledged.
\end{acknowledgments}

\section*{Funding} 
SKK acknowledges support from Pacific Northwest National Laboratory (PNNL) as part of an internship program. SVA and SKK gratefully acknowledge funding from US Depart of Energy, Office of Basic Energy Sciences (Geosciences) under award number DE-SC0021626 as well as US National Science Foundation award \#205324. The computing resources used were made available under NSF's Leadership Resources Allocation (LRAC) award. XH and TDS acknowledge funding from the DOE Office of Biological and Environmental Research, Subsurface Biogeochemical Research program, through the PNNL Subsurface Science Scientific Focus Area project (http://sbrsfa.pnnl.gov/). 

\section*{Declaration of Interests} 
The authors report no conflict of interest.

\section*{APPENDIX A. Verification Studies and Grid Resolution} 
The solver has been thoroughly verified and validated for a range of cases~\citep{AptePatankar} and has also been used for large-scale, parallel simulations of oscillatory flow over a layer of sediment particles~\citep{ghodke2016dns,ghodke2018roughness} and flow through porous media~\citep{finn2013relative,he2019characteristics}.
\begin{table}
\centering
\caption{Grid refinement study for flow over an isolated sphere with non-isotropic, rectilinear grids similar to those used in the present study.}
\label{tab:drag}
\begin{tabular}{@{}l c c c}\hline
$Re_p$& 50 & 150 & 350 \\ \hline
& \multicolumn{3}{c}{\underline{Drag Coefficient, $C_D$}} \\ 
Isotropic, $D_p/\Delta y = 100$ & 1.54 & 0.9 & 0.65 \\ 
$\Delta x =  \Delta z =  3 \Delta y$ & 1.58 & 0.91 & 0.66 \\
$\Delta x = \Delta z =  4\Delta y$ &1.56 & 0.9 & 0.66 \\  \hline
\end{tabular}
\end{table}

In turbulent boundary layer over a sediment bed, there is need to use non-isotropic and high-aspect ratio grids to minimize the total control volumes and yet provide sufficient resolution needed to capture all scales of turbulence. Specifically, for DNS of boundary layers, the resolution near the sediment bed and in the bed-normal direction should be such that $\Delta y^{+} < 1$, where $\Delta y^{+}$ represents resolution in wall units. The code was used to predict flow over an isolated sphere at different Reynolds numbers and using isotropic and non-isotropic, rectilinear grids. The drag force was compared with published data~\cite{AptePatankar} and is given in Table~\ref{tab:drag}. It is observed that the high-aspect ratio grids are capable of predicting the drag forces accurately for $Re_p$ up to 350. The particle Reynolds numbers for the present work are much smaller (38.5 and 77) which do not result in vortex shedding. However, to compare the effectiveness of the non-isotropic grids in capturing vortex shedding, the Strouhal number for vortex shedding at $Re_p=350$ was 0.131, which compared reasonably well with the range of values between 0.135 - 0.14 predicted on much finer, isotropic grids in literature~\citep{mittal2008versatile,mittal1999fourier,bagchi2001direct}.

\bibliographystyle{jfm}
\bibliography{hyporheicDNS}

\begin{thebibliography}{53}
\expandafter\ifx\csname natexlab\endcsname\relax\def\natexlab#1{#1}\fi
\def\au#1{#1} \def\ed#1{#1} \def\yr#1{#1}\def\at#1{#1}\def\jt#1{\textit{#1}}
  \def\bt#1{#1}\def\bvol#1{\textbf{#1}} \def\vol#1{#1} \def\pg#1{#1}
  \def\publ#1{#1}\def\arxiv#1{#1}\def\org#1{#1}\def\st#1{\textit{#1}}

\bibitem[Anderson {\em et~al.\/}(2008)]{anderson2008groundwater}
{\sc \au{Anderson, Mary~P} \& \au{others}} \yr{2008} {\em Groundwater:
  Selection, Introduction and Commentary.\/}.  \publ{IAHS Press}.

\bibitem[Apte \& Patankar(2008)]{AptePatankar}
{\sc \au{Apte, SV} \& \au{Patankar, NA}} \yr{2008}  \at{{A formulation for
  fully resolved simulation (FRS) of particle--turbulence interactions in
  two-phase flows}}.  \jt{International Journal of Numerical Analysis and
  Modeling}  \bvol{5, Suppl},  \pg{1--16}.

\bibitem[Apte {\em et~al.\/}(2009)Apte, Martin \& Patankar]{apte2009frs}
{\sc \au{Apte, S.~V.}, \au{Martin, M.} \& \au{Patankar, N.~A.}} \yr{2009}
  \at{A numerical method for fully resolved simulation ({FRS}) of rigid
  particle-flow interactions in complex flows}.  \jt{Journal of Computational
  Physics}  \bvol{228}~(8),  \pg{2712--2738}.

\bibitem[Bagchi {\em et~al.\/}(2001)Bagchi, Ha \&
  Balachandar]{bagchi2001direct}
{\sc \au{Bagchi, Prosenjit}, \au{Ha, MY} \& \au{Balachandar, S}} \yr{2001}
  \at{Direct numerical simulation of flow and heat transfer from a sphere in a
  uniform cross-flow}.  \jt{J. Fluids Eng.}  \bvol{123}~(2),  \pg{347--358}.

\bibitem[Bencala {\em et~al.\/}(1983)Bencala, Rathbun, Jackman, Kennedy,
  Zellweger \& Avanzino]{bencala1983rhodamine}
{\sc \au{Bencala, KE}, \au{Rathbun, RE}, \au{Jackman, A~P}, \au{Kennedy, VC},
  \au{Zellweger, GW} \& \au{Avanzino, RJ}} \yr{1983}  \at{Rhodamine wt dye
  losses in a mountain stream environment}.  \jt{Water Resources Bulletin}
  \bvol{19}~(6),  \pg{943--950}.

\bibitem[Bomminayuni \& Stoesser(2011)]{bomminayuni2011turbulence}
{\sc \au{Bomminayuni, Sandeep} \& \au{Stoesser, Thorsten}} \yr{2011}
  \at{Turbulence statistics in an open-channel flow over a rough bed}.
  \jt{Journal of Hydraulic Engineering}  \bvol{137}~(11),  \pg{1347--1358}.

\bibitem[Breugem {\em et~al.\/}(2006)Breugem, Boersma \&
  Uittenbogaard]{breugem2006influence}
{\sc \au{Breugem, WP}, \au{Boersma, BJ} \& \au{Uittenbogaard, RE}} \yr{2006}
  \at{The influence of wall permeability on turbulent channel flow}.
  \jt{Journal of Fluid Mechanics}  \bvol{562},  \pg{35}.

\bibitem[Briggs {\em et~al.\/}(2009)Briggs, Gooseff, Arp \&
  Baker]{briggs2009method}
{\sc \au{Briggs, Martin~A}, \au{Gooseff, Michael~N}, \au{Arp, Christopher~D} \&
  \au{Baker, Michelle~A}} \yr{2009}  \at{A method for estimating surface
  transient storage parameters for streams with concurrent hyporheic storage}.
  \jt{Water Resources Research}  \bvol{45}~(4).

\bibitem[Chen {\em et~al.\/}(2018)Chen, Cardenas \& Chen]{chen2018hyporheic}
{\sc \au{Chen, Xiaobing}, \au{Cardenas, M~Bayani} \& \au{Chen, Li}} \yr{2018}
  \at{Hyporheic exchange driven by three-dimensional sandy bed forms:
  Sensitivity to and prediction from bed form geometry}.  \jt{Water Resources
  Research}  \bvol{54}~(6),  \pg{4131--4149}.

\bibitem[Chen {\em et~al.\/}(2022)Chen, Bao, Fang, Perkins, Ren, Song, Duan,
  Hou, He \& Scheibe]{chen2022modeling}
{\sc \au{Chen, Yunxiang}, \au{Bao, Jie}, \au{Fang, Yilin}, \au{Perkins,
  William~A}, \au{Ren, Huiying}, \au{Song, Xuehang}, \au{Duan, Zhuoran},
  \au{Hou, Zhangshuan}, \au{He, Xiaoliang} \& \au{Scheibe, Timothy~D}}
  \yr{2022}  \at{Modeling of streamflow in a 30 km long reach spanning 5 years
  using openfoam 5. x}.  \jt{Geoscientific Model Development}  \bvol{15}~(7),
  \pg{2917--2947}.

\bibitem[D'angelo {\em et~al.\/}(1993)D'angelo, Webster, Gregory \&
  Meyer]{d1993transient}
{\sc \au{D'angelo, DJ}, \au{Webster, JR}, \au{Gregory, SV} \& \au{Meyer, JL}}
  \yr{1993}  \at{Transient storage in appalachian and cascade mountain streams
  as related to hydraulic characteristics}.  \jt{Journal of the North American
  Benthological Society}  \pg{pp. 223--235}.

\bibitem[Diaz-Daniel {\em et~al.\/}(2017)Diaz-Daniel, Laizet \&
  Vassilicos]{diaz2017wall}
{\sc \au{Diaz-Daniel, Carlos}, \au{Laizet, Sylvain} \& \au{Vassilicos,
  J~Christos}} \yr{2017}  \at{Wall shear stress fluctuations: Mixed scaling and
  their effects on velocity fluctuations in a turbulent boundary layer}.
  \jt{Physics of Fluids}  \bvol{29}~(5),  \pg{055102}.

\bibitem[Dye {\em et~al.\/}(2013)Dye, McClure, Miller \&
  Gray]{dye2013description}
{\sc \au{Dye, Amanda~L}, \au{McClure, James~E}, \au{Miller, Cass~T} \&
  \au{Gray, William~G}} \yr{2013}  \at{Description of non-darcy flows in porous
  medium systems}.  \jt{Physical Review E}  \bvol{87}~(3),  \pg{033012}.

\bibitem[Fang {\em et~al.\/}(2018)Fang, Xu, He \& Dey]{fang2018influence}
{\sc \au{Fang, Hongwei}, \au{Xu, Han}, \au{He, Guojian} \& \au{Dey, Subhasish}}
  \yr{2018}  \at{Influence of permeable beds on hydraulically macro-rough
  flow}.  \jt{Journal of Fluid Mechanics}  \bvol{847},  \pg{552--590}.

\bibitem[Finn \& Apte(2013)]{finn2013relative}
{\sc \au{Finn, Justin} \& \au{Apte, Sourabh~V}} \yr{2013}  \at{Relative
  performance of body fitted and fictitious domain simulations of flow through
  fixed packed beds of spheres}.  \jt{International journal of multiphase flow}
   \bvol{56},  \pg{54--71}.

\bibitem[Ghodke \& Apte(2016)]{ghodke2016dns}
{\sc \au{Ghodke, Chaitanya~D} \& \au{Apte, Sourabh~V}} \yr{2016}  \at{Dns study
  of particle-bed--turbulence interactions in an oscillatory wall-bounded
  flow}.  \jt{Journal of Fluid Mechanics}  \bvol{792},  \pg{232--251}.

\bibitem[Ghodke \& Apte(2018)]{ghodke2018roughness}
{\sc \au{Ghodke, Chaitanya~D} \& \au{Apte, Sourabh~V}} \yr{2018}  \at{Roughness
  effects on the second-order turbulence statistics in oscillatory flows}.
  \jt{Computers \& Fluids}  \bvol{162},  \pg{160--170}.

\bibitem[Goharzadeh {\em et~al.\/}(2005)Goharzadeh, Khalili \&
  J{\o}rgensen]{goharzadeh2005transition}
{\sc \au{Goharzadeh, Afshin}, \au{Khalili, Arzhang} \& \au{J{\o}rgensen,
  Bo~Barker}} \yr{2005}  \at{Transition layer thickness at a fluid-porous
  interface}.  \jt{Physics of Fluids}  \bvol{17}~(5),  \pg{057102}.

\bibitem[Grant {\em et~al.\/}(2018)Grant, Gomez-Velez \&
  Ghisalberti]{grant2018hypor}
{\sc \au{Grant, Stanley~B.}, \au{Gomez-Velez, Jesus~D.} \& \au{Ghisalberti,
  Marco}} \yr{2018}  \at{Modeling the effects of turbulence on hyporheic
  exchange and local-to-global nutrient processing in streams}.  \jt{Water
  Resources Research}  \bvol{54}~(9),  \pg{5883--5889}.

\bibitem[Harvey {\em et~al.\/}(1996)Harvey, Wagner \&
  Bencala]{harvey1996evaluating}
{\sc \au{Harvey, Judson~W}, \au{Wagner, Brian~J} \& \au{Bencala, Kenneth~E}}
  \yr{1996}  \at{Evaluating the reliability of the stream tracer approach to
  characterize stream-subsurface water exchange}.  \jt{Water Resources
  Research}  \bvol{32}~(8),  \pg{2441--2451}.

\bibitem[He {\em et~al.\/}(2018)He, Apte, Schneider \& Kadoch]{he2018angular}
{\sc \au{He, Xiaoliang}, \au{Apte, Sourabh}, \au{Schneider, Kai} \& \au{Kadoch,
  Benjamin}} \yr{2018}  \at{Angular multiscale statistics of turbulence in a
  porous bed}.  \jt{Physical Review Fluids}  \bvol{3}~(8),  \pg{084501}.

\bibitem[He {\em et~al.\/}(2019)He, Apte, Finn \& Wood]{he2019characteristics}
{\sc \au{He, Xiaoliang}, \au{Apte, Sourabh~V}, \au{Finn, Justin~R} \& \au{Wood,
  Brian~D}} \yr{2019}  \at{Characteristics of turbulence in a face-centred
  cubic porous unit cell}.  \jt{Journal of Fluid Mechanics}  \bvol{873},
  \pg{608--645}.

\bibitem[Hester {\em et~al.\/}(2017)Hester, Cardenas, Haggerty \&
  Apte]{hester2017importance}
{\sc \au{Hester, Erich~T}, \au{Cardenas, M~Bayani}, \au{Haggerty, Roy} \&
  \au{Apte, Sourabh~V}} \yr{2017}  \at{The importance and challenge of
  hyporheic mixing}.  \jt{Water Resources Research}  \bvol{53}~(5),
  \pg{3565--3575}.

\bibitem[Jackson(1981)]{jackson1981displacement}
{\sc \au{Jackson, PS}} \yr{1981}  \at{On the displacement height in the
  logarithmic velocity profile}.  \jt{Journal of fluid mechanics}  \bvol{111},
  \pg{15--25}.

\bibitem[Jackson {\em et~al.\/}(2013)Jackson, Haggerty \&
  Apte]{jackson2013fluid}
{\sc \au{Jackson, TR}, \au{Haggerty, R} \& \au{Apte, SV}} \yr{2013}  \at{A
  fluid-mechanics based classification scheme for surface transient storage in
  riverine environments: quantitatively separating surface from hyporheic
  transient storage}.  \jt{Hydrol. Earth Syst. Sci}  \bvol{17},
  \pg{2747--2779}.

\bibitem[Jackson {\em et~al.\/}(2015)Jackson, Apte, Haggerty \&
  Budwig]{jackson2015flow}
{\sc \au{Jackson, Tracie~R}, \au{Apte, Sourabh~V}, \au{Haggerty, Roy} \&
  \au{Budwig, Ralph}} \yr{2015}  \at{Flow structure and mean residence times of
  lateral cavities in open channel flows: influence of bed roughness and
  shape}.  \jt{Environmental Fluid Mechanics}  \bvol{15}~(5),  \pg{1069--1100}.

\bibitem[Jeon {\em et~al.\/}(1999)Jeon, Choi, Yoo \& Moin]{jeon1999space}
{\sc \au{Jeon, Sejeong}, \au{Choi, Haecheon}, \au{Yoo, Jung~Yul} \& \au{Moin,
  Parviz}} \yr{1999}  \at{Space--time characteristics of the wall shear-stress
  fluctuations in a low-reynolds-number channel flow}.  \jt{Physics of fluids}
  \bvol{11}~(10),  \pg{3084--3094}.

\bibitem[Jim{\'e}nez(2004)]{jimenez2004turbulent}
{\sc \au{Jim{\'e}nez, Javier}} \yr{2004}  \at{Turbulent flows over rough
  walls}.  \jt{Annu. Rev. Fluid Mech.}  \bvol{36},  \pg{173--196}.

\bibitem[Kim {\em et~al.\/}(2020)Kim, Blois, Best \&
  Christensen]{kim2020experimental}
{\sc \au{Kim, Taehoon}, \au{Blois, Gianluca}, \au{Best, James~L} \&
  \au{Christensen, Kenneth~T}} \yr{2020}  \at{Experimental evidence of
  amplitude modulation in permeable-wall turbulence}.  \jt{Journal of Fluid
  Mechanics}  \bvol{887}.

\bibitem[Kuwata \& Suga(2016)]{kuwata2016lattice}
{\sc \au{Kuwata, Y} \& \au{Suga, K}} \yr{2016}  \at{Lattice boltzmann direct
  numerical simulation of interface turbulence over porous and rough walls}.
  \jt{International Journal of Heat and Fluid Flow}  \bvol{61},  \pg{145--157}.

\bibitem[Li {\em et~al.\/}(2020)Li, Liu, Kaufman, Turetcaia, Chen \&
  Cardenas]{li2020flexible}
{\sc \au{Li, Bing}, \au{Liu, Xiaofeng}, \au{Kaufman, Matthew~H}, \au{Turetcaia,
  Anna}, \au{Chen, Xingyuan} \& \au{Cardenas, M~Bayani}} \yr{2020}
  \at{Flexible and modular simultaneous modeling of flow and reactive transport
  in rivers and hyporheic zones}.  \jt{Water Resources Research}
  \bvol{56}~(2),  \pg{e2019WR026528}.

\bibitem[Ma {\em et~al.\/}(2021)Ma, Alam{\'e} \& Mahesh]{ma2021direct}
{\sc \au{Ma, Rong}, \au{Alam{\'e}, Karim} \& \au{Mahesh, Krishnan}} \yr{2021}
  \at{Direct numerical simulation of turbulent channel flow over random rough
  surfaces}.  \jt{Journal of Fluid Mechanics}  \bvol{908}.

\bibitem[Manes {\em et~al.\/}(2009)Manes, Pokrajac, McEwan \&
  Nikora]{manes2009turbulence}
{\sc \au{Manes, Costantino}, \au{Pokrajac, Dubravka}, \au{McEwan, Ian} \&
  \au{Nikora, Vladimir}} \yr{2009}  \at{Turbulence structure of open channel
  flows over permeable and impermeable beds: A comparative study}.  \jt{Physics
  of Fluids}  \bvol{21}~(12),  \pg{125109}.

\bibitem[Mignot {\em et~al.\/}(2009)Mignot, Barth{\'e}lemy \&
  Hurther]{mignot2009double}
{\sc \au{Mignot, Emmanuel}, \au{Barth{\'e}lemy, Eric} \& \au{Hurther, David}}
  \yr{2009}  \at{Double-averaging analysis and local flow characterization of
  near-bed turbulence in gravel-bed channel flows}.  \jt{Journal of Fluid
  Mechanics}  \bvol{618},  \pg{279--303}.

\bibitem[Mittal(1999)]{mittal1999fourier}
{\sc \au{Mittal, Rajat}} \yr{1999}  \at{A fourier--chebyshev spectral
  collocation method for simulating flow past spheres and spheroids}.
  \jt{International journal for numerical methods in fluids}  \bvol{30}~(7),
  \pg{921--937}.

\bibitem[Mittal {\em et~al.\/}(2008)Mittal, Dong, Bozkurttas, Najjar, Vargas \&
  Von~Loebbecke]{mittal2008versatile}
{\sc \au{Mittal, Rajat}, \au{Dong, Haibo}, \au{Bozkurttas, Meliha}, \au{Najjar,
  FM}, \au{Vargas, Abel} \& \au{Von~Loebbecke, Alfred}} \yr{2008}  \at{A
  versatile sharp interface immersed boundary method for incompressible flows
  with complex boundaries}.  \jt{Journal of computational physics}
  \bvol{227}~(10),  \pg{4825--4852}.

\bibitem[Moser {\em et~al.\/}(1999)Moser, Kim \& Mansour]{moser1999direct}
{\sc \au{Moser, Robert~D}, \au{Kim, John} \& \au{Mansour, Nagi~N}} \yr{1999}
  \at{Direct numerical simulation of turbulent channel flow up to re $\tau$=
  590}.  \jt{Physics of fluids}  \bvol{11}~(4),  \pg{943--945}.

\bibitem[Nikora {\em et~al.\/}(2002)Nikora, Koll, McLean, Ditrich \&
  Aberle]{nikora2002zero}
{\sc \au{Nikora, Vladimir~Ivanovich}, \au{Koll, K}, \au{McLean, SR},
  \au{Ditrich, A} \& \au{Aberle, J}} \yr{2002}  \at{Zero-plane displacement for
  rough-bed open-channel flows.}  \bt{In {\em Fluvial Hydraulics River Flow
  2002\/}},  \pg{pp. 83--92}.

\bibitem[{\"O}rl{\"u} \& Schlatter(2011)]{orlu2011fluctuating}
{\sc \au{{\"O}rl{\"u}, Ramis} \& \au{Schlatter, Philipp}} \yr{2011}  \at{On the
  fluctuating wall-shear stress in zero pressure-gradient turbulent boundary
  layer flows}.  \jt{Physics of fluids}  \bvol{23}~(2),  \pg{021704}.

\bibitem[Panton {\em et~al.\/}(2017)Panton, Lee \&
  Moser]{panton2017correlation}
{\sc \au{Panton, Ronald~L}, \au{Lee, Myoungkyu} \& \au{Moser, Robert~D}}
  \yr{2017}  \at{Correlation of pressure fluctuations in turbulent wall
  layers}.  \jt{Physical Review Fluids}  \bvol{2}~(9),  \pg{094604}.

\bibitem[Raupach {\em et~al.\/}(1991)Raupach, Antonia \&
  Rajagopalan]{raupach1991rough}
{\sc \au{Raupach, MR}, \au{Antonia, RA} \& \au{Rajagopalan, S}} \yr{1991}
  \at{Rough-wall turbulent boundary layers}.  \jt{Applied Mechanics Reviews}
  \bvol{44}~(1),  \pg{1}.

\bibitem[Raupach \& Thom(1981)]{raupach1981turbulence}
{\sc \au{Raupach, MR} \& \au{Thom, A~St}} \yr{1981}  \at{Turbulence in and
  above plant canopies}.  \jt{Annual Review of Fluid Mechanics}  \bvol{13}~(1),
   \pg{97--129}.

\bibitem[Raupach \& Shaw(1982)]{raupach1982averaging}
{\sc \au{Raupach, Michael~R} \& \au{Shaw, RH}} \yr{1982}  \at{Averaging
  procedures for flow within vegetation canopies}.  \jt{Boundary-Layer
  Meteorology}  \bvol{22}~(1),  \pg{79--90}.

\bibitem[Shen {\em et~al.\/}(2020)Shen, Yuan \& Phanikumar]{shen2020direct}
{\sc \au{Shen, Guangchen}, \au{Yuan, Junlin} \& \au{Phanikumar, Mantha~S}}
  \yr{2020}  \at{Direct numerical simulations of turbulence and hyporheic
  mixing near sediment--water interfaces}.  \jt{Journal of Fluid Mechanics}
  \bvol{892}.

\bibitem[Singh {\em et~al.\/}(2007)Singh, Sandham \&
  Williams]{singh2007numerical}
{\sc \au{Singh, KM}, \au{Sandham, ND} \& \au{Williams, JJR}} \yr{2007}
  \at{Numerical simulation of flow over a rough bed}.  \jt{Journal of Hydraulic
  Engineering}  \bvol{133}~(4),  \pg{386--398}.

\bibitem[Valett {\em et~al.\/}(1996)Valett, Morrice, Dahm \&
  Campana]{valett1996parent}
{\sc \au{Valett, H~Maurice}, \au{Morrice, John~A}, \au{Dahm, Clifford~N} \&
  \au{Campana, Michael~E}} \yr{1996}  \at{Parent lithology,
  surface--groundwater exchange, and nitrate retention in headwater streams}.
  \jt{Limnology and oceanography}  \bvol{41}~(2),  \pg{333--345}.

\bibitem[Voermans {\em et~al.\/}(2017)Voermans, Ghisalberti \&
  Ivey]{voermans2017variation}
{\sc \au{Voermans, JJ}, \au{Ghisalberti, M} \& \au{Ivey, GN}} \yr{2017}
  \at{The variation of flow and turbulence across the sediment--water
  interface}.  \jt{Journal of Fluid Mechanics}  \bvol{824},  \pg{413--437}.

\bibitem[Voermans {\em et~al.\/}(2018)Voermans, Ghisalberti \&
  Ivey]{voermans2018model}
{\sc \au{Voermans, Joey~J}, \au{Ghisalberti, Marco} \& \au{Ivey, Gregory~N}}
  \yr{2018}  \at{A model for mass transport across the sediment-water
  interface}.  \jt{Water Resources Research}  \bvol{54}~(4),  \pg{2799--2812}.

\bibitem[Williams \& Philipse(2003)]{williams2003random}
{\sc \au{Williams, SR} \& \au{Philipse, AP}} \yr{2003}  \at{Random packings of
  spheres and spherocylinders simulated by mechanical contraction}.
  \jt{Physical Review E}  \bvol{67}~(5),  \pg{051301}.

\bibitem[Wilson {\em et~al.\/}(2008)Wilson, Huettel \& Klein]{wilson2008grain}
{\sc \au{Wilson, Alicia~M}, \au{Huettel, Markus} \& \au{Klein, Stephen}}
  \yr{2008}  \at{Grain size and depositional environment as predictors of
  permeability in coastal marine sands}.  \jt{Estuarine, Coastal and Shelf
  Science}  \bvol{80}~(1),  \pg{193--199}.

\bibitem[Zagni \& Smith(1976)]{zagni1976channel}
{\sc \au{Zagni, Anthony~FE} \& \au{Smith, Kenneth~VH}} \yr{1976}  \at{Channel
  flow over permeable beds of graded spheres}.  \jt{Journal of the hydraulics
  division}  \bvol{102}~(2),  \pg{207--222}.

\bibitem[Zhou {\em et~al.\/}(1999)Zhou, Adrian, Balachandar \&
  Kendall]{zhou1999mechanisms}
{\sc \au{Zhou, Jigen}, \au{Adrian, Ronald~J}, \au{Balachandar,
  Sivaramakrishnan} \& \au{Kendall, TM1693393}} \yr{1999}  \at{Mechanisms for
  generating coherent packets of hairpin vortices in channel flow}.
  \jt{Journal of fluid mechanics}  \bvol{387},  \pg{353--396}.

\bibitem[Zippe \& Graf(1983)]{zippe1983turbulent}
{\sc \au{Zippe, Hans~J} \& \au{Graf, Walter~H}} \yr{1983}  \at{Turbulent
  boundary-layer flow over permeable and non-permeable rough surfaces}.
  \jt{Journal of Hydraulic research}  \bvol{21}~(1),  \pg{51--65}.

\end{thebibliography}

\end{document}